\documentclass[sigconf, nonacm]{acmart}
\usepackage{amsmath,amsfonts}
\usepackage{algorithmic}
\usepackage{graphicx}
\usepackage{textcomp}
\usepackage{xcolor}
\usepackage{booktabs}
\usepackage{xcolor}
\usepackage{pgfplots}
\usepackage{subcaption}
\usepackage[usenames, dvipsnames]{xcolor}
\usepackage{colortbl}
\usepackage{bbding}
\usepackage{braket}
\usepackage{epsfig,latexsym,graphics}
\usepackage[font={small}]{caption}
\usepackage{setspace}
\usepackage{enumitem}
\pgfplotsset{compat=1.16}
\usepackage{cmd}
\usepackage{svg}
\usepackage[spacesep,definevectors]{easyvector}

\newcommand\vldbdoi{XX.XX/XXX.XX}
\newcommand\vldbpages{XXX-XXX}
\newcommand\vldbvolume{19}
\newcommand\vldbissue{4}
\newcommand\vldbyear{2025}

\newcommand\vldbtitle{\shorttitle} 
\newcommand\vldbavailabilityurl{https://github.com/mingyu-hkustgz/LabelANN}
\newcommand\vldbpagestyle{plain} 

\begin{document}
\title{Elastic Index Selection for Label-Hybrid AKNN Search}

\author{Mingyu Yang}
\affiliation{%
\institution{HKUST (GZ) \& HKUST}
}
\email{myang250@connect.hkust-gz.edu.cn}

\author{Wenxuan Xia}
\affiliation{%
\institution{HKUST (GZ)}
}
\email{wxia248@connect.hkust-gz.edu.cn}

\author{Wentao Li}
\affiliation{%
\institution{University of Leicester}
}
\email{wl226@leicester.ac.uk}

\author{Raymond Chi-Wing Wong}
\affiliation{%
\institution{HKUST
}
}
\email{raywong@cse.ust.hk}

\author{Wei Wang}
\authornote{Corresponding author.}
\affiliation{%
\institution{HKUST (GZ) \& HKUST}
}
\email{weiwcs@ust.hk}

\begin{abstract}
Real-world vector embeddings often carry additional label attributes, such as keywords and tags.
In this context, \textbf{label-hybrid approximate $k$-nearest neighbor (AKNN)} search retrieves the top-$k$ approximate nearest vectors to a query, subject to the constraint that their labels fully contain the query-label set.
A naive solution builds a separate index for every query-label set, but the exponential growth of such sets makes this approach storage-prohibitive. 
To overcome this, we propose selectively indexing only a subset of query-label sets while still ensuring efficient processing for all queries.
This is made possible by a key insight into label containment: an index built for a label set $L$ can also serve any query whose label set $L'$ is a superset of $L$, with query cost bounded by the elastic factor: the ratio between the number of vectors matching $L$ and those matching $L'$.
We formalize the index-selection task as a constrained optimization problem that chooses which label sets to index to satisfy space and query efficiency constraints.
We prove the problem is NP-complete and propose efficient greedy algorithms for its efficiency- and space-constrained variants.
Extensive experiments on real-world datasets show that our method achieves 10$\times$–800$\times$ speedups over state-of-the-art techniques.
Moreover, our approach is index-agnostic and can be seamlessly integrated into existing vector database systems.
\end{abstract}

\maketitle

\pagestyle{\vldbpagestyle}
\begingroup\small\noindent\raggedright\textbf{PVLDB Reference Format:}\\
Mingyu Yang, Wenxuan Xia, Wentao Li, Raymond Chi-Wing Wong, Wei Wang. \vldbtitle. PVLDB, \vldbvolume(\vldbissue): \vldbpages, \vldbyear.\\
\href{https://doi.org/\vldbdoi}{doi:\vldbdoi}
\endgroup
\begingroup
\renewcommand\thefootnote{}\footnote{\noindent
This work is licensed under the Creative Commons BY-NC-ND 4.0 International License. Visit \url{https://creativecommons.org/licenses/by-nc-nd/4.0/} to view a copy of this license. For any use beyond those covered by this license, obtain permission by emailing \href{mailto:info@vldb.org}{info@vldb.org}. Copyright is held by the owner/author(s). Publication rights licensed to the VLDB Endowment. \\
\raggedright Proceedings of the VLDB Endowment, Vol. \vldbvolume, No. \vldbissue\ %
ISSN 2150-8097. \\
\href{https://doi.org/\vldbdoi}{doi:\vldbdoi} \\
}\addtocounter{footnote}{-1}\endgroup

\ifdefempty{\vldbavailabilityurl}{}{
\vspace{.3cm}
\begingroup\small\noindent\raggedright\textbf{PVLDB Artifact Availability:}\\
The source code, data, and/or other artifacts have been made available at \url{https://github.com/mingyu-hkustgz/LabelANN}.
\endgroup
}

\section{Introduction}\label{sec:intro}
The \textbf{$k$-nearest neighbor (KNN)} search over high-dimensional vectors is a core operation in modern data systems, underpinning a wide range of applications such as recommendation systems~\cite{CF-2007-recommender-sys}, data mining~\cite{NN-datamining-1967-TIT}, face recognition~\cite{AnalyticDB-VLDB-2020}, product search~\cite{AnalyticDB-VLDB-2020}, and retrieval-augmented generation (RAG) for large language models~\cite{LLM-RAG-NIPS-2020}.
In real-world deployments, vector embeddings are often associated with \textbf{label attributes}, such as product brands, keywords, and geolocations. 
For example, in an e-commerce setting, a user may search for items similar to a given photo while specifying label constraints like brand name and release year.

\begin{figure}[!t]
    \centering
    \includegraphics[width=0.9\columnwidth]{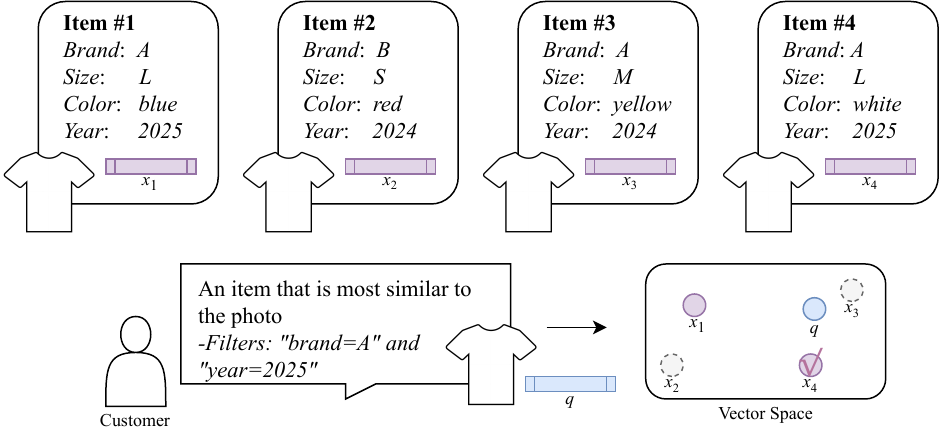}
    \vspace{-2ex}\caption{Example of label-hybrid AKNN search in an online shopping scenario ($k=1$).  
    Each item is represented by an embedding vector $x_i$ ($i \in [1,4]$) and associated with label attributes such as brand and year.  
    A customer submits a reference photo $q$ together with a query-label set $L_q$.  
    The task is to find the item most similar to $q$ that also satisfies the label constraint.
    Thus, $x_2$ and $x_3$ are discarded for label mismatch, and $x_4$ is returned as the nearest neighbor.}\vspace{-2ex}
    \label{fig:example}
\end{figure}

In this context, we introduce the problem of \textbf{label-hybrid $k$-nearest neighbor (KNN)} search, illustrated in Fig.~\ref{fig:example}. 
Unlike traditional KNN search, this variant considers both the vector embeddings and their associated label attributes.   
Formally, each entry in the database $S$ consists of a vector embedding and a label set $L$. 
Given a query vector $q$ and the label constraint (expressed as a query-label set $L_q$), the goal is to retrieve the top-$k$ nearest neighbors of $q$ in $S$, where each returned vector with label set $L$ must contain all labels in the query-label set $L_q$, i.e., $L_q \subseteq L$.
However, due to the curse of dimensionality~\cite{Curse-of-dim-1998}, computing exact label-hybrid KNN in high-dimensional spaces is computationally expensive. 
Recent work therefore studies the \textbf{label-hybrid approximate KNN (AKNN)} problem, which efficiently returns the top-$k$ approximate neighbors satisfying the label constraint, trading a small amount of accuracy for substantial performance gains~\cite{UNG-SIGMOD-2025,Acorn-SIGMOD-2024}.

\stitle{Existing Solutions.}
To support label-hybrid AKNN search, existing approaches often employ graph-based indexes~\cite{HNSW-PAMI-2020,Diskann-NIPS-2019,Filtered-diskann-WWW-2023,NSG-VLDB-2019-deng-cai,SSG-PAMI-2022-deng-cai,HVS-VLDB-2021-kejing-lu,NHQ-NIPS-2022-mengzhao-wang,SymphonyQG-SIGMOD-2025,FANNG:harwood2016fanng,tMRNG:journals/pacmmod/PengCCYX23,ANNSurvey-TKDE-2020-Wei-Wang,Graph-ANNS-Survey-VLDB-2021-mengzhao,DEG-SIGMOD-2025} combined with filter-based search strategies, due to their good search efficiency.
Specifically, the graph-based index is constructed by adding base vectors in database $S$ as nodes and connecting each node to its nearby neighbors to form edges, which are carefully selected to support efficient navigation.
At query time, on top of the graph-based index, filter-based search strategies~\cite{Filtered-diskann-WWW-2023} are applied to enforce label constraints and retrieve the final results.
In particular, two strategies—$\PRE$ and $\POST$—are integrated into the graph-based index.
The $\PRE$ strategy prunes nodes (i.e., base vectors) that do not satisfy the query label constraints, removing both the nodes and their connections to neighbors during the search.
In contrast, the $\POST$ strategy retains all nodes and connections, but excludes any label-mismatched nodes from being added to the results.

A base vector \textbf{matches} the query if its label set contains the query-label set, and the selectivity of a query is the number of its label-matched vectors.
The performance of both strategies drops sharply when selectivity is low.
(1) $\PRE$ suffers from reduced accuracy as removing many label-mismatched nodes sparsifies the graph, making it difficult to reach the target vectors.
(2) While $\POST$ retains all nodes, it still requires computing distances between the query and many mismatched vectors, causing inefficiency.  
To address these challenges, existing solutions such as Milvus~\cite{Milvus-SIGMOD-2021}, ADB~\cite{AnalyticDB-VLDB-2020}, VBASE~\cite{Vbase-OSDI-2023}, and CHASE~\cite{CHASE-arxiv-Sean-Wang} dynamically select between $\PRE$ and $\POST$ based on the query workload and cost estimation. Yet, the limitations of these search strategies persist.
Some heuristic approaches, including $\NHQ$~\cite{NHQ-NIPS-2022-mengzhao-wang} and HQANN~\cite{HQANN-CIKM-2022}, propose fusion distances that combine both vector similarity and label matching between the query and base vectors.
These methods require manual tuning of vector–label weight parameters and still perform well below the state-of-the-art.

\begin{figure}[!t]
    \centering
    \includegraphics[width=0.95\columnwidth]{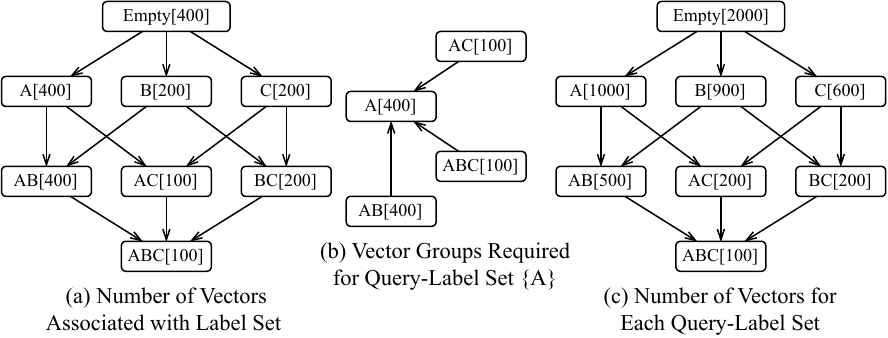}
    \caption{
    Motivating example where database $S$ and queries draw labels from the alphabet $\{A,B,C\}$.  
    Base vectors in $S$ are grouped by their associated label sets $L$, each group denoted as $\{L\}$[\textit{count}].  
    Arrows connect a group $L$ to groups with its minimal supersets.  
    For instance, Fig.~\ref{fig:motivate-exmaple}(a) shows $400$ vectors labeled only $\{A\}$, written as $A[400]$.  
    Given a AKNN query with $L_q=\{A\}$, Fig.~\ref{fig:motivate-exmaple}(b) highlights all relevant groups---those whose label sets are supersets of $\{A\}$---namely $\{A,AB,AC,ABC\}$, totaling $1,000$ matching vectors that must be indexed under $L_q=\{A\}$.  
    Fig.~\ref{fig:motivate-exmaple}(c) summarizes the index sizes required for each query-label set.  
    Supporting all possible query-label sets in the workload requires indexing $5,400$ vectors---about $2.75\times$ the original dataset size.}\vspace{-4ex}
    \label{fig:motivate-exmaple}
\end{figure}

\stitle{State-of-the-Art.}
To support label-hybrid AKNN search, $\ACORN$~\cite{Acorn-SIGMOD-2024} and $\UNG$~\cite{UNG-SIGMOD-2025} represent the state-of-the-art.
$\ACORN$ extends the $\PRE$ strategy to address connectivity issues under low selectivity by introducing a parameter $\gamma$ that increases the graph density---each node has $\gamma$ times more outgoing edges than in a standard index.
This denser graph improves robustness when filtering out label-mismatched nodes during query processing.
However, $\ACORN$ ignores label information during index construction, leading to potential result incompleteness.

To address this, $\UNG$ exploits label containment during graph construction.  
It partitions base vectors into groups by their label sets and builds a subgraph for each group.  
Subgraphs for a label set $L_q$ are connected to those of its minimal supersets via cross-group edges, ensuring that any vector with a superset of $L_q$ is reachable from the group of $L_q$, thereby guaranteeing completeness.  
However, both $\ACORN$ and $\UNG$ are limited to graph-based indexes and thus lack \emph{index flexibility}.  
Their performance also degrades when query selectivity is low.  
Experiments show significant deterioration as the size of the alphabet grows, and neither method can \emph{flexibly adjust the index structure} to meet tight storage constraints.

\stitle{Motivation.}
Given an alphabet $\Sigma$ from which both the database $S$ and queries draw labels, we may face an exponential explosion of possible label sets.
Specifically, let $\mathcal{L}$ be the set of query-label sets in the workload (a subset of the power set of $\Sigma$).  
A naive strategy for AKNN queries is to build a separate index for each query-label set $L_q \in \mathcal{L}$, where the index for $L_q$ stores all vectors whose label sets contain $L_q$.  
Under this scheme, a vector with label set $L$ must be inserted into every index whose query-label set is a subset of $L$.  
As illustrated in Fig.~\ref{fig:motivate-exmaple}, a vector labeled $\{ABC\}$ must appear in eight indexes: $\{\emptyset, A, B, C, AB, AC, BC, ABC\}$, where $\emptyset$ denotes the top index with no label constraint ($L_q=\emptyset$).
This replication creates substantial overhead: empirical studies show that for $|\Sigma|=6$–$10$, the total index entries can be $64\times$–$1024\times$ larger than the original database, leading to prohibitive indexing and storage costs.

\begin{table}[!t]
\begin{small}
  \centering
\caption{We compare our method with existing solutions along three dimensions.  
\textbf{Search Performance} measures accuracy and efficiency, validated experimentally.  
\textbf{Index Flexibility} captures the ability to support label-hybrid AKNN search without dependence on a specific index structure.  
\textbf{Space–Performance Trade-offs} assess how well the method balances indexing cost and storage efficiency.
}\vspace{-2ex}
\resizebox{8.5cm}{!}{
\begin{tabular}{l|c|cccc}
    \toprule
    Feature         & $\ELI$ (Our) & $\NHQ$ & $\UNG$ & $\ACORN$ & Filtered  \\ 
    \midrule
    Search Performance      & $\star\star\star$  & $\star$      & $\star^\uparrow$  & $\star^\uparrow$         & $\star$     \\ 
    Index Flexibility       & \checkmark         & $\times$     & $\times$          & $\times$                 & \checkmark  \\ 
    Space–Performance Trade-offs     & \checkmark         & $\times$     & $\times$          & $\times$  & $\times$ \\ 
    \bottomrule
  \end{tabular}}
  \label{tab:features}
\end{small}
\end{table}

\stitle{Our Solution.}
Indexing every label set in $\mathcal{L}$ is impractical due to the sheer number of indexes and the resulting storage cost.  
A natural question arises: can we \emph{index only a subset of} $\mathcal{L}$ while still enabling efficient label-hybrid AKNN search?  
This is non-trivial---if a query-label set $L\in\mathcal{L}$ is not indexed, queries with label set $L$ may be inefficient to process.  
Fortunately, label containment provides a key insight: an index built for $L$ can also serve any query whose label set $L'$ is a superset of $L$.  
We further show that the query latency is bounded by an \emph{elastic factor}---the ratio of vectors matching $L'$ to those matching $L$.  
Note that a higher elastic factor implies better search efficiency.
For example, in Fig.~\ref{fig:motivate-exmaple}(c), all vectors matching $\{AB\}$ are contained in those matching $\{A\}$, with an overlap ratio (or elastic factor) of at least $0.5$.  
Thus, the index on $\{A\}$ can efficiently answer queries for $\{AB\}$.  
The index on $\{B\}$ has an even higher overlap with $\{AB\}$, yielding greater efficiency.
We therefore formalize the problem of \textbf{index selection} as choosing a subset of $\mathcal{L}$ to index while bounding both space usage and the elastic factor.

We first show that the decision version of the index-selection problem is NP-complete.  
We then formulate two optimization variants:  
(1) an \textbf{efficiency-constrained} version that minimizes index space while guaranteeing a lower bound on the elastic factor (search efficiency), and  
(2) a \textbf{space-constrained} version that maximizes the elastic factor under a fixed space budget.  
We propose a greedy algorithm for the first variant and extend it to the second, and we further describe how to update indexes in dynamic settings.  
Overall, our methods selectively index label sets to control space while maintaining high elastic factors for efficient label-hybrid AKNN search.  
Compared with existing approaches (see Table~\ref{tab:features}), our solution provides greater flexibility, higher search efficiency, and superior space–performance trade-offs.

\stitle{Contribution.}
We summarize our main contributions as follows:

\sstitle{Problem Analysis (\S~\ref{sec:problem}).}
We identify the limitations of existing label-hybrid AKNN solutions: limited flexibility in the underlying indexes and suboptimal query performance.  
To address these issues, we measure search performance using the \emph{elastic factor}, which quantifies how an index built on a query-label set $L$ can support queries over its supersets, motivating our novel index-selection problem.

\sstitle{Index Selection Problem Formulation (\S~\ref{sec:problem}).}
We formally define the decision version of the index-selection problem: given a query workload with label sets $\mathcal{L}$, can we choose a subset of $\mathcal{L}$ so that the total index space stays within a specified bound and the elastic factor exceeds a given threshold?  
We prove this problem is NP-complete, showing that exact solutions are intractable.

\sstitle{Index Selection Problem Solutions (\S~\ref{sec:elastic-index}).}
Building on the decision version, we define two practical optimization variants of the index selection problem: (1) the efficiency-constrained variant, which minimizes space given a lower bound on the elastic factor; and (2) the space-constrained variant, which maximizes the elastic factor under a space limit. 
These formulations enable users to balance space and query efficiency.
We design a greedy algorithm for the first variant and extend it to solve the second.

\sstitle{Extensive Experiments (\S~\ref{sec:exp}).}
We evaluate our algorithm on multiple datasets with diverse label distributions associated with the base vectors.  
Experiments show that our approach attains near-optimal search efficiency with only a $100\%$ space overhead and outperforms competitors on large-scale datasets with large alphabets, achieving $10\times$–$800\times$ speedups over state-of-the-art baselines.

Due to space constraints, some proofs and experiments are omitted, which can be found in our appendix.

\section{Preliminary}\label{sec:preliminary}
We formally define the label-hybrid approximate $k$-nearest neighbor search problem in \S~\ref{subsec:defn}.
We then review existing solutions in \S~\ref{subsec:index}.  We list the commonly used notations in the table~\ref{tab:notation}.

\subsection{Problem Statement}\label{subsec:defn}
Each entry $(x_i, L_i)$ in the dataset $S$ consists of a $d$-dimensional base vector $x_i \in \mathbb{R}^d$ and an associated label set $L_i$, which may be empty, where all possible labels are from the alphabet $\Sigma$.
A label-hybrid search is denoted by a query $(q, L_q)$, where $q$ is the query vector and $L_q$ is the query-label set.
When $L_q$ is specified, we first filter the dataset $S$ to obtain a candidate subset $S(L_q) \subseteq S$, defined as: $S(L_q) = \{(x_i, L_i) \in S \mid L_q \subseteq L_i \}$, i.e., the label-matched vectors whose label sets contain all labels in $L_q$.
The label-hybrid $k$-nearest neighbor search is performed over this candidate subset.
Formally,

\begin{definition}[Label-Hybrid KNN Search]
    Given a dataset $S$ and a query $(q, L_q)$, label-hybrid KNN search returns a set $S' \subseteq S(L_q)$ of $k$ entries such that for every $(x_i, L_i) \in S'$ and every $(x_j, L_j) \in S(L_q)$,  $\delta(x_i, q) \le \delta(x_j, q)$, where $\delta(\cdot,\cdot)$ denotes the vector distance (We use the Euclidean distance as an example).
\end{definition}

Exact $k$-nearest neighbor search often suffers from the \textit{curse of dimensionality}~\cite{Curse-of-dim-1998}, causing traditional methods~\cite{Cover-Tree-2006-ICML,R*-tree-SIGMOD-1990,SA-tree-VLDBJ-2002-navarro} that perform well in low-dimensional spaces to degrade greatly in high-dimensional settings. 
As a result, approximate $k$-nearest neighbor (AKNN) search has been extensively studied~\cite{LSH-1999,RPLSH-1995,LSH-Pstable-2004,PMLSH-bolong-2020,ADSampling:journals/sigmod/GaoL23,PQ-fast-scan-VLDB-2015,OPQ-PAMI-2014,LOPQ-2014-CVPR,DeltaPQ-VLDB-2020,HNSW-PAMI-2020,Rabitq-SIGMOD-2024,NSG-VLDB-2019-deng-cai,SSG-PAMI-2022-deng-cai,Starling-SIGMOD-2024-Mengzhao,Diskann-NIPS-2019,ANNSurvey-TKDE-2020-Wei-Wang,MRQ-RESQ-arxiv-2024-Mingyu,BSA-DDC-ICDE-2024-Mingyu} for its ability to improve efficiency at the cost of some accuracy.
This challenge also arises in the context of label-hybrid search. Therefore, we focus on the label-hybrid approximate $k$-nearest neighbor (AKNN) problem. 
To evaluate the result quality, we use \textbf{recall} as the metric, defined as $\text{recall} = |\hat{S} \cap S'| / |S'|$, where $S'$ is the exact result set and $\hat{S}$ is the approximate result set with $|\hat{S}|=k$. 
An effective label-hybrid AKNN solution should balance high search efficiency with high recall.

\begin{table}[!t]
  \caption{Summary of Notations}\label{tab:notation}\vspace{-2ex}
  \small
  \begin{tabular*}{\linewidth}{@{\extracolsep{\fill}} p{18mm} | p{68mm}}
    \toprule
    Notation & Description \\
    \midrule
    $S$            & Set of vectors, each with an associated label set \\
    $L_i, L_q$     & Base and query label sets \\
    $S(L_q)$       & Subset of $S$ whose label sets contain $L_q$ \\
    $|L|$          & Cardinality of label set $L$ \\
    $\mathcal{L}$  & Collection of all label sets \\
    $|\mathcal{L}|$& Number of label sets in $\mathcal{L}$ \\
    $\mathcal{I}$  & Collection of indexes \\
    $\mathbb{L}$   & The selected indexes from $\mathcal{L}$ \\
    $\Sigma$       & Label alphabet (all possible labels) \\
    $N$            & Cardinality of $S$ \\
    $q$            & Query vector \\
    $e$            & Elastic factor \\
    $d$            & Dimensionality of vectors in $S$ \\
    $\delta(u,v)$  & Distance (or similarity) between $u$ and $v$ \\
    $I$            & AKNN search index (e.g., $\HNSW$) \\
    $\tau$         & Space budget \\
    $c$            & Elastic-factor threshold \\
    \bottomrule
  \end{tabular*}
\end{table}

\stitle{Remark.}
Label-hybrid AKNN search can be viewed as a special case of filtered nearest-neighbor search~\cite{Filtered-diskann-WWW-2023},  
where our filtering requires the label set $L_i$ of a base vector to contain the query-label set $L_q$.  
Other variants---such as set intersection ($L_q \cap L_i \neq \emptyset$) or set equality ($L_q = L_i$)---can be addressed using the solutions developed for our studied AKNN search problem.
A detailed analysis appears in our appendix.

\subsection{Existing Solutions}\label{subsec:index}
To address the label-hybrid AKNN search problem, existing approaches typically combine different types of AKNN indexes with filter-based search strategies. Among these, graph-based indexes~\cite{HNSW-PAMI-2020,FANNG:harwood2016fanng,NSG-VLDB-2019-deng-cai,tMRNG:journals/pacmmod/PengCCYX23,Diskann-NIPS-2019,HVS-VLDB-2021-kejing-lu} are widely adopted due to their state-of-the-art search efficiency. In such indexes, base vectors in the dataset $S$ are treated as nodes in a graph, where each node is carefully connected to its nearby vectors, enabling efficient navigation.

\stitle{Graph Index.}
When a graph is used as an index, search starts from a designated entry node and iteratively explores neighbors that are progressively closer to the query until the top-1 result is found.  
A key index design component is the edge-occlusion strategy, which guides the search toward the query while keeping the graph sparse and the node degree bounded by a constant~\cite{NSG-VLDB-2019-deng-cai,Worse-case-NIPS-2023}. 
For AKNN search---where more than one result is required---beam search is typically employed.  
It maintains the top-$m$ closest candidates during traversal, where $m$ is the beam width.  
Under ideal indexing conditions (i.e., high index quality and the query present in the dataset)~\cite{NSG-VLDB-2019-deng-cai,Worse-case-NIPS-2023}, only one extra expansion step on the graph is needed to retrieve the next neighbor, yielding an overall search complexity logarithmic in the dataset size.

\begin{figure}[!t]
    \centering
    \includegraphics[width=0.45\columnwidth]{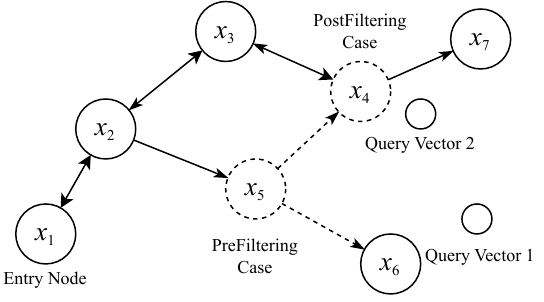}
    \caption{The Example of Filter-Based Search}
    \label{fig:graph-search}
\end{figure}

\begin{figure*}[!thb]
    \centering
    \includegraphics[width=1.8\columnwidth]{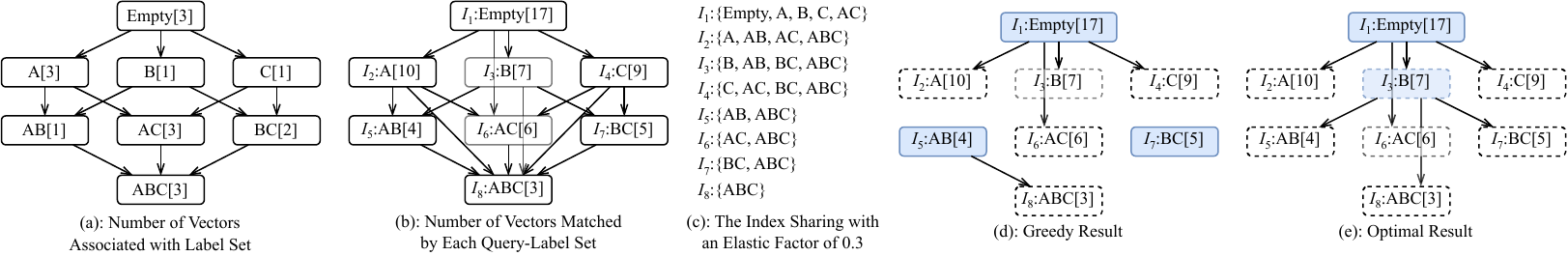}
    \caption{The Running Example of Elastic Factor and Greedy Methods}\label{fig:run-example}
\end{figure*}

\stitle{Filter-Based Search.}
On top of the graph index, we introduce two filter-based search strategies—$\PRE$ and $\POST$---to support label-hybrid AKNN search. 
Both strategies maintain the original index structure and require only label-based filtering during graph traversal.
(1) $\PRE$ strategy: Given a query $(q, L_q)$, $\PRE$ prunes unmatched nodes and their neighbors during traversal. As shown in Fig.~\ref{fig:graph-search}, suppose nodes $x_4$ and $x_5$ do not satisfy the label constraint in query~1. 
In this case, $\PRE$ removes the outgoing edges of $x_5$, thereby making the true nearest neighbor $x_6$ unreachable from the entry node $x_1$.
(2) $\POST$ strategy: This method allows traversal through unmatched nodes but excludes them from the result set. 
For instance, when processing query~2 in Fig.~\ref{fig:graph-search}, $\POST$ visits nodes $x_1, x_2, x_3, x_4,$ and $x_7$. 
Although $x_4$ is label-unmatched, its outgoing edge is still used for routing, enabling the algorithm to eventually reach and return $x_7$.

For top-$k$ queries, $\POST$ traverses incrementally, visiting successive nearest neighbors until $k$ matching results are found.
However, if most nodes are label-mismatched, the algorithm may still need to examine up to $O(N)=O(|S|)$ nodes to obtain $k$ results, because these filtered-out nodes must still be visited for $\POST$.
$\PRE$ faces a similar challenge---under low selectivity, it may fail to reach any valid neighbors from the entry node, making its correctness and efficiency difficult to guarantee.

\section{Elastic Index Selection Problem}\label{sec:problem}
This section formally defines the index selection problem, aiming to overcome the exhaustive cost in indexing space and time caused by exponential label combinations. 
In particular, we first introduce the elastic factor, which captures how an index built on a query-label set can be reused for its supersets. 
We then formally define the index selection problem as a decision problem: determining whether it is possible to select a subset of label sets for indexing such that both space usage and query efficiency remain within given bounds. 
Finally, we prove that the problem is computationally hard.

\subsection{Elastic Factor for Index-Sharing}\label{subsec:index-share}

Given an alphabet $\Sigma$, one could in principle build an index for every label set in the power set of $\Sigma$, but the number of such sets is exponential.
Fortunately, index sharing can dramatically reduce the required indexes.  
Specifically, an index built on a query-label set $L_2$ can serve as a substitute for another query-label set $L_1$ if it contains all base vectors associated with $L_1$.
This holds when $L_2 \subseteq L_1$, because every vector matching $L_1$ also matches its subset $L_2$.
For example, the group labeled $\{A\}$ includes all entries from its superset group $\{AB\}$, since any vector labeled $\{AB\}$ also contains label $\{A\}$.
Thus, a query with label set $\{AB\}$ can be answered using an index built on $\{A\}$, allowing indexes for some query-label sets to be omitted and reducing storage costs using label containment.

\stitle{Elastic Factor.}
We next analyze the query efficiency provided by shared indexes.  
For simplicity, assume we already have a subset of indexes $\mathbb{I} = \{I_1,\ldots,I_m\}$ built on a selected subset of possible query-label sets (we discuss how to choose this subset later).  
Each index $I_i \in \mathbb{I}$ contains the base vectors in $S$ that match some query-label set $L_i$; we sometimes use $I_i$ to refer directly to these label-matched vectors.
An index can answer a query with label set $L_q$ only if it contains all base vectors matching $L_q$, following the label-containment rule described earlier. 
Intuitively, if the set $S(L_q)$ of vectors matching the query-label set $L_q$ is fully contained in some index $I \in \mathbb{I}$ and overlaps heavily with $I$, query efficiency is high because few label-mismatched vectors remain.  
This intuition motivates the \emph{elastic factor}: the maximum overlap ratio, across all indexes in $\mathbb{I}$ that can answer query $q$, between $S(L_q)$ and the index used.

\begin{definition}[Elastic Factor]
Given a label-hybrid dataset $S$, a query $(q, L_q)$, and an index set $\mathbb{I}=\{I_1,\dots,I_m\}$ where each $I_i \subseteq S$, the \emph{elastic factor} of $\mathbb{I}$ with respect to $L_q$ is defined as:

\begin{equation*}
    e(S(L_q),\mathbb{I}) = \max_{S(L_q) \subseteq I_i} \frac{|S(L_q)|}{|I_i|}.
\end{equation*}
\end{definition}

We only use indexes $I_i$ satisfying $S(L_q) \subseteq I_i$, as they contain all vectors matching $L_q$ and can thus answer query $q$ without omission.  
The \emph{elastic factor} is the maximum overlap ratio $|S(L_q)|/|I_i|$ between $S(L_q)$ and any such index $I_i \in \mathbb{I}$.
When $\mathbb{I}$ contains a single index $I_i$, we also denote the elastic factor between $L_q$ and $I_i$ as $e(S(L_q), I_i)$.
To guarantee a non-zero elastic factor, we include a top index $I_{\top}$ built over all base vectors in $S$ without label filtering, ensuring $S(L_q) \subseteq I_{\top}$.  
The elastic factor lies in $(0,1]$: it is $1$ when an index is built exactly on the query-label set $L_q$ and decreases toward $0$ when the chosen index contains more label-mismatched vectors.

\begin{exmp}
Figure~\ref{fig:run-example}(a) lists the vector groups with their associated label sets in the database $S$.  
Three vectors have no labels ({Empty}[3]) and three have the label set $\{ABC\}[3]$.  
Figure~\ref{fig:run-example}(b) shows the number of vectors matched by each possible query–label set.  
For example, a query with label set $L_q=\{A\}$ matches all vectors labeled $\{A\}$, $\{AB\}$, $\{AC\}$, or $\{ABC\}$, totaling 10 vectors.  
Building index $I_2$ on these 10 vectors (label set $\{A\}$) supports not only $L_q=\{A\}$ but also queries such as $L_q=\{AB\}$, since it contains all required base vectors. 
Figure~\ref{fig:run-example}(c) illustrates index sharing with an elastic factor of $0.3$.  
Index $I_2$ can answer the query $L_q=\{ABC\}$ because the overlap ratio is $3/10 = 0.3$,  
whereas index $I_1$ (label set $\emptyset$) cannot, as its overlap ratio $3/17$ is below the threshold.
\end{exmp}

\stitle{Connection to Query Efficiency.}
In the extreme, one could perform filter-based search (e.g., $\POST$) on the top index $I_{\top}$, which contains all entries and whose label set ($\emptyset$) is a subset of every query-label set.  
However, this is inefficient for queries with non-empty $L_q$ because the search must traverse many label-mismatched nodes in the graph index.  
The inefficiency arises because the elastic factor of $I_{\top}$ is typically very low for such queries.  
To raise the elastic factor, additional indexes must be added to $\mathbb{I}$.
The key question then becomes: when the elastic factor of $\mathbb{I}$ for a query $q$ is at least $c$, what is the resulting query cost?

To answer this, we first analyze the expected cost of KNN search (ignoring label constraints).  
To retrieve the $k$ nearest neighbors, we first use the index to locate the top-1 result and then incrementally continue the search from each previously discovered neighbor until all $k$ results are obtained (e.g., extending from top-1 to top-2, and so on).
Thus, the efficiency of KNN search depends on the expected number of steps required to obtain $k$ valid results.
When using an index built over the entire dataset $S$ (as in $\POST$), the expected number of steps $\mathbb{E}(r)$ can grow to $O(N)$ under low query selectivity, where $N = |S|$, explaining its inefficiency.  
In contrast, if the query is answered using an index with a constant elastic factor $c$, the expected number of steps to obtain $k$ results is bounded by $O(k/c)$~\cite{Acorn-SIGMOD-2024,ESG-arxiv-2025-Mingyu}.  
This bound holds because, in graph-based indexes, retrieving the next nearest-neighbor/result typically requires only one additional expansion step after locating the current result, under ideal conditions (e.g., slow growth or when the query is contained in the dataset).  
Extending this reasoning to our label-hybrid AKNN search, we still first locate the top-1 nearest neighbor and then continue the search until all $k$ results \textit{satisfying the label constraint} are retrieved.  
We show that, apart from the cost of finding the first neighbor, the additional cost of retrieving the remaining results remains $O(k/c)$, as formalized in Lemma~\ref{lem:filter-search-complexity}.  

\begin{figure}[!t]
\centering
\begin{footnotesize}
\begin{tikzpicture}
    \begin{customlegend}[legend columns=4,
        legend entries={$e=0.001$,$e=0.01$,$e=0.1$,$e=1$},
        legend style={at={(0.45,1.15)},anchor=north,draw=none,font=\scriptsize,column sep=0.1cm}]
    \addlegendimage{line width=0.15mm,color=violate,mark=o,mark size=0.5mm}
    \addlegendimage{line width=0.15mm,color=navy,mark=triangle,mark size=0.5mm}
    \addlegendimage{line width=0.15mm,color=forestgreen,mark=square,mark size=0.5mm}
    \addlegendimage{line width=0.15mm,color=orange,mark=star,mark size=0.5mm}
    \end{customlegend}
\end{tikzpicture}

\begin{tikzpicture}[scale=0.9]
\begin{axis}[
height=\columnwidth/2.50,
width=\columnwidth/1.80,
xlabel=recall@10(\%),
ylabel=Qps,
ymode=log,
xmin=90,
xmax=100.2,
label style={font=\scriptsize},
tick label style={font=\scriptsize},
title style={font=\scriptsize},
title style={yshift=-2.5mm},
ymajorgrids=true,
xmajorgrids=true,
grid style=dashed,
]
\addplot[line width=0.15mm,color=violate,mark=o,mark size=0.5mm,smooth]
plot coordinates {
(98.176, 23.3746)
(99.17, 15.0974)
(99.739, 7.82372)
(99.876, 5.49328)
(99.908, 4.42268)
};
\addplot[line width=0.15mm,color=navy,mark=triangle,mark size=0.5mm,smooth]
plot coordinates {
(94.398, 217.538)
(97.017, 137.626)
(98.964, 73.2294)
(99.504, 48.3191)
(99.667, 38.4367)
(99.881, 19.9605)
(99.907, 17.0642)
};
\addplot[line width=0.15mm,color=forestgreen,mark=square,mark size=0.5mm,smooth]
plot coordinates {
(85.173, 2883.52)
(90.623, 1493.05)
(95.762, 907.571)
(97.651, 671.497)
(98.325, 547.621)
(99.393, 289.157)
(99.534, 249.28)
(99.777, 148.694)
(99.824, 124.799)
(99.913, 77.7668)
};
\addplot[line width=0.15mm,color=orange,mark=star,mark size=0.5mm,smooth]
plot coordinates {
(83.486, 4520.16)
(94.336, 3004.63)
(97.737, 1818.32)
(98.866, 1013.97)
(99.523, 555.184)
};

\end{axis}
\end{tikzpicture}\hspace{0.5mm}
\begin{tikzpicture}[scale=0.9]
\begin{axis}[
height=\columnwidth/2.50,
width=\columnwidth/1.80,
xlabel=recall@20(\%),
ylabel=Qps,
ymode=log,
xmin=90,
xmax=100.2,
label style={font=\scriptsize},
tick label style={font=\scriptsize},
title style={font=\scriptsize},
title style={yshift=-2.5mm},
ymajorgrids=true,
xmajorgrids=true,
grid style=dashed,
]
\addplot[line width=0.15mm,color=violate,mark=o,mark size=0.5mm,smooth]
plot coordinates {
(98.701, 12.5537)
(99.4925, 8.27389)
(99.7715, 5.42746)
(99.8555, 4.27924)
(99.9375, 2.18954)
};
\addplot[line width=0.15mm,color=navy,mark=triangle,mark size=0.5mm,smooth]
plot coordinates {
(96.022, 117.74)
(98.0535, 75.2734)
(99.058, 49.5773)
(99.3875, 39.1214)
(99.823, 20.4763)
(99.8615, 17.5496)
(99.9385, 10.5468)
};
\addplot[line width=0.15mm,color=forestgreen,mark=square,mark size=0.5mm,smooth]
plot coordinates {
(88.5985, 1027.12)
(93.2615, 763.967)
(96.144, 429.043)
(97.2225, 395.119)
(99.044, 198.427)
(99.243, 166.53)
(99.679, 103.964)
(99.744, 89.4377)
(99.8835, 53.7758)
};
\addplot[line width=0.15mm,color=orange,mark=star,mark size=0.5mm]
plot coordinates {
(79.5525, 9267.03)
(91.5185, 1522.64)
(96.557, 1312.58)
(98.537, 985.984)
(99.2945, 592.522)
(99.5775, 440.769)
(99.7205, 338.419)
(99.811, 245.327)
(99.822, 235.164)
(99.8495, 201.133)
(99.867, 179.963)
};

\end{axis}
\end{tikzpicture}\hspace{0.5mm}

\caption{Effect of the elastic factor on query efficiency ($k=10$: left, $k=20$: right).  
We randomly generate label sets for both base and query vectors and build $\HNSW$ indexes on the SIFT100M dataset.  
Queries are grouped by elastic factor---0.001, 0.01, 0.1, and 1---with $e=1$ serving as the optimal baseline.}\label{fig:var-elastic}\vspace{-4ex}
\end{footnotesize}
\end{figure}

\begin{lemma}\label{lem:filter-search-complexity}
Given a dataset $S$, a query $(q, L_q)$, and a selected index set $\mathbb{I}$, let $O(C)$ denote the expected time to retrieve the top-1 neighbor from a graph index. 
If the elastic factor satisfies $e\big(S(L_q), \mathbb{I}\big) \ge c$ for some constant $c \in (0,1]$, then the expected time to obtain the top-$k$ label-hybrid AKNN results with $\POST$ is $O\left(C + \tfrac{k}{c}\right)$.
\end{lemma}

\emph{Proof sketch.}
We make two assumptions.
(i) After incurring a cost of $O(C)$ to locate the top-1 nearest neighbor, each successive nearest neighbor can be retrieved in amortized constant time.
Note that we do not require the visited neighbors to satisfy the query’s label constraints, and the property holds for most graph indexes.
(ii) Conditioned on distance ordering, at least a fraction $c$ of the visited nearest neighbors satisfy the query’s label constraints, guaranteed by $e(\cdot,\cdot) \ge c$.
Since $\POST$ outputs a vector only if it matches the query label set, more than $k$ neighbors may need to be examined.
Let $T$ denote the number of (steps to examine) neighbors until $k$ matches are found. 
$T$ follows a negative-hypergeometric (or geometric) distribution with success probability at least $c$, which yields $\mathbb{E}[T] \le k/c$.
Including the cost $O(C)$ for locating the first neighbor, the total expected cost is bounded by $O(C + k/c)$.

\noindent\textbf{Remark on Lemma~\ref{lem:filter-search-complexity}.}
The above result provides an \emph{expected} bound under two standard assumptions:  
(1) locating the top-1 nearest neighbor costs $O(C)$, and enumerating additional neighbors requires amortized constant work;  
(2) Label-matched and mismatched vectors are assumed to be uniformly distributed in the graph and independent of the labels, ensuring that at least a $c$-fraction of visited vectors satisfy the label constraint (as guaranteed by the elastic factor).
This bound is not a worst-case guarantee, as these assumptions may not always hold for graph indexes such as HNSW.  
Nevertheless, the elastic factor $c$ remains a useful control knob for expected performance, because the additional time cost scales directly with $k/c$.  
Our empirical studies (Fig.~\ref{fig:var-elastic}) confirm this monotonic relationship: larger elastic factors yield higher query throughput (Qps) at fixed recall, and the observed performance drop as $c$ decreases.

\sstitle{Empirical studies.}
To further validate Lemma~\ref{lem:filter-search-complexity}, we measure the practical efficiency of $\POST$ search under varying elastic factors on the 100M subset of the SIFT1B dataset~\cite{ann-benchmakrs}, denoted SIFT100M.  
In the classical $\POST$ strategy, the elastic factor is fixed as it uses only the top index containing all base vectors.  
Here, we add more indexes to $\mathbb{I}$ to vary the elastic factor and report results in Fig.~\ref{fig:var-elastic}.  
As shown, higher elastic factors yield better efficiency (measured in Qps), consistent with the lower time complexity predicted by theory.  
When the factor reaches $1$, search achieves optimal efficiency---equivalent to indexing only the label-matched data.  
Efficiency decreases sublinearly as the elastic factor drops: for example, reducing the factor to $\tfrac{1}{10}$ lowers throughput by only $2\times$ at 98\% recall with $k{=}10$.  
This occurs because the added cost scales with $k$, while the top-1 search still runs in $O(C)$, usually $O(\log N)$.  
The gap widens for larger $k$ (e.g., a $3\times$ slowdown at $k{=}20$ and 98\% recall), but in most applications $k$ is small relative to $N$.  
Thus, as long as the elastic factor remains constant, the overall time cost stays bounded.

\subsection{Problem Definition}\label{sub:pp}

We have analyzed the impact of the elastic factor on search efficiency for a \textit{given} index set $\mathbb{I}$.  
When using $\mathbb{I}$, the expected query cost is directly tied to the elastic factor of $\mathbb{I}$ for each query.  
If the elastic factor for every query in the workload is guaranteed to exceed a constant lower bound $c$, then the AKNN search cost is bounded by an additional $O(k/c)$.  
This establishes a clear relationship between index cost (the total size of indexes in $\mathbb{I}$) and query efficiency (captured by the elastic factor).  
Building on this, among \textit{all subsets of indexes} that can correctly answer the query workload, we seek a subset $\mathbb{I}$ that is both small and yields a large elastic factor, ensuring strong query performance.  
This motivates the formal definition of the \textbf{Elastic Index Selection} ($\EIS$) problem.


\begin{definition}[Fixed Efficiency Index Selection ($\EIS$)]\label{defn:EIS}
\ 
\begin{itemize}[leftmargin=8\labelsep]

   \item[\textbf{Input}]  
   A label-hybrid dataset $S$, a query workload with label sets $\mathcal{L} = \{L_1, \dots, L_n\}$, and a universal index set $\mathcal{I} = \{I_1, \dots, I_n\}$,  
   where each index $I_i = S(L_i)$ is built over the base vectors matching the query-label set $L_i$ with cost $|I_i|$.  
   Let $\tau$ and $c$ be non-negative real numbers.

   \item[\textbf{Output}]  
   Decide whether there exists a subset $\mathbb{I} \subseteq \mathcal{I}$ such that  
   $e\big(S(L_i), \mathbb{I}\big) \ge c$ for every $L_i \in \mathcal{L}$ and the total cost satisfies  
   $\sum_{I \in \mathbb{I}} |I| \le \tau$.
\end{itemize}
\end{definition}

The $\EIS$ problem is a \emph{decision} problem: determine whether a subset of indexes from the universal set $\mathcal{I}$ can be chosen such that  
(1) the total space usage is within a user-specified bound $\tau$, and  
(2) the elastic factor—which governs query efficiency—is at least a given threshold $c$.  
In Section~\ref{sec:elastic-index}, we present the \emph{optimization} version, where the goal is either to minimize space cost while satisfying a required elastic-factor threshold or to maximize the elastic factor subject to a space budget.

\begin{figure}[!t]
    \centering
    \includegraphics[width=0.9\columnwidth]{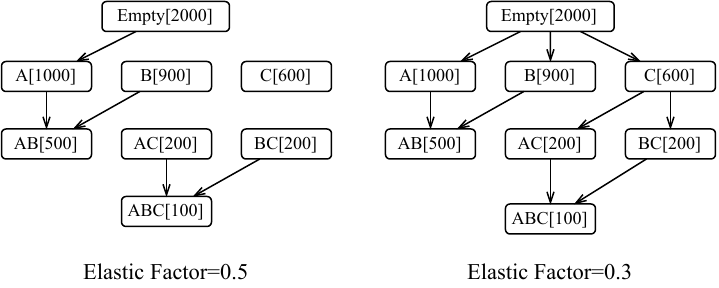}
    \caption{Example of the $\EIS$ Problem}\label{fig:diff-elastic}\vspace{-2ex}
\end{figure}

\begin{exmp}
The $\EIS$ problem under different elastic-factor thresholds, $c=0.5$ and $c=0.3$ is in Fig.~\ref{fig:diff-elastic}.
When $c=0.5$, the top index $I_{\top}$ (built on the empty label set) can serve queries with $L_q=\{A\}$ and $L_q=\emptyset$, since their overlaps exceed 0.5.  
It cannot, however, support $L_q=\{B\}$, whose overlap is only 0.45.  
With a more relaxed threshold of $c=0.3$, $I_{\top}$ also supports $L_q=\{B\}$ and $L_q=\{C\}$,  
but it still cannot answer $L_q=\{AB\}$, whose overlap is 0.25, unless $c<0.25$.  
Finally, when $c=0$, the top index can serve all possible queries.
\end{exmp}

\stitle{Remark on Query Workload.}
In the problem definition, we assume a query workload with query-label sets  
$\mathcal{L}_q = \{L_1, \dots, L_n\}$.  
This assumption is justified for two reasons:
(1) It generalizes the traditional setting where $\mathcal{L}_q$ includes all subsets of the alphabet $\Sigma$ (i.e., the power set).  
(2) It enables a workload-driven approach: $\mathcal{L}_q$ can reflect historical query patterns, allowing index selection and optimization to match realistic query behavior.

Although the total number of possible query-label sets can reach $2^{|\Sigma|}$ in the worst case, the effective number is typically far smaller due to mutual exclusivity among certain labels.  
For example, a product has only one brand name, making brand labels mutually exclusive.  
Our formulation explicitly captures such cases, ensuring that incompatible label combinations---like different brand names---do not co-occur in $\mathcal{L}_q$.
We also assume the workload always includes the empty set $\emptyset$, representing queries with no label constraints and thus matching all base vectors.  
For simplicity, we treat the cost of this top index as zero, i.e., $|I_{\text{top}}| = 0$.

\stitle{Remark on Index Size/Cost}.
The cost of each index $I_i$ (built for query-label set $L_i$) reflects the space it occupies.  
For graph-based indexes, this cost can be approximated by the number of vectors it contains, since node degree is typically bounded by a constant~\cite{Worse-case-NIPS-2023,NSG-VLDB-2019-deng-cai}.  
In practice, each node maintains $M$ edges to support efficient memory access, where $M$ is a user-defined parameter~\cite{hnswlib}.  
Thus, the overall space usage of an index can be estimated as the product of its vector count and $M$.  
In our study, however, we measure index size/cost $|I|$ simply by the number of vectors it contains.

\stitle{Index Selection at Query Time.}
At query time, a label set $L_q$ may be served by multiple indexes in the selected set $\mathbb{I}$.  
For example, a query with $L_q=\{AB\}$ can be answered by the index for $\{A\}$ or for $\{B\}$ if both are present in $\mathbb{I}$.  
In such cases, our method selects the index that yields the maximum elastic factor with $S(L_q)$.
This choice improves query efficiency because greater overlap means fewer label-mismatched vectors in the index, reducing the cost of scanning irrelevant entries.

\subsection{Problem Hardness}
We prove that the $\EIS$ problem is NP-complete in Theorem~\ref{thm:np-hard}, underscoring the computational challenges inherent in solving it.

\begin{theorem}\label{thm:np-hard}
    The $\EIS$ problem is NP-complete.
\end{theorem}

\emph{Proof sketch.}
We show $\EIS$ is NP-complete by reducing from the 3-Set Cover (3-$\SC$) problem~\cite{k-set-cover-STOC-1997,Karp-NP-NPC-NP-Reduce}. $\EIS$ is in NP as validity is verifiable in polynomial time.
Given a 3-$\SC$ instance with elements $\mathcal{U}$, subsets $\mathcal{S}$, and budget $k$, we construct an $\EIS$ instance with a universal index set $\mathcal{I}$ containing:
(1) an element index $I_{u_i}$ for each $u_i \in \mathcal{U}$;
(2) a subset index $I_{s_j}$ for each $s_j \in \mathcal{S}$;
(3) a top index $I_{\top}$ covering all base vectors; and
(4) a bottom index $I_{\mathrm{all}}$.

Let $b > 3$. We set the index sizes to $|I_{u_i}| = A = b+1$, $|I_{s_j}| = B = 2b$, $|I_{\top}| = N > 4b$, and $|I_{\mathrm{all}}| = b$. We set the elastic factor $c = 2b/N$ and the cost budget $\tau = k \cdot B$.
The construction ensures the following overlap properties:
(1) $I_{\top}$ covers every subset index $I_{s_j}$ with ratio $B/N = c$, but fails to cover element indexes $I_{u_i}$ sufficiently (ratio $A/N < c$).
(2) A subset index $I_{s_j}$ covers an element index $I_{u_i}$ with ratio $A/B > c$ if and only if $u_i \in s_j$.
(3) To enforce cost constraints, we introduce duplicate elements $u'_i$ such that covering a specific element $u_i$ directly via element indexes requires selecting both $I_{u_i}$ and $I_{u'_i}$ (total cost $2A = 2b+2$). However, covering $u_i$ via a relevant subset index $I_{s_j}$ costs only $B = 2b$. Since $2A > B$, it is strictly cheaper to cover elements using subset indexes.
The detailed proof is available in our appendix.

The $\EIS$ problem is a decision problem with a yes/no answer. 
In practice, however, one often seeks to optimize either space usage or query time, as these are the key factors in deploying AKNN search.  
We therefore study two optimization variants of $\EIS$:
(1) \textbf{Efficiency-constrained variant} that minimizes total space while enforcing a lower bound on the elastic factor and (2) \textbf{Space-constrained variant} that maximizes the elastic factor subject to a given space budget.  
Because $\EIS$ is NP-complete, both variants are NP-hard.
We next show how to address these variants.

\section{Problem Solution}\label{sec:elastic-index}
We first solve the efficiency-constrained $\EIS$ problem, then extend it to the space-constrained variant, and conclude with a discussion.

\subsection{Method for Efficiency-constrained Variant}
This optimization version of the $\EIS$ problem resembles Definition~\ref{defn:EIS}, but aims to compute the following output:

\begin{itemize}[leftmargin=8\labelsep]
    \item[Output] 
    A subset $\mathbb{I} \subseteq \mathcal{I}$ such that $e(S(L_i), \mathbb{I}) \ge c$ for all $L_i \in \mathcal{L}$, and the total cost $\sum_{I \in \mathbb{I}} |I|$ is minimized.
\end{itemize}

Since this variant of the $\mathsf{EIS}$ problem is NP-hard, we employ a greedy algorithm to obtain an approximate solution.  
The basic idea is as follows.  
Given a collection of indexes $\mathcal{I}$ (corresponding to a query workload with label sets $\mathcal{L}$),  
each index $I \in \mathcal{I}$ has an associated space cost $|I|$, measured by the number of matched vectors it contains.  
We define a benefit function for each index and iteratively select the index with the highest benefit.  
By definition of the elastic factor, adding more indexes improves it and eventually covers all label sets in $\mathcal{L}$ (if all are selected).  
Yet, to minimize total cost, selection must be made carefully to ensure solution quality.
A key preprocessing is that we always include the top index $I_{\top}$ in the solution.  
As noted in $\S$~\ref{sub:pp}, this guarantees that every query has at least one index covering all of its matched vectors, ensuring correctness.  
Thus the answer set $\mathbb{I}$ is initialized with $I_{\top}$.  
The benefit of adding a candidate index $I \in \mathcal{I} \setminus \mathbb{I}$ to the current solution $\mathbb{I}$ is defined as $B(I,\mathbb{I})$, which is defined below.

\begin{definition}
\label{def:benefit}
Let $\mathbb{C} \subset \mathcal{I}$ be the set of indexes already covered by the selected set $\mathbb{I}$  
with overlap ratio at least $c$, where $c$ is the elastic–factor threshold: $\mathbb{C} = \bigl\{\, I_i \in \mathcal{I} \;\big|\; \exists\, I_j \in \mathbb{I} :
   (I_i \subseteq I_j) \wedge \bigl(|I_i|/|I_j| \ge c\bigr) \bigr\}$.
Let $\mathbb{C}'$ be the set of indexes newly covered by adding $I$, also with overlap ratio at least $c$: $\mathbb{C}' = \bigl\{\, I_i \in \mathcal{I} \setminus \mathbb{C} :
   (I_i \subseteq I) \wedge \bigl(|I_i|/|I| \ge c\bigr) \bigr\}$.
The benefit of adding $I$ to $\mathbb{I}$ is $B(I,\mathbb{I}) = \sum_{I_i \in \mathbb{C}'} \frac{|I_i|}{|I|}$.
\end{definition}

The benefit captures two aspects:  
(i) the number of previously uncovered indexes that the candidate index $I$ can cover  
(with overlap ratio at least $c$, i.e., the set $\mathbb{C}'$), and  
(ii) the actual overlap ratio between $I$ and each newly covered index $I_i \in \mathbb{C}'$.  
In other words, the benefit is the total overlap ratio of $I$ with all indexes in $\mathbb{C}'$.

\begin{algorithm}[!t]
	\caption{Greedy Algorithm}
	\label{algo:greedy-algorithm}
	\begin{small}
	\KwIn{The Universal Index Set $\mathcal{I}$, The Elastic Factor $c$}
    \KwOut{The Selected Index Set $\mathbb{I}$}
    $\mathbb{I} \gets \{I_{\top}\}$\;
    \While{Stop Condition Not Met}
    {
        $I\gets I \in \mathcal{I}\setminus\mathbb{I}$ such that $B(I, \mathbb{I})$ is maximum\;
        $\mathbb{I} \gets\mathbb{I} \cup \{I\}$\;
        \textbf{if} { \textbf{For} \textbf{all} $L_q \in \mathcal{L}$, $e(S(L_q),\mathbb{I}) \ge c)$} \textbf{then}{
            \textbf{Break}\;
        }
    }
    \textbf{return} $\mathbb{I}$\tcp*{Selected Index Set}
\end{small}
\end{algorithm}

\stitle{Algorithm.}
We implement the greedy method in Algorithm~\ref{algo:greedy-algorithm}.  
First, we add the top index $I_{\top}$ to the answer set $\mathbb{I}$ (Line~1).  
At each iteration, we select the next candidate index—each candidate $I$ corresponds to a query–label set $L_q$ in the workload—based on its benefit score (Lines~3–4).  
The process stops when all label sets in $\mathcal{L}$ are covered, yielding a valid solution (Line~5).

\begin{exmp}
Using the example dataset in Fig.~\ref{fig:run-example}, we evaluate the benefit of different index choices,  
with Table~\ref{tab:benefit} reporting the benefit ratio for each candidate index.  
Note that the top index $I_1 = I_{\top}$ is always selected first, regardless of its benefit.  
Selecting $I_1$ covers the vectors in $I_2$, $I_3$, $I_4$, $I_6$, and itself 
(with overlap ratio at least the elastic–factor threshold $c = 0.3$),  
for a total of $17 + 10 + 7 + 9 + 6 = 49$ vectors at a cost of $17$,  
yielding a benefit of $49/17 \approx 2.88$.  
After $I_1$ is selected, the benefits of the remaining candidates change.  
Initially, $I_2$ covers $I_2$, $I_5$, $I_6$, and $I_8$,  
i.e., $23$ vectors at a cost of $10$, for a benefit of $2.3$.  
In the second round, as $I_1$ already covers $I_2$ and $I_6$,  
$I_2$ can now cover only $I_5$ and $I_8$, reducing its benefit to $0.7$.  
The greedy algorithm next selects the index with the highest updated benefit 
(Fig.~\ref{fig:run-example}(d)), which is $I_5$.  
At this point, $I_1$ and $I_5$ cover all vectors except those in $I_7$.  
The third round selects $I_7$, with a cost of $5$.  
The greedy solution therefore has a total cost of $17 + 4 + 5 = 26$, which is not optimal.  
As shown in Fig.~\ref{fig:run-example}(e), the optimal solution instead selects $I_3$ in the second round.  
Although $I_3$ covers only $I_5$, $I_7$, and $I_8$ with a lower benefit of $1.71$ (due to overlap with $I_1$),  
it achieves full coverage using just two indexes, $I_1$ and $I_3$,  
for a total cost of $17 + 7 = 24$, outperforming the greedy approach.
\end{exmp}

\begin{table}[!t]
\begin{footnotesize}
\centering
\caption{The Benefits of Each Candidate Index in Each Round}\label{tab:benefit}\vspace{-2ex}
\begin{tabular}{l|r|r|r}
    \toprule
        & Init Round & Second Round & Third Round\\ 
    \midrule
    $I_1$ & 49 / 17 = 2.88 &               &   \\ 
    $I_2$ & 23 / 10 = 2.30 & 7 / 10 = 0.70 & 0 / 10 = 0.00 \\
    $I_3$ & 19 / 7 = 2.71  & 12 / 7 = 1.71 & 5 / 7 = 0.71 \\ 
    $I_4$ & 23 / 9 = 2.55  & 14 / 9 = 1.55 & 5 / 9 = 0.55 \\ 
    $I_5$ & 7 / 4 = 1.75   & 7 / 4 = 1.75  &   \\ 
    $I_6$ & 9 / 6 = 1.50   & 3 / 6 = 0.50  & 0 / 6 = 0.00 \\
    $I_7$ & 8 / 5 = 1.60   & 8 / 5 = 1.60  & 5 / 5 = 1.00 \\ 
    $I_8$ & 3 / 3 = 1.00   & 3 / 3 = 1.00  & 0 / 3 = 0.00 \\ 
    \bottomrule
  \end{tabular}
\end{footnotesize}\vspace{0.5em}
\end{table}

\stitle{Implementation Details and Cost Analysis.}  
To make the greedy method practical, we first create an index in $\mathcal{I}$ for each query–label set in $\mathcal{L}$.  
In theory, there can be up to $2^{|\Sigma|}$ query–label sets---and thus the same number of indexes---in the worst case.  
However, given a query workload $\mathcal{L}$, we only need to compute the sizes of  
$O\!\left(\min\!\left(2^{|\Sigma|},\,|\mathcal{I}|\right)\right)$ indexes during preprocessing, where $|\mathcal{I}| = |\mathcal{L}|$
When the labels in the alphabet $\Sigma$ follow a power–law distribution (as in many real-world datasets),  this preprocessing cost is empirically a small fraction of the total runtime.  
For other label distributions, however, the overhead can be substantial.  
To handle such cases, following \cite{Impl-data-cube-SIGMOD-1996},  
we estimate the size of some index $I$ in $\mathcal{I}$ under a query–label set $L_q$ by sampling or advanced cardinality–estimation techniques~\cite{cardinality-estimation-survey-VLDB-2017}, without fully materializing $I$.  
Once index sizes are approximated, we can efficiently compute the benefit of each candidate and apply the greedy algorithm.  
To further improve performance, we maintain a max-heap to track the index with the highest benefit at each iteration.  
As selecting an index can affect the benefits of up to $|\mathcal{I}|$ related indexes,  
we check whether the heap’s top entry has become stale; if so, we update its benefit and reinsert it.  
This strategy keeps the practical runtime well below $O(|\mathcal{I}|^2)$ --- the cost incurred by the greedy method.


\subsection{Method for Space-Constrained Variant}
In the previous section, we addressed the efficiency-constrained index-selection problem.  
We now turn to selecting indexes that maximize query efficiency while satisfying a space constraint.  
This optimization variant of the $\mathsf{EIS}$ problem follows Definition~\ref{defn:EIS},  
but focuses on computing the following output:
\begin{itemize}[leftmargin=8\labelsep]
\item[Output]
A subset $\mathbb{I} \subseteq \mathcal{I}$ such that the total cost $\sum_{I \in \mathbb{I}} |I| \le \tau$, and the minimum elastic factor $\min(e(S(L_q),e(\mathbb{I}))$ over all $L_q \in \mathcal{L}$ is maximized.
\end{itemize}
\stitle{Insight.}
We observe that the elastic–factor threshold $c$ is monotone for the $\mathsf{EIS}$ problem.  
For example, if an index set $\mathbb{I}$ satisfies a threshold of $0.5$ for a query workload $\mathcal{L}$, it also satisfies any smaller threshold $c' < 0.5$.  
This holds because the threshold $c$ bounds the overlap between the indexes in $\mathbb{I}$ and any label set in $\mathcal{L}$,  
requiring the overlap to be at least $c$.  
Thus, if the \textit{stricter} bound $c$ is met, any \textit{looser} bound $c'$ is automatically satisfied.  

\stitle{Method.}
Leveraging this property, we reduce the space–constrained problem to its efficiency–constrained counterpart  
and apply a binary search to find a set $\mathbb{I}$ whose elastic factor exceeds a given $c$  
while keeping the total cost within the budget $\tau$.  
We perform the binary search over $c \in (0,1]$ in decreasing order.  
At each step, for a candidate value of $c$, we invoke Algorithm~\ref{algo:greedy-algorithm}  
to check if the returned index set $\mathbb{I}$ has total cost at most $\tau$.  
The search stops once such a set is found, yielding the maximum achievable elastic factor under the space constraint.  
Overall, this procedure requires $O\!\left(\log \tfrac{1}{\varepsilon}\right)$ calls to Algorithm~\ref{algo:greedy-algorithm},  
where $\varepsilon$ is the resolution for the binary search (set to $0.001$ in our paper).  
In practice, the overhead of combining binary search with the greedy method is negligible---typically under $1\%$ of the index-construction time,  
or about $1$–$2$ seconds even for large workloads $\mathcal{L}$.

\subsection{Discussions}

\stitle{Limitation.}  
For either efficiency- or space-constrained $\EIS$ problem,  
our greedy methods output a subset of indexes $\mathbb{I}$ from the universal index set $\mathcal{I}$,  
where each index corresponds to a query–label set in the workload $\mathcal{L}$.  
The efficiency of our approach for AKNN search stems from controlling either space or query time  
through index sharing based on label-containment relationships.
In extreme cases, however, index sharing may be ineffective.  
For example, if the query workload consists of label sets that are completely disjoint, there is no containment relationship between them.  
We must then either create a separate index for each disjoint query–label set  
(resulting in high space cost)  
or build a single index covering all vectors in the database  
(resulting in a small elastic factor).  
Fortunately, in real-world scenarios, when query–label sets are disjoint, their base vectors are typically disjoint as well,  
so building an index for each query–label set remains affordable,  
requiring only $O(N)$ space.  
By contrast, algorithms such as $\mathsf{UNG}$, $\mathsf{ACORN}$,  
and \textsc{FilterDiskANN} are unaffected by this limitation  
as they do not rely on index sharing and may perform better in such cases.

\stitle{Extensions to Dynamic Scenarios.}  
In real-world deployments, both the database vectors and the query workload may evolve over time.  
We describe how our method accommodates such updates.

\sstitle{Database vector dynamics.}  
Vectors can be inserted into or deleted from the database.  
We focus on insertions, since deletions can be handled by marking a vector as ignored  
and continuing the search until all results are found~\cite{hnswlib,Diskann-NIPS-2019,SSG-PAMI-2022-deng-cai}.  
When new vectors arrive, we insert each into every index $I_i$ (built on query–label set $L_i$)  
whose labels match the vector.  
For queries, we continue using the original selected index set $\mathbb{I}$, but periodically check for data changes and update the indexes once 10–20\% new vectors have been added.

\sstitle{Query workload dynamics.}  
When new label sets $L_q$ appear in the query workload,  
we first collect all matching vectors for each $L_q$. 
If fewer than $4,000$ (a threshold determined by experiments) are collected, we scan them directly to answer the new query.  
Otherwise, we select indexes from the set $\mathbb{I}$ built on label sets that are subsets of $L_q$.  
Note that the top index $I_{\top}$ is always available (as $\emptyset$ is a subset of any query-label set).  
If multiple indexes qualify, we choose the one with the maximum overlap with the collected vectors to answer the query.
We selectively create an index for the new label set in the query workload (e.g., when the number of collected vectors exceeds 4,000), but adjust set $\mathbb{I}$ only periodically to maintain efficiency.

\sstitle{Index updates.}  
For both types of dynamics, we postpone structural updates to the current selected index set $\mathbb{I}$ and process them in batches.  
(1) For the efficiency-constrained variant:
We maintain the current set $\mathbb{I}$ as long as it satisfies the elastic–factor constraint.  
If vector insertions or newly created indexes (from new queries) violate this constraint, we reapply Algorithm~\ref{algo:greedy-algorithm} to add the necessary indexes to enlarge $\mathbb{I}$.
(2) For the space-constrained variant:
We use the existing $\mathbb{I}$ and its current elastic factor as the starting point.  
If additional space is available, we attempt to increase the target elastic factor.  
Otherwise, we lower the target, prune redundant indexes to free space,  
and continue the binary-search process.

Throughout these updates, we support non-blocking operations:  
The index set need not be rebuilt from scratch,  
and incoming queries are routed dynamically to the index with the highest elastic factor,  
ensuring uninterrupted service.  
Empirical results in Section~\ref{sec:exp} confirm the efficiency of this update strategy.

\input{figures/result-small}

\section{Experiment}\label{sec:exp}
\subsection{Experiment Settings}

\begin{table}[!t] 
\centering 
\caption{The Statistics of Datasets}\vspace{-2ex}
\label{tab:dataset_details} 
\begin{footnotesize}
\begin{tabular}{c|c c c c c} 
\toprule
\textbf{Dataset}& \textbf{Dimension} & \textbf{Size} & \textbf{Query Size} & \textbf{Type} & \textbf{Label}\\ 
\midrule
SIFT    & 128  & 1,000,000 & 10000 & Image & Synthetic \\
MSMARCO & 1024 & 1,000,000 & 1000 &  Text & Synthetic \\
LAION   & 512  & 1,182,243 & 1000 & Image & Meta \\ 
OpenAI-1536 & 1536 & 999,000   & 1000 & Text & Meta \\
PAPER       & 200  & 2,029,997 & 10000 & Text & Real \\ 
TripClick   & 768  & 1,055,976 & 1000 & Text & Real \\
Arxiv   & 4000 & 100,000     & 1000 & Text & Real \\
DEEP    & 96   & 100,000,000 & 1000 & Image & Synthetic\\ 
\bottomrule
\end{tabular} 
\end{footnotesize}
\end{table}

\stitle{Datasets.}
We evaluate on $8$ publicly available datasets (see Table~\ref{tab:dataset_details}) that are standard benchmarks for AKNN search, as well as datasets generated by state-of-the-art embedding models such as MSMARCO, OpenAI-1536~\cite{MRQ-RESQ-arxiv-2024-Mingyu}, TripClick, and Arxiv~\cite{Arxiv-Dataset-2025-Filter-Search}.
To examine scalability, we also sample 100M vectors (denoted as DEEP) from the Deep1B~\cite{MRQ-RESQ-arxiv-2024-Mingyu,ExRaBitQ-arxiv-2024}\footnote{www.tensorflow.org/datasets/catalog/deep1b}.
For label annotations, datasets PAPER, LAION, TripClick, OpenAI-1536, and Arxiv supply real labels or metadata.
For others, we synthesize labels following prior work~\cite{UNG-SIGMOD-2025,Filtered-diskann-WWW-2023}.
Label sources are grouped into three categories:

\noindent (1) \textbf{Real} labels (Paper, TripClick, Arxiv):
For Paper, each vector is labeled by the paper’s venue, research field, and institution type. 
For TripClick, we use the 28 most frequent clinic categories, consistent with prior works~\cite{Acorn-SIGMOD-2024,UNG-SIGMOD-2025}.
For Arxiv, we assign the main subject categories of each vector as its label.

\noindent (2) \textbf{Metadata-derived} labels (LAION, OpenAI-1536):
Using vector-associated metadata (e.g., text or images), we follow the procedure of~\cite{Acorn-SIGMOD-2024} to construct a candidate label set and then assign to each vector its top-3 most relevant labels.

\noindent (3) \textbf{Synthetic} labels (SIFT, MSMARCO, DEEP):
Since keywords and tags in real systems typically follow heavy-tailed distributions, prior work models label frequency using Zipf (power-law) distributions~\cite{Filtered-diskann-WWW-2023,UNG-SIGMOD-2025}. 
We reuse publicly available code to generate labels for both base and query vectors under this distribution. 
In subsequent experiments, we also evaluate alternative distributions (Uniform, Poisson, and Multinomial) and vary the alphabet size $|\Sigma| \in \{8,12,24,32\}$ to assess robustness~\cite{technicalreport}.

\stitle{Metrics.}
The evaluation covers both efficiency and accuracy.
For the search efficiency metric, we use Queries Per Second(Qps), which indicates the number of queries processed by the algorithm per second, as it is most commonly used in benchmarks. For the search accuracy, we use \textit{recall} defined in~\S~\ref{subsec:defn} as the metric to align with the baselines~\cite{Acorn-SIGMOD-2024,UNG-SIGMOD-2025}.
All results are reported as averages.
We also report the \textbf{average selectivity} of a query workload. 
For each query $(q, L_q)$, the \textbf{selectivity} is defined as $|S(L_q)|/|S|$, i.e., the fraction of base vectors whose label sets contain $L_q$. 
We compute the average selectivity (denoted \emph{Ave.\ Select}) across all test queries in the workload and annotate it at the top of each figure to highlight this value.

\stitle{Algorithms.}  
We compare the following algorithms:

\noindent
$\bullet$ \textbf{$\ELI$-0.2:} Our greedy method for the efficiency-constrained variant, where $0.2$ denotes the elastic–factor threshold.

\noindent
$\bullet$ \textbf{$\ELI$-2.0:} Our method for the space-constrained variant, where $2.0$ indicates the space bound (at most twice the original dataset size).

\noindent
$\bullet$ \textbf{$\UNG$:} Unified Navigating Graph for AKNN search~\cite{UNG-SIGMOD-2025}.

\noindent
$\bullet$ \textbf{$\ACORN$}-1: ANN Constraint-Optimized Retrieval Network with low construction overhead~\cite{Acorn-SIGMOD-2024}.

\noindent
$\bullet$ \textbf{$\ACORN$-$\gamma$:} ANN Constraint-Optimized Retrieval Network for high-efficiency search~\cite{Acorn-SIGMOD-2024}.

\noindent
$\bullet$ \textbf{Brute-Force:} Scan all label-matched vectors to find the results.

\stitle{Implementation Details.}
All code was implemented in C++ and compiled with GCC 9.4.0 using the \texttt{-Ofast} optimization flag.  
Experiments were conducted on a workstation equipped with Intel(R) Xeon(R) Platinum 8352V CPUs @ 2.10\,GHz and 512\,GB of memory.  
We employed multi-threading (144 threads) for index construction and a single thread for search evaluation.  
For the underlying index, we used $\mathsf{HNSW}$~\cite{HNSW-PAMI-2020} with parameters $M = 16$ and \textit{efconstruct} = 200.  
For $\ELI$-0.2 ($<1.0$), we applied the index-selection method to achieve a fixed elastic factor of $0.2$.  
For $\ELI$-2.0 ($>1.0$), we used the fixed-space method, allowing at most double the original database size to maximize efficiency.  
For other baselines, such as $\ACORN$ and $\UNG$, we adopted the default parameters from their respective papers:  
$\alpha = 1.2$, $L = 100$ for $\UNG$;  
$\gamma = 80$, $M = 128$, $M_B = 128$ for $\ACORN$-$\gamma$ on the LAION and TripClick datasets;  
and $\gamma = 12$, $M = 32$, $M_B = 64$ for the remaining datasets.

\subsection{Experiment Results}

\stitle{Exp-1: Comparison of Query Efficiency.}
We begin by evaluating query efficiency (Qps) across different methods.  
Fig.~\ref{fig:recall-small} compares $\UNG$, $\ACORN$-1, $\ACORN$-$\gamma$, and our methods ($\ELI$-0.2, $\ELI$-2.0) under varying alphabet sizes $|\Sigma|$ and label distributions. 
Here, \textit{Real}, \textit{Meta}, and \textit{Zipf} denote the real, meta, and synthetic Zipf-based label distributions, respectively.
Points toward the upper right indicate better performance.  
Our methods achieve the best trade-off between search efficiency and accuracy.  
For example, at 95\% recall with $|\Sigma| = 12$, $\ELI$-0.2 delivers a $4\times$ speedup over $\UNG$ on SIFT  
and a $10\times$ speedup on MSMARCO.  
Performance remains stable across different $|\Sigma|$ values and real label distributions,  
whereas $\UNG$'s retrieval efficiency drops sharply as $|\Sigma|$ grows.  

Thanks to the elastic factor, $\ELI$-0.2 sustains nearly $12\times$ higher query speed than $\UNG$  
when $|\Sigma| = 32$ on the OpenAI-1536 dataset.  
In contrast, $\ACORN$ plateaus in accuracy for large $|\Sigma|$ due to limitations in its $\PRE$ strategy.
For example, on Fig.~8(a), as $|\Sigma|$ increases, $\ACORN$ accuracy declines and $\UNG$ efficiency degrades,  
while $\ELI$-0.2 maintains consistently high efficiency.  
Although larger $|\Sigma|$ lowers average query selectivity, our design keeps runtime stable.  
A modest gap appears between $\ELI$-0.2 and $\ELI$-2.0.
For example, on Fig.~8(b), under a Zipf distribution, the elastic factor of $\ELI$-2.0 is typically below $0.2$ for large $|\Sigma|$, so $\ELI$-0.2 often performs slightly better due to its higher elastic-factor threshold.

\begin{figure*}[!t]
\centering
\begin{footnotesize}
\begin{tikzpicture}
    \begin{customlegend}[legend columns=7,
        legend entries={$\UNG$,$\ELI$-0.2,$\ELI$-2.0,$\ACORN$-1,$\ACORN$-$\gamma$, $\HNSW$-$\POST$, Brute-Force},
        legend style={at={(0.45,1.15)},anchor=north,draw=none,font=\scriptsize,column sep=0.1cm}]
    \addlegendimage{line width=0.15mm,color=norm1,mark=o,mark size=0.5mm}
    \addlegendimage{line width=0.15mm,color=norm2,mark=triangle,mark size=0.5mm}
    \addlegendimage{line width=0.15mm,color=norm3,mark=square,mark size=0.5mm}
    \addlegendimage{line width=0.15mm,color=norm4,mark=otimes,mark size=0.5mm}
    \addlegendimage{line width=0.15mm,color=norm5,mark=star,mark size=0.5mm}
    \addlegendimage{line width=0.15mm,color=black,mark=diamond,mark size=0.5mm}
    \addlegendimage{only marks, mark=pentagon*, mark size=3pt, color=dark3}
    \end{customlegend}
\end{tikzpicture}

\begin{tikzpicture}[scale=1]
\begin{axis}[
height=\columnwidth/2.60,
width=\columnwidth/1.80,
xlabel=recall@10(\%),
ylabel=Qps,
ymode=log,
title={ No Filter 100\% Selectivity},
label style={font=\scriptsize},
tick label style={font=\scriptsize},
title style={font=\scriptsize},
title style={yshift=-2.5mm},
ymajorgrids=true,
xmajorgrids=true,
grid style=dashed,
]
\addplot[line width=0.15mm,color=norm1,mark=o,mark size=0.5mm]
plot coordinates {
( 85.04, 127.23 )
( 98.25, 46.72 )
( 99.79, 14.66 )
( 99.79, 11.30 )
( 99.81, 8.37 )
( 99.80, 7.81 )
( 99.79, 5.85 )
};
\addplot[line width=0.15mm,color=norm4,mark=otimes,mark size=0.5mm]
plot coordinates {
( 29.75, 1326.26 )
( 41.32, 744.60 )
( 59.10, 479.62 )
( 74.41, 227.07 )
( 86.30, 143.12 )
( 93.12, 93.48 )
( 93.51, 87.47 )
( 96.87, 56.34 )
( 98.49, 38.67 )
( 98.75, 35.27 )
};
\addplot[line width=0.15mm,color=norm5,mark=star,mark size=0.5mm]
plot coordinates {
( 64.68, 1308.90 )
( 74.96, 1267.43 )
( 85.17, 772.80 )
( 91.88, 427.90 )
( 96.33, 259.47 )
( 98.26, 149.48 )
( 98.42, 139.33 )
( 99.26, 84.65 )
( 99.55, 55.44 )
( 99.62, 51.77 )
};
\addplot[line width=0.15mm,color=black,mark=diamond,mark size=0.5mm]
plot coordinates {
( 73.66, 3546.10 )
( 82.58, 2439.02 )
( 91.96, 1426.53 )
( 97.30, 714.80 )
( 99.38, 550.36 )
( 99.88, 299.76 )
( 99.90, 284.58 )
};

\end{axis}
\end{tikzpicture}\hspace{0.5mm}
\begin{tikzpicture}[scale=1]
\begin{axis}[
height=\columnwidth/2.60,
width=\columnwidth/1.80,
xlabel=recall@10(\%),
ylabel=Qps,
ymode=log,
title={ 30\% Selectivity},
label style={font=\scriptsize},
tick label style={font=\scriptsize},
title style={font=\scriptsize},
title style={yshift=-2.5mm},
ymajorgrids=true,
xmajorgrids=true,
grid style=dashed,
]
\addplot[only marks, mark=pentagon*, mark size=3pt, color=dark3]
coordinates {(100, 17.8) }; 
\addplot[line width=0.15mm,color=norm1,mark=o,mark size=0.5mm]
plot coordinates {
    ( 86.451, 694.245 )
    ( 94.871, 164.606 )
    ( 96.243, 121.307 )
    ( 98.368, 73.72 )
    ( 98.653, 65.424 )
    ( 99.741, 36.227 )
};
\addplot[line width=0.15mm,color=norm2,mark=triangle,mark size=0.5mm]
plot coordinates {
    ( 89.171, 386.774 )
    ( 95.052, 252.453 )
    ( 98.679, 141.082 )
    ( 99.767, 87.687 )
    ( 99.974, 71.06 )
};
\addplot[line width=0.15mm,color=norm3,mark=square,mark size=0.5mm]
plot coordinates {
    ( 79.922, 2776.98 )
    ( 90.155, 1746.61 )
    ( 96.062, 992.288 )
    ( 98.705, 543.662 )
    ( 99.56, 189.03 )
    ( 99.585, 210.584 )
    ( 99.948, 148.977 )
};
\addplot[line width=0.15mm,color=norm4,mark=otimes,mark size=0.5mm]
plot coordinates {
    ( 75.052, 309.791 )
    ( 85.933, 219.943 )
    ( 92.772, 165.098 )
    ( 96.425, 100.338 )
    ( 96.684, 96.428 )
    ( 98.549, 61.968 )
    ( 99.404, 42.315 )
    ( 99.482, 38.845 )
};
\addplot[line width=0.15mm,color=norm5,mark=star,mark size=0.5mm]
plot coordinates {
    ( 72.409, 842.795 )
    ( 86.166, 423.71 )
    ( 93.161, 264.746 )
    ( 97.176, 201.672 )
    ( 98.834, 143.175 )
    ( 98.886, 140.058 )
    ( 99.585, 87.39 )
    ( 99.741, 58.734 )
    ( 99.767, 53.805 )
};
\addplot[line width=0.15mm,color=black,mark=diamond,mark size=0.5mm]
plot coordinates {
    ( 89.171, 265.475 )
    ( 95.052, 232.811 )
    ( 98.679, 155.897 )
    ( 99.767, 103.624 )
    ( 99.974, 69.953 )
};

\end{axis}
\end{tikzpicture}\hspace{0.5mm}
\begin{tikzpicture}[scale=1]
\begin{axis}[
height=\columnwidth/2.60,
width=\columnwidth/1.80,
xlabel=recall@10(\%),
ylabel=Qps,
ymode=log,
title={ 10\% Selectivity},
label style={font=\scriptsize},
tick label style={font=\scriptsize},
title style={font=\scriptsize},
title style={yshift=-2.5mm},
ymajorgrids=true,
xmajorgrids=true,
grid style=dashed,
]
\addplot[only marks, mark=pentagon*, mark size=3pt, color=dark3]
coordinates {(100, 33.4) }; 
\addplot[line width=0.15mm,color=norm1,mark=o,mark size=0.5mm]
plot coordinates {
    ( 81.174, 565.392 )
    ( 98.292, 148.285 )
    ( 99.288, 143.075 )
    ( 99.787, 82.891 )
    ( 99.893, 78.317 )
    ( 100.0, 48.955 )
};
\addplot[line width=0.15mm,color=norm2,mark=triangle,mark size=0.5mm]
plot coordinates {
    ( 80.285, 4072.46 )
    ( 89.181, 2650.94 )
    ( 95.623, 1535.52 )
    ( 98.256, 841.317 )
    ( 99.288, 433.642 )
    ( 99.324, 414.454 )
    ( 99.715, 241.409 )
    ( 99.929, 166.865 )
};
\addplot[line width=0.15mm,color=norm3,mark=square,mark size=0.5mm]
plot coordinates {
    ( 79.217, 4257.58 )
    ( 89.039, 2754.9 )
    ( 95.409, 1596.59 )
    ( 98.078, 869.969 )
    ( 99.324, 451.768 )
    ( 99.466, 410.819 )
    ( 99.751, 152.469 )
    ( 99.964, 131.925 )
};
\addplot[line width=0.15mm,color=norm4,mark=otimes,mark size=0.5mm]
plot coordinates {
    ( 78.541, 366.362 )
    ( 89.858, 203.476 )
    ( 95.125, 119.422 )
    ( 97.9, 104.345 )
    ( 97.972, 102.145 )
    ( 98.932, 66.18 )
    ( 99.502, 45.572 )
};
\addplot[line width=0.15mm,color=norm5,mark=star,mark size=0.5mm]
plot coordinates {
    ( 70.213, 1518.92 )
    ( 84.199, 900.641 )
    ( 91.922, 531.191 )
    ( 96.299, 300.214 )
    ( 98.47, 178.299 )
    ( 98.683, 131.678 )
    ( 99.787, 63.748 )
    ( 99.858, 65.685 )
};
\addplot[line width=0.15mm,color=black,mark=diamond,mark size=0.5mm]
plot coordinates {
    ( 95.587, 56.768 )
    ( 98.221, 42.351 )
    ( 99.253, 31.741 )
    ( 99.822, 23.552 )
    ( 99.964, 18.067 )
};

\end{axis}
\end{tikzpicture}\hspace{0.5mm}
\begin{tikzpicture}[scale=1]
\begin{axis}[
height=\columnwidth/2.60,
width=\columnwidth/1.80,
xlabel=recall@10(\%),
ylabel=Qps,
ymode=log,
title={ 5\% Selectivity},
label style={font=\scriptsize},
tick label style={font=\scriptsize},
title style={font=\scriptsize},
title style={yshift=-2.5mm},
ymajorgrids=true,
xmajorgrids=true,
grid style=dashed,
]
\addplot[only marks, mark=pentagon*, mark size=3pt, color=dark3]
coordinates {(100, 108.930) }; 
\addplot[line width=0.15mm,color=norm1,mark=o,mark size=0.5mm]
plot coordinates {
    ( 76.036, 403.148 )
    ( 97.087, 189.097 )
    ( 98.739, 145.733 )
    ( 99.099, 135.476 )
    ( 99.91, 109.109 )
};
\addplot[line width=0.15mm,color=norm2,mark=triangle,mark size=0.5mm]
plot coordinates {
    ( 86.546, 695.198 )
    ( 93.243, 636.711 )
    ( 96.516, 571.184 )
    ( 98.438, 468.354 )
    ( 99.46, 361.564 )
    ( 99.519, 345.436 )
    ( 99.64, 263.867 )
    ( 99.7, 217.363 )
};
\addplot[line width=0.15mm,color=norm3,mark=square,mark size=0.5mm]
plot coordinates {
    ( 85.405, 689.441 )
    ( 92.072, 646.602 )
    ( 96.396, 571.184 )
    ( 98.228, 485.423 )
    ( 99.039, 369.18 )
    ( 99.099, 357.68 )
    ( 99.369, 269.636 )
    ( 99.429, 183.673 )
    ( 99.49, 148.066 )
    ( 99.519, 197.041 )
};
\addplot[line width=0.15mm,color=norm4,mark=otimes,mark size=0.5mm]
plot coordinates {
    ( 77.087, 611.009 )
    ( 88.198, 372.9 )
    ( 94.024, 229.181 )
    ( 96.697, 120.13 )
    ( 96.937, 93.618 )
    ( 97.898, 86.787 )
    ( 98.168, 65.888 )
    ( 98.228, 60.922 )
};
\addplot[line width=0.15mm,color=norm5,mark=star,mark size=0.5mm]
plot coordinates {
    ( 78.589, 940.678 )
    ( 90.54, 579.13 )
    ( 95.466, 355.39 )
    ( 97.658, 220.969 )
    ( 97.928, 152.822 )
    ( 98.949, 146.309 )
    ( 99.129, 104.03 )
    ( 99.159, 94.228 )
};
\addplot[line width=0.15mm,color=black,mark=diamond,mark size=0.5mm]
plot coordinates {
    ( 94.174, 24.28 )
    ( 96.967, 21.817 )
    ( 98.559, 17.263 )
    ( 99.279, 13.799 )
    ( 99.73, 11.113 )
    ( 99.91, 9.076 )
};

\end{axis}
\end{tikzpicture}\hspace{0.5mm}
\caption{ The Effect of the Query Selectivity (on Arxiv)}\label{fig:var-group}\vspace{-2ex}
\end{footnotesize}
\end{figure*}

\stitle{Exp-2: Effect of Query Selectivity.}  
To further compare methods, we vary query selectivity and measure its impact on query efficiency.  
We evaluate on Arxiv with four settings: \emph{No Filter} (100\% selectivity),  
30\%, 10\%, and 5\% selectivity, as shown in Fig.~\ref{fig:var-group}.  
In the No-Filter setting, the problem reduces to a standard AKNN search without label constraints.  
Here, $\ELI$-0.2 and $\ELI$-2.0 match the performance of $\HNSW$-$\POST$ (i.e., $\HNSW$ with the $\POST$ strategy) and outperform other baselines.  
At 30\% selectivity, $\ELI$-2.0 consistently surpasses $\UNG$ and $\ACORN$ across the full recall range,  
while $\HNSW$-$\POST$ performs nearly the same as $\ELI$-0.2.
When selectivity drops to 10\% and 5\%, the advantage of our methods widens: they maintain high Qps even at high recall and consistently outperform all other methods across different recall levels.
These results validate the elastic-factor cost model:  
with sufficiently large $c$ (e.g., $c \!\ge\! 0.2$), the additional computation scales as $k/c$.  
We also compare with brute-force search, which becomes competitive when very few vectors match the labels  
(e.g., fewer than $4,000$).  
In practice, we therefore build indexes only for query–label sets containing at least about $4,000$ vectors and use brute force otherwise.

\begin{table}[!t]
\begin{footnotesize}
    \centering
    \caption{The Comparison of Indexing Time (s)}\label{tab:index-time}\vspace{-2ex}
    \begin{tabular}{l|ccccc}
        \toprule
        ~ & $\UNG$ & $\ACORN$-1 & $\ACORN$-$\gamma$ & $\ELI$-0.2 & $\ELI$-2.0 \\ 
        \midrule
        SIFT & 405 & 7  & 62  & 34  & 25 \\
        MSMARCO & 459 & 29 & 290 & 148 & 87  \\
        PAPER & 65 & 18 & 128  & 39  & 64  \\  
        LAION & 27 & 105 & 3238  & 121  & 63  \\  
        TripClick & 283 & 137 & 4289  & 127  & 81  \\  
        OpenAI-1536 & 218 & 47 & 568 & 409 & 166  \\ 
         Arxiv       & 116  & 9  & 50 & 23 & 25 \\ 
        DEEP  & -   &  912 & 11328   & 8864  & 3991\\
        \bottomrule
    \end{tabular}
\end{footnotesize}
\end{table}

\begin{table}[!t]
\begin{footnotesize}
    \centering
    \caption{The Comparison of Index Size (Mb)}\label{tab:index-size}\vspace{-2ex}
    \begin{tabular}{l|cccccc}
        \toprule
        ~ & Base & $\UNG$ & $\ACORN$-1 & $\ACORN$-$\gamma$ & $\ELI$-0.2 & $\ELI$-2.0 \\ 
        \midrule
        SIFT & 488 & 186 & 442 & 485 & 634 & 382 \\ 
        MSMARCO & 3906 & 233 & 442 & 485 & 634 & 382 \\ 
        PAPER & 1548 & 372 & 898 & 986 & 699 & 650 \\
        LAION & 2309 & 204 & 2038 & 1820 & 871 & 405 \\
        TripClick & 3093 & 192 & 1820 & 1625 & 814 & 540 \\ 
        OpenAI-1536 & 5853 & 261 & 442 & 485 & 782 & 342 \\
         Arxiv & 1525 & 20 & 80 & 85 & 26 & 30 \\ 
        DEEP & 36621 & - & 44262 & 48591 & 85293 & 37902 \\ 
        \bottomrule
    \end{tabular}
\end{footnotesize}
\end{table}

\begin{figure}[!t]
    \centering
    \includegraphics[width=0.9\linewidth]{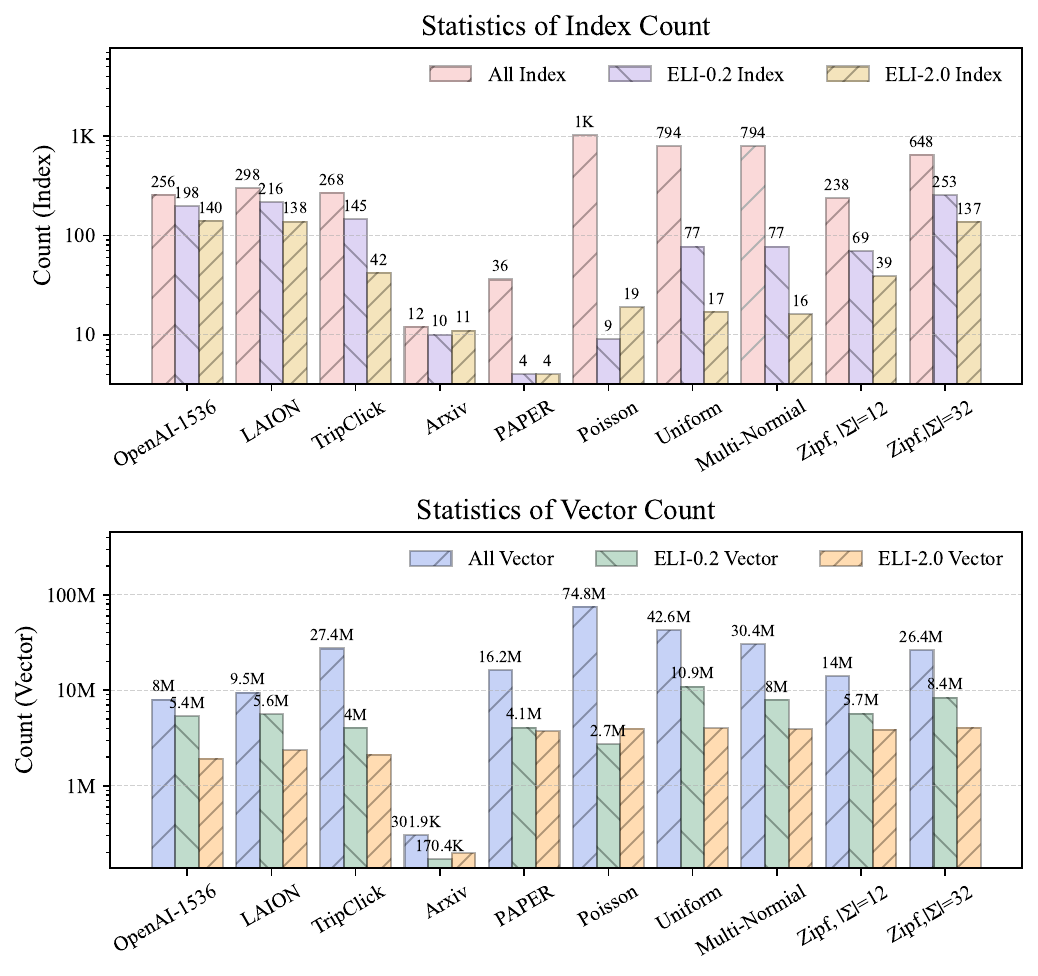}
    \caption{The Statistics of Index Selection by Our Methods}\vspace{-4ex}
    \label{fig:Index-Selection}
\end{figure}

\stitle{Exp-3: Index Selection Statistics.}  
Recall that our methods aim to select a subset of indexes $\mathbb{I}$ from the universal index set $\mathcal{I}$.  
To illustrate how the selection works, we report both the number of indexes chosen and the total number of vectors covered  
by the indexes selected by our methods ($\ELI$-0.2 and $\ELI$-2.0).  
We evaluate on real datasets (\textsc{OpenAI}-1536, \textsc{Laion}, TripClick, Arxiv, and \textsc{Paper}).  
For \textsc{Paper}, we also vary the label distribution (e.g., Poisson) as an additional test.  
The results are presented in Fig.~\ref{fig:Index-Selection}.
The top panel of Fig.~\ref{fig:Index-Selection} presents the total number of candidate indexes $|\mathcal{I}|$  
as well as the number selected by $\ELI$-0.2 and $\ELI$-2.0.  
The bottom panel shows the total size of all candidate indexes (i.e., total number of vectors across indexes)  
and the size of the indexes actually selected by $\ELI$-0.2 and $\ELI$-2.0.

From Fig.~\ref{fig:Index-Selection}, we observe:
(1) \textbf{Compact label-set selection.}  
Both methods select up to an order of magnitude fewer indexes than the full set $\mathcal{I}$,  
verifying the effectiveness of index sharing in reducing index count.
(2) \textbf{Significant reduction in indexed size.}  
The bottom panel shows that $\ELI$-0.2 indexes only a small fraction of the total base vectors,  
while $\ELI$-2.0 indexes roughly twice the number of base vectors.
(3) \textbf{Robustness across distributions.}  
Under various label distributions for \textsc{Paper} (Uniform, Poisson, Multinomial, and Zipf),  
our methods always achieve substantial reductions in both the number of selected indexes and their total size.

\stitle{Exp-4: Indexing Time and Space.}  
We compare indexing time in Table~\ref{tab:index-time} and index size in Table~\ref{tab:index-size}.  
For datasets with real labels, results are reported under their native label distributions.  
For synthetic datasets, we report results with $|\Sigma|=32$, consistent with the cmdefficiency comparison in Fig.~\ref{fig:recall-small}.  
(1) \textbf{Indexing time.}  
Table~\ref{tab:index-time} shows that our methods (especially $\ELI$-2.0) are far more efficient than $\UNG$ and $\ACORN$-$\gamma$.  
Specifically, $\UNG$ requires nearly $20\times$ the indexing time on SIFT and $5\times$ on MSMARCO,  
while $\ACORN$-$\gamma$ averages about $2\times$ the indexing time of our methods.  
Although $\ACORN$-1 builds indexes quickly, its omission of label information results in poorer search efficiency.
(2) \textbf{Index size.}  
Table~\ref{tab:index-size} reports index sizes, where \textit{Base} denotes the size of the raw vectors.  
On the OpenAI-1536 dataset, all methods require only about one-tenth of the total storage space.  
Our methods achieve the best search performance with comparable index size.  
In particular, both $\ACORN$ and $\ELI$-2.0 use roughly twice the space of the original vectors,  
yet $\ELI$-2.0 achieves strong search performance with much smaller additional space.  
$\UNG$ saves the most space, but in high-dimensional settings its index size is only a small fraction of the base vector dataset,  
leading to lower search efficiency.

\stitle{Exp-5: Varying Label Distributions.}  
Label distributions influence label-hybrid AKNN search, so we evaluate performance under different distributions---Zipf, Uniform, Poisson, and Multinomial---with $|\Sigma| = 12$.  
Experiments are conducted on the \textsc{Paper} dataset, and results are shown in Fig.~\ref{fig:recall-var-label-dis}.  
(1) \textbf{Stable performance.}  
Our methods perform consistently across all tested distributions.  
In contrast, the search accuracy of other methods varies significantly with the distribution:  
under Uniform and Multinomial settings, $\ACORN$ fails to reach 80\% recall, while $\UNG$ and our methods maintain high accuracy.
(2) \textbf{High efficiency.}  
Across all distributions, our methods remain highly competitive.  
They achieve recall comparable to $\UNG$ at high thresholds and deliver $3\times$–$5\times$ better search efficiency.

\begin{figure*}[!t]
\centering
\begin{footnotesize}
\begin{tikzpicture}
    \begin{customlegend}[legend columns=5,
        legend entries={$\UNG$,$\ELI$-0.2,$\ELI$-2.0,$\ACORN$-1,$\ACORN$-$\gamma$},
        legend style={at={(0.45,1.15)},anchor=north,draw=none,font=\scriptsize,column sep=0.1cm}]
    \addlegendimage{line width=0.15mm,color=norm1,mark=o,mark size=0.5mm}
    \addlegendimage{line width=0.15mm,color=norm2,mark=triangle,mark size=0.5mm}
    \addlegendimage{line width=0.15mm,color=norm3,mark=square,mark size=0.5mm}
    \addlegendimage{line width=0.15mm,color=norm4,mark=otimes,mark size=0.5mm}
    \addlegendimage{line width=0.15mm,color=norm5,mark=star,mark size=0.5mm}
    \end{customlegend}
\end{tikzpicture}
\\[-\lineskip]
\begin{tikzpicture}[scale=1]
\begin{axis}[
height=\columnwidth/2.60,
width=\columnwidth/1.80,
xlabel=recall@10(\%),
ylabel=Qps,
ymode=log,
title={Ave. Select 18\% \textbf{Zipf}},
label style={font=\scriptsize},
tick label style={font=\scriptsize},
title style={font=\scriptsize},
title style={yshift=-2.5mm},
ymajorgrids=true,
xmajorgrids=true,
grid style=dashed,
]
\addplot[line width=0.15mm,color=norm1,mark=o,mark size=0.5mm]
plot coordinates {
    ( 81.308, 6920.4 )
    ( 93.311, 1647.2 )
    ( 95.362, 1074.3 )
    ( 97.358, 612.6 )
    ( 97.756, 529.4 )
    ( 99.236, 242.9 )
};
\addplot[line width=0.15mm,color=norm2,mark=triangle,mark size=0.5mm]
plot coordinates {
    ( 96.95, 3110.4 )
    ( 99.192, 3003.9 )
    ( 99.676, 2144.5 )
    ( 99.792, 1776.8 )
    ( 99.929, 1043.3 )
    ( 99.949, 921.4 )
    ( 99.974, 607.5 )
    ( 99.973, 540.0 )
};
\addplot[line width=0.15mm,color=norm3,mark=square,mark size=0.5mm]
plot coordinates {
    ( 97.088, 4759.6 )
    ( 99.187, 2908.7 )
    ( 99.788, 1376.7 )
    ( 99.919, 967.2 )
};
\addplot[line width=0.15mm,color=norm4,mark=otimes,mark size=0.5mm]
plot coordinates {
    ( 81.19, 1751.3 )
    ( 90.76, 771.6 )
    ( 94.83, 585.8 )
    ( 96.77, 464.5 )
    ( 97.35, 246.2 )
    ( 97.39, 229.9 )
    ( 97.66, 126.7 )
    ( 97.79, 81.3 )
    ( 97.8, 63.0 )
};
\addplot[line width=0.15mm,color=norm5,mark=star,mark size=0.5mm]
plot coordinates {
    ( 82.95, 5952.4 )
    ( 89.92, 3460.2 )
    ( 93.02, 1972.4 )
    ( 94.46, 1114.8 )
    ( 94.96, 600.2 )
    ( 95.18, 212.6 )
    ( 95.22, 162.4 )
    ( 95.23, 174.4 )
};

\end{axis}
\end{tikzpicture}\hspace{0.5mm}
\begin{tikzpicture}[scale=1]
\begin{axis}[
height=\columnwidth/2.60,
width=\columnwidth/1.80,
xlabel=recall@10(\%),
ylabel=Qps,
ymode=log,
title={ Ave. Select 6.8\% \textbf{Uniform}},
label style={font=\scriptsize},
tick label style={font=\scriptsize},
title style={font=\scriptsize},
title style={yshift=-2.5mm},
ymajorgrids=true,
xmajorgrids=true,
grid style=dashed,
]
\addplot[line width=0.15mm,color=norm1,mark=o,mark size=0.5mm]
plot coordinates {
    ( 88.301, 1984.5 )
    ( 92.912, 1334.9 )
    ( 96.951, 768.3 )
    ( 97.676, 619.9 )
    ( 99.505, 315.0 )
};
\addplot[line width=0.15mm,color=norm2,mark=triangle,mark size=0.5mm]
plot coordinates {
    ( 98.808, 2602.1 )
    ( 99.708, 1566.4 )
    ( 99.936, 929.5 )
};
\addplot[line width=0.15mm,color=norm3,mark=square,mark size=0.5mm]
plot coordinates {
    ( 99.227, 1441.1 )
    ( 99.791, 910.5 )
    ( 99.949, 531.5 )
};
\addplot[line width=0.15mm,color=norm4,mark=otimes,mark size=0.5mm]
plot coordinates {
    ( 70.79, 920.0 )
    ( 72.72, 493.1 )
    ( 73.74, 259.3 )
    ( 73.83, 240.7 )
    ( 74.48, 99.3 )
    ( 74.54, 74.6 )
    ( 74.55, 65.0 )
};
\addplot[line width=0.15mm,color=norm5,mark=star,mark size=0.5mm]
plot coordinates {
};
\end{axis}
\end{tikzpicture}\hspace{0.5mm}
\begin{tikzpicture}[scale=1]
\begin{axis}[
height=\columnwidth/2.60,
width=\columnwidth/1.80,
xlabel=recall@10(\%),
ylabel=Qps,
ymode=log,
title={ Ave. Select 20\% \textbf{Poisson}},
label style={font=\scriptsize},
tick label style={font=\scriptsize},
title style={font=\scriptsize},
title style={yshift=-2.5mm},
ymajorgrids=true,
xmajorgrids=true,
grid style=dashed,
]
\addplot[line width=0.15mm,color=norm1,mark=o,mark size=0.5mm]
plot coordinates {
    ( 78.838, 6939.6 )
    ( 92.904, 1216.1 )
    ( 95.113, 1023.6 )
    ( 97.118, 576.3 )
    ( 97.568, 493.8 )
    ( 99.102, 225.6 )
};
\addplot[line width=0.15mm,color=norm2,mark=triangle,mark size=0.5mm]
plot coordinates {
    ( 96.568, 2650.4 )
    ( 98.016, 1885.4 )
    ( 98.381, 1090.6 )
    ( 98.465, 627.5 )
    ( 98.481, 356.6 )
    ( 98.483, 341.2 )
    ( 98.486, 206.5 )
    ( 98.489, 131.1 )
    ( 98.49, 118.0 )
};
\addplot[line width=0.15mm,color=norm3,mark=square,mark size=0.5mm]
plot coordinates {
    ( 96.048, 3732.7 )
    ( 97.734, 2209.0 )
    ( 98.151, 1286.5 )
    ( 98.255, 745.4 )
    ( 98.303, 404.4 )
    ( 98.305, 406.1 )
    ( 98.322, 251.7 )
    ( 98.329, 176.2 )
};
\addplot[line width=0.15mm,color=norm4,mark=otimes,mark size=0.5mm]
plot coordinates {
    ( 88.99, 2849.0 )
    ( 95.4, 1543.2 )
    ( 97.66, 822.4 )
    ( 98.75, 339.2 )
    ( 99.26, 121.9 )
    ( 99.31, 99.0 )
    ( 99.43, 63.1 )
};
\addplot[line width=0.15mm,color=norm5,mark=star,mark size=0.5mm]
plot coordinates {
    ( 89.22, 5494.5 )
    ( 95.31, 3496.5 )
    ( 97.07, 1923.1 )
    ( 97.77, 1083.4 )
    ( 98.12, 587.2 )
    ( 98.25, 290.0 )
    ( 98.28, 108.3 )
};

\end{axis}
\end{tikzpicture}\hspace{0.5mm}
\begin{tikzpicture}[scale=1]
\begin{axis}[
height=\columnwidth/2.60,
width=\columnwidth/1.80,
xlabel=recall@10(\%),
ylabel=Qps,
ymode=log,
title={ Ave. Select 5.3\% \textbf{\text{multinormal}}},
label style={font=\scriptsize},
tick label style={font=\scriptsize},
title style={font=\scriptsize},
title style={yshift=-2.5mm},
ymajorgrids=true,
xmajorgrids=true,
grid style=dashed,
]
\addplot[line width=0.15mm,color=norm1,mark=o,mark size=0.5mm]
plot coordinates {
    ( 92.199, 1634.0 )
    ( 95.554, 1141.5 )
    ( 98.117, 682.4 )
    ( 98.539, 577.8 )
    ( 99.703, 297.1 )
};
\addplot[line width=0.15mm,color=norm2,mark=triangle,mark size=0.5mm]
plot coordinates {
    ( 98.807, 2063.1 )
    ( 99.635, 1318.9 )
    ( 99.813, 807.3 )
    ( 99.839, 438.6 )
    ( 99.856, 279.2 )
    ( 99.855, 265.5 )
    ( 99.857, 161.7 )
    ( 99.856, 114.7 )
    ( 99.857, 104.6 )
    ( 99.856, 94.6 )
};
\addplot[line width=0.15mm,color=norm3,mark=square,mark size=0.5mm]
plot coordinates {
    ( 99.28, 1037.6 )
    ( 99.679, 593.2 )
    ( 99.779, 362.1 )
    ( 99.817, 208.8 )
    ( 99.834, 119.5 )
    ( 99.836, 112.4 )
    ( 99.846, 67.6 )
    ( 99.849, 47.6 )
    ( 99.85, 43.4 )
    ( 99.851, 39.2 )
};
\addplot[line width=0.15mm,color=norm4,mark=otimes,mark size=0.5mm]
plot coordinates {
    ( 76.36, 1451.4 )
    ( 79.62, 817.7 )
    ( 81.21, 445.8 )
    ( 81.6, 241.0 )
    ( 81.64, 225.4 )
    ( 81.8, 127.7 )
    ( 81.87, 65.0 )
};
\addplot[line width=0.15mm,color=norm5,mark=star,mark size=0.5mm]
plot coordinates {
    ( 71.12, 2012.1 )
    ( 73.94, 1107.4 )
    ( 75.09, 599.5 )
    ( 75.13, 565.9 )
    ( 75.44, 317.7 )
    ( 75.69, 203.1 )
    ( 75.75, 180.0 )
    ( 75.82, 119.5 )
};

\end{axis}
\end{tikzpicture}\hspace{0.5mm}

\caption{Varying Label Distributions}\label{fig:recall-var-label-dis}\vspace{-2ex}

\end{footnotesize}
\end{figure*}

\begin{figure}[h]
\centering
\begin{footnotesize}
\begin{tikzpicture}
    \begin{customlegend}[legend columns=4,
        legend entries={$\ELI$-0.2,$\ELI$-2.0,$\ELI'$-0.2,$\ELI'$-2.0 },
        legend style={at={(0.45,1.15)},anchor=north,draw=none,font=\scriptsize,column sep=0.1cm}]
    \addlegendimage{line width=0.15mm,color=norm2,mark=triangle,mark size=0.5mm}
    \addlegendimage{line width=0.15mm,color=blue1,mark=pentagon,mark size=0.5mm}
    \addlegendimage{line width=0.15mm,color=forestgreen,mark=o,mark size=0.5mm}
    \addlegendimage{line width=0.15mm,color=amber,mark=square,mark size=0.5mm}
    \end{customlegend}
\end{tikzpicture}

\subfloat[SIFT]{\vgap
\begin{tikzpicture}[scale=1]
\begin{axis}[
height=\columnwidth/2.50,
width=\columnwidth/1.80,
xlabel=recall@10(\%),
ylabel=Qps,
ymode=log,
label style={font=\scriptsize},
tick label style={font=\scriptsize},
title style={font=\scriptsize},
title style={yshift=-2.5mm},
ymajorgrids=true,
xmajorgrids=true,
grid style=dashed,
]
\addplot[line width=0.15mm,color=norm2,mark=triangle,mark size=0.5mm]
plot coordinates {
(85.241, 8278.15)
(91.245, 6369.43)
(96.77, 3895.6)
(99.1, 2320.72)
(99.812, 1285.18)
(99.974, 752.049)
};
\addplot[line width=0.15mm,color=blue1,mark=pentagon,mark size=0.5mm]
plot coordinates {
(88.565, 3573.98)
(93.523, 2526.53)
(97.671, 1486.55)
(99.381, 838.434)
(99.876, 512.847)
(99.98, 307.702)
};

\addplot[line width=0.15mm,color=forestgreen,mark=square,mark size=0.5mm]
plot coordinates {
(85.36, 7423.9)
(91.328, 5302.23)
(96.797, 3276.54)
(99.16, 1993.22)
(99.821, 1123.97)
(99.97, 679.209)
};

\addplot[line width=0.15mm,color=amber,mark=o,mark size=0.5mm]
plot coordinates {
(88.312, 3859.51)
(93.27, 2732.24)
(97.511, 1575.3)
(99.336, 915.835)
(99.857, 529.129)
(99.98, 334.359)
};

\end{axis}
\end{tikzpicture}\hspace{0.5mm}
}
\subfloat[Arxiv]{\vgap
\begin{tikzpicture}[scale=1]
\begin{axis}[
height=\columnwidth/2.50,
width=\columnwidth/1.80,
xlabel=recall@10(\%),
ylabel=Qps,
ymode=log,
label style={font=\scriptsize},
tick label style={font=\scriptsize},
title style={font=\scriptsize},
title style={yshift=-2.5mm},
ymajorgrids=true,
xmajorgrids=true,
grid style=dashed,
]
\addplot[line width=0.15mm,color=norm2,mark=triangle,mark size=0.5mm]
plot coordinates {
(92.26, 617.665)
(95.48, 550.055)
(98.42, 391.543)
(99.5, 251.004)
(99.94, 145.243)
};
\addplot[line width=0.15mm,color=blue1,mark=pentagon,mark size=0.5mm]
plot coordinates {
(92.72, 606.428)
(95.85, 504.286)
(98.62, 346.861)
(99.58, 212.495)
(99.92, 127.226)
};

\addplot[line width=0.15mm,color=forestgreen,mark=square,mark size=0.5mm]
plot coordinates {
(91.57, 549.451)
(95.07, 504.032)
(98.25, 377.216)
(99.44, 247.463)
(99.89, 160.514)
(99.96, 98.1162)

};

\addplot[line width=0.15mm,color=amber,mark=o,mark size=0.5mm]
plot coordinates {
(92.41, 529.661)
(95.78, 438.212)
(98.52, 338.868)
(99.58, 215.332)
(99.95, 133.049)
};

\end{axis}
\end{tikzpicture}\hspace{0.5mm}
}

\caption{Query Performance Drop Caused by Incremental Updates}\label{fig:var-dynamic}\vspace{-2ex}
\end{footnotesize}
\end{figure}

\begin{figure}[htbp]
    \centering
    \includegraphics[width=0.99\linewidth]{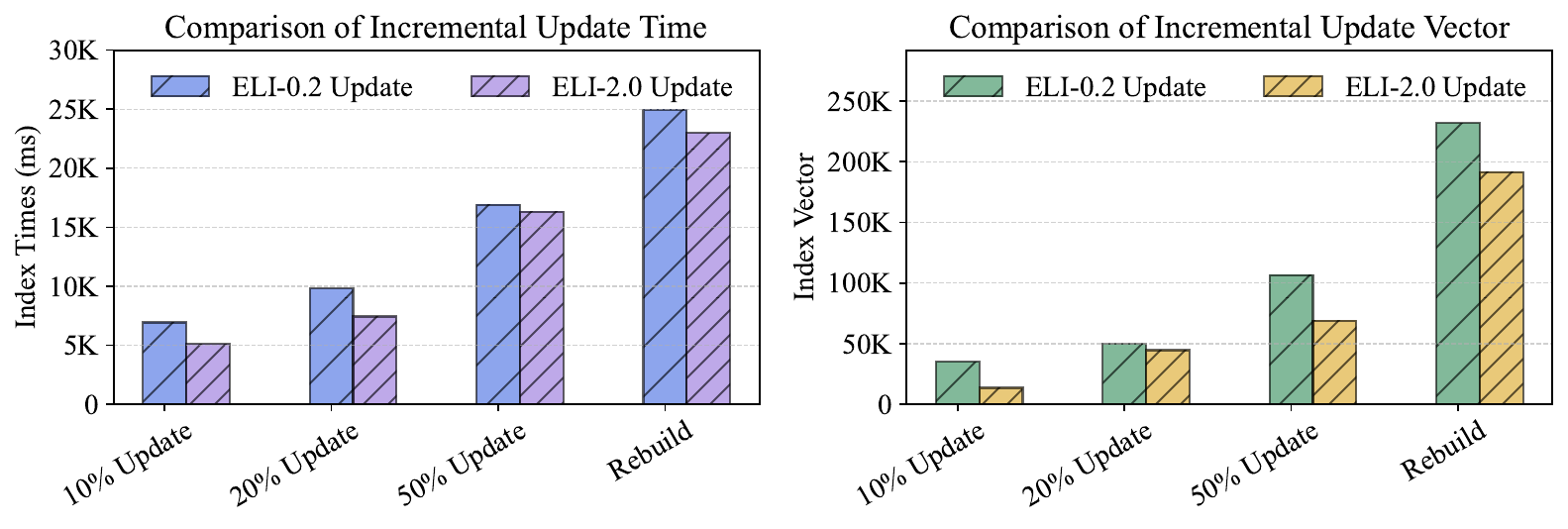}
\caption{The Time and Space Cost of Incremental Updates}\vspace{-2ex}
    \label{fig:dynamic-static}
\end{figure}

\stitle{Exp-6: Index Update Evaluation.}  
We evaluate the effectiveness of our index-update strategy on the Arxiv and \textsc{SIFT} datasets.  
Each dataset is partitioned into two halves.  
The first 50\% of the data, with label-set size $|\Sigma| = 12$, is used to construct the initial index set $\mathbb{I}$.  
The remaining 50\%, featuring an expanded label set of $|\Sigma| = 24$, simulates the insertion of both new vectors  
(50\% more vectors) and an enlarged query workload (12 additional labels).

\sstitle{Query efficiency.}  
Fig.~\ref{fig:var-dynamic} reports query efficiency.  
$\ELI$-0.2 and $\ELI$-2.0 denote the methods rebuilt from scratch using the full dataset,  
whereas $\ELI'$-0.2 and $\ELI'$-2.0 represent our batch-update strategy,  
which first builds indexes on the initial 50\% of data and then incrementally updates them.  
The results show that periodic updates incur less than a 5\% drop in query efficiency compared with full re-computation,  
demonstrating that the batch update mechanism preserves practical performance.

\sstitle{Update cost.}  
Figure~\ref{fig:dynamic-static} compares the update cost of $\ELI'$-0.2 and $\ELI'$-2.0 against rebuilding from scratch.  
The left panel shows update time: the rightmost bar corresponds to full re-computation on 100\% of the data,  
while the three bars to the left represent incremental insertions of 10\%, 20\%, and 50\% new vectors  
into the initial index built on the first 50\% of the data.  
Update time naturally increases with the number of inserted vectors but remains well below the cost of rebuilding,  
even when 50\% of the data is added.  
The right panel reports space usage (in units of updated vectors).  
As more vectors are inserted, space usage grows slightly but remains smaller than that of full re-computation.  
Notably, both our methods, especially $\ELI'$-2.0, use significantly less space than their rebuilt counterparts.

Overall, these experiments show that our incremental update strategy  
achieves significantly lower update time and space overhead compared with rebuilding from scratch(Fig.~\ref{fig:dynamic-static}),  
while incurring only negligible loss in query efficiency (Fig.~\ref{fig:var-dynamic}).

\section{Related Work}

\stitle{Attribute-Hybrid AKNN Search.}
When vectors carry label attributes (as defined in this paper) or numerical attributes (e.g.,~\cite{iRangeGraph-SIGMOD-2025}), an \textbf{attribute-hybrid} AKNN search can be defined to return approximate $k$-nearest neighbors whose attributes satisfy given constraints.  
Such searches can be handled using either attribute-free or attribute-dependent solutions.

\sstitle{Attribute-Free Solutions}. 
These methods ignore attributes during index construction and consider them only at query time.  
Specifically, the $\PRE$ and $\POST$ search strategies are widely used, each performing differently under varying query workloads.  
Systems such as ADB~\cite{AnalyticDB-VLDB-2020}, VBASE~\cite{Vbase-OSDI-2023}, CHASE~\cite{CHASE-arxiv-Sean-Wang}, and Milvus~\cite{Milvus-SIGMOD-2021} choose between these strategies using cost models.  
$\ACORN$ also follows the $\PRE$ strategy but builds a denser graph index to mitigate its connectivity issues.  
Because these approaches ignore attributes during index construction, their search accuracy and efficiency lag behind methods that incorporate attributes directly.

\sstitle{Attribute-Dependent Solutions}. 
These methods build indexes by considering the attribute type.  
(1) For attribute-hybrid AKNN search with \emph{numerical} attributes, recent work~\cite{iRangeGraph-SIGMOD-2025,Window-Filter-ICML-2024,ESG-arxiv-2025-Mingyu} organizes query-specified numeric intervals using segment trees, while SeRF~\cite{SeRF-SIGMOD-2024} further compresses indexes across overlapping intervals.  
(2) For attribute-hybrid AKNN search with \emph{label} attributes, $\UNG$~\cite{UNG-SIGMOD-2025} exploits label-set inclusion, adding cross-group edges so the graph index searches only filtered vectors.  
NHQ~\cite{NHQ-NIPS-2022-mengzhao-wang} incorporates attribute similarity into vector similarity, requiring manual tuning of vector and attribute weights.

\stitle{Data Cube Computation.}
Existing studies on OLAP and data-cube computation select group-by aggregates over the subset lattice of dimensions, exploiting the fact that coarser aggregates (supersets) can be derived from finer-grained materializations (subsets)~\cite{Impl-data-cube-SIGMOD-1996,View-Select-SIGMOD-2012,DATA-Cube-KDD}.  
Our setting shares this lattice structure and the reuse principle: an index built for a query-label set can serve queries with superset labels via label containment.  
However, key differences remain:  
(1) Cube computation relies on algebraic properties of aggregates (e.g., distributive or algebraic measures) to derive superset results exactly from subsets, whereas our reuse is containment-based, yielding a query cost bounded by $O(C + k/c)$ on graph-based indexes, where $c$ is the elastic factor and $k/c$ is the additional cost.
(2) Cube/view selection minimizes aggregate-query latency subject to storage limits, while our Elastic Index Selection ($\EIS$) problem selects a subset of query-label sets to index so as to satisfy both space budgets and workload-wide elastic-factor coverage, ensuring predictable AKNN efficiency under label constraints.
(3) Although both problems are NP-hard and typically approached with greedy heuristics, our benefit function and feasibility constraints are tailored to vector-search semantics and the label-containment relationship.

\stitle{Frequent Itemset Mining.}
Our work is conceptually related to Frequent Itemset Mining (FIM), where labels correspond to items and label sets to transactions~\cite{Data-Mining-Concepts-Jiawei-Han,Mining-Frequent-Patterns-SIGMOD-2000-Jiawei-Han,Scalable-TKDE-Item-Mining}.  
Both exploit the anti-monotone property of support (the Apriori principle), with classic algorithms such as Apriori and FP-Growth widely used for FIM~\cite{Mining-Frequent-Patterns-SIGMOD-2000-Jiawei-Han,Data-Mining-Concepts-Jiawei-Han}.  
The objectives, however, differ fundamentally.  
FIM finds all itemsets whose support exceeds a threshold for pattern discovery, whereas our $\ELI$ task is an optimization problem: selecting a subset of query-label sets on which to build indexes.  
This selection is guided not by absolute support but by a novel \emph{elastic factor} modeling query efficiency.  
Thus, while FIM outputs frequent patterns, our method produces a concrete indexing plan for AKNN search, making the formulation and goals distinct.

\section{Conclusion}
This paper investigates the label-hybrid AKNN search problem and shows that  
an index built for a query–label set remains effective for all its supersets with bounded query time.  
Building on this insight, we formalize the \textbf{Elastic Index Selection} ($\mathsf{EIS}$) problem,  
prove that $\mathsf{EIS}$ is NP-complete, and introduce two optimization variants solved by efficient greedy algorithms.  
Extensive experiments demonstrate the effectiveness and superiority of our approach.  
As future work, we plan to explore hardware acceleration for further speedups.

\begin{acks}
We are grateful to the anonymous reviewers for their constructive comments. Wentao Li is supported by NSFC (Grant No.62302417). Raymond Chi-Wing Wong is supported by PRP/004/25FX.
Wei Wang is supported by HKUST(GZ)-IEIP-RoP (G01RF000256), Guangdong Provincial Key Lab of Integrated Communication, Sensing and Computation for Ubiquitous Internet of Things (No.2023B1212010007, SL2023A03J00934), Guangzhou Municipal Science and Technology Project (No. 2023A03J0003, 2023A03J0013, and 2024A03J0621). This work is also supported by Ant Group.
\end{acks}

\balance
\bibliographystyle{ACM-Reference-Format}
\bibliography{sample}

\newpage
\nobalance
\section*{A Appendix}

\subsection*{A.1 Proofs}

\stitle{Proof of Lemma~\ref{lem:filter-search-complexity}.}
We analyze the expected search time of the $\POST$ strategy on a graph index by decomposing it into two phases:  
(1) the initial search to locate the first nearest neighbor (top-1 result), and  
(2) iterative expansion to collect $k$ neighbors satisfying the query’s label constraint.

\sstitle{Phase 1: Initial Search}  
Let $C$ denote the expected cost of finding the top-1 neighbor of query $q$ on the underlying graph index.  
This abstracts the entry cost of popular structures such as HNSW or NSG, typically analyzed as $O(\log N)$ under standard assumptions~\cite{Acorn-SIGMOD-2024,NSG-VLDB-2019-deng-cai}, where $N$ is the number of indexed points.

\sstitle{Phase 2: Iterative Expansion}  
After the first match, $\POST$ inspects candidates in increasing distance order until $k$ items satisfying $L_q$ are found.  
Efficiency depends on the probability that a candidate satisfies the label constraint.  
We assume (i) label attributes are independent of vector positions, and (ii) label-matched points are uniformly distributed in the space and the number is larger than $k$.  
If the elastic factor satisfies $e\big(S(L_q), \mathbb{I}\big) \ge c$, then at least a $c$-fraction of candidates within the selected index set $\mathbb{I}$ match $L_q$.  
Under these assumptions, the probability that any visited candidate matches $L_q$ is $p \ge c$.  
Finding $k$ matches is therefore a sequence of Bernoulli trials with success probability $p$ without replacement, so the number of trials $T$ follows a negative-hypergeometric distribution with expectation $\mathbb{E}[T] = k\frac{N+1}{pN+1} < k/c$.  
Modern graph-based indexes~\cite{hnswlib,SymphonyQG-SIGMOD-2025,Diskann-NIPS-2019,NSG-VLDB-2019-deng-cai} retrieve the next neighbor in $O(1)$ amortized time because each node maintains a bounded-degree neighbor list.  
Hence, the expected cost of this phase is $O(k/c)$.

\sstitle{Total Cost}  
Summing both phases, the expected time to obtain the top-$k$ label-hybrid AKNN results is  $O\!\left(C + \frac{k}{c}\right)$.
This matches and generalizes prior analyses of post-filtering strategies, such as HNSW-based filtered search~\cite{Acorn-SIGMOD-2024}, within our elastic-factor framework.  
As with HNSW and NSG, the bound captures \emph{expected} complexity and does not guarantee worst-case recall, but it aligns with recent NSG analyses~\cite{ESG-arxiv-2025-Mingyu} that report the same $O(C + k/c)$ cost under similar conditions.

\stitle{Proof of Theorem~\ref{thm:np-hard}.}
Given a dataset $S$, a query workload with label sets $\mathcal{L} = \{L_1,\dots,L_n\}$,  and a universal index set $\mathcal{I} = \{I_1,\dots,I_n\}$  (where each $I_i \in \mathcal{I}$ is built on a label set $L_i \in \mathcal{L}$),  the $\EIS$ problem asks whether there exists a subset  $\mathbb{I} \subseteq \mathcal{I}$ such that for every query–label set  $L_q \in \mathcal{L}$ we can find an index $I \in \mathbb{I}$ with  elastic factor $e(S(L_q), I)$ greater than a threshold $c$, while the total cost satisfies  $\sum_{I \in \mathbb{I}} |I| \le \tau$.

The problem is in NP as we can verify in polynomial time that  (1) the total cost $\sum_{I \in \mathbb{I}} |I|$ does not exceed the budget $\tau$, and  (2) each query–label set is covered by some selected index with elastic factor at least $c$. 
    To show NP-completeness, we reduce from the NP-complete  \textbf{3-Set Cover (3-$\SC$)} problem~\cite{Karp-NP-NPC-NP-Reduce,k-set-cover-STOC-1997}.

\begin{definition}[3-Set Cover (3-$\SC$)]\label{defn:3-SC}
\begin{itemize}[leftmargin=8\labelsep]
    \item[\textbf{Input}]  A universal set $\mathcal{U}=\{u_1,u_2,\ldots,u_p\}$ of $p$ elements and a collection  $\mathcal{S}=\{s_1,s_2,\ldots,s_l\}$ of $l$ subsets, where each $s_i$ has size at most $3$ and  $\bigcup_{i=1}^{l} s_i = \mathcal{U}$. Let the selection budget be $k$.
    \item[\textbf{Output}] Determine if there exists a subset $\mathbb{S}\subseteq \mathcal{S}$ of size at most $k$ such that $\bigcup_{s_i\in \mathbb{S}} s_i = \mathcal{U}$.
\end{itemize}
\end{definition}

\begin{figure}[!t]
    \centering
    \includegraphics[width=0.89\columnwidth]{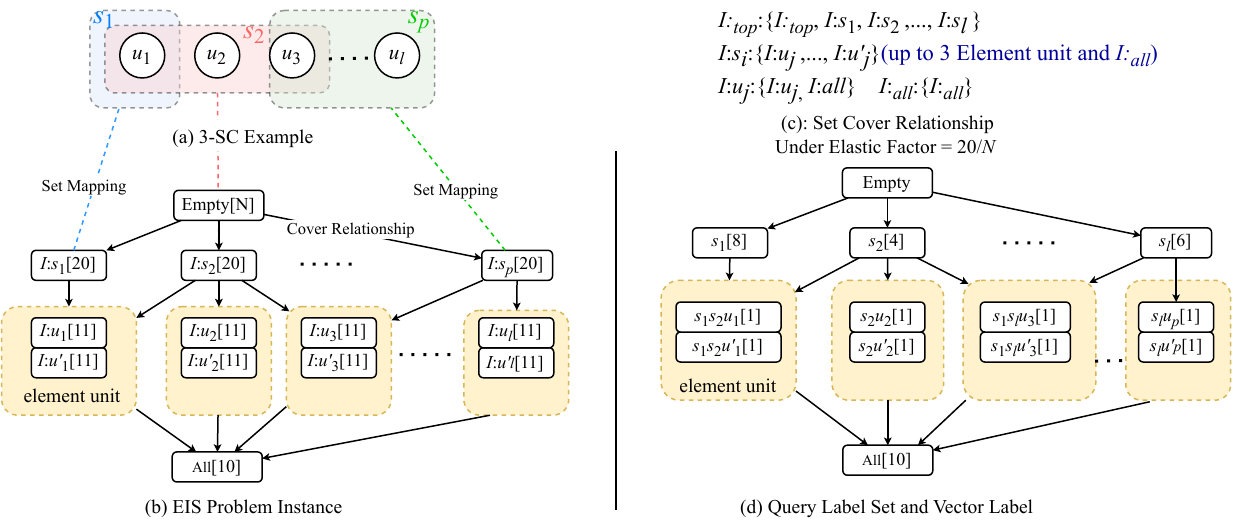}
    \caption{
    \textbf{Proof of NP-completeness.}  
    Set $D = b = 10$, so $A = b + 1 = 11$, $B = 2b = 20$, and $N = 50 > 4b$.
    Then $c = \frac{B}{N} = \frac{20}{50} = 0.4$.
    Thus we have $10$ vectors in $I_{\mathrm{all}}$ and $50$ vectors in $I_{\top}$.  
    Each element index $I_{u_i}$ is duplicated with a dummy index $I_{u'_i}$, and each subset index $I_{s_j}$ connects to at most three element indexes $I_{u_i}$ or $I_{u'_i}$.  Moreover, $|I_{s_j}|/|I_{\top}| = 0.4 = c, |I_{u_i}|/|I_{\top}| = \frac{11}{50} = 0.22 < c$, $|I_{u_i}|/|I_{s_i}|=\frac{11}{20}>c$.
    }\label{fig:np-hard-appendix}
\end{figure}

\stitle{Construction.}
Given a 3-$\mathsf{SC}$ instance $(\mathcal{U}, \mathcal{S}, k)$, we create a corresponding $\EIS$ instance (see Fig.~\ref{fig:np-hard-appendix}). 
Let the alphabet be $\Sigma = \mathcal{U} \cup \mathcal{S}$, and define four types of indexes as follows:
(1) for each element $u_i \in \mathcal{U}$, an \textbf{element index} $I_{u_i}$ of size $|I_{u_i}| = A$; 
(2) for each subset $s_j \in \mathcal{S}$, a \textbf{subset index} $I_{s_j}$ of size $|I_{s_j}| = B$; 
together with two special indexes: 
(3) a \textbf{top index} $I_{\top}$ (Empty[N] in Fig.~\ref{fig:np-hard-appendix}) containing all base vectors, with size $|I_{\top}| = N$; and 
(4) a \textbf{bottom index} $I_{\mathrm{all}}$ (All[D] in Fig.~\ref{fig:np-hard-appendix}) containing vectors matching all labels in $\Sigma$, with size $|I_{\mathrm{all}}| = D$. 
The top index $I_{\top}$, required for AKNN search, always exists, and we omit its cost for simplicity. 
The costs $A, B, D$ and the elastic-factor threshold $c$ are set to satisfy the following conditions.

\noindent(1)~Each subset index $I_{s_j}$ covers an element index $I_{u_i}$ with overlap ratio at least $c$  
(i.e., $I_{u_i} \subseteq I_{s_j}$, $|I_{u_i}|/|I_{s_j}| \ge c$) if and only if $u_i \in s_j$.  

\noindent(2)~The top index $I_{\top}$ covers every subset index $I_{s_j}$ with overlap ratio $c$,  
while its overlap ratio with any element index $I_{u_i}$ is less than $c$.

\noindent(3)~To cover element/label set $u_i$, selecting the element index $I_{u_i}$ incurs higher cost than selecting any subset index $I_{s_j}$, where $u_i \in s_j$.

\sstitle{Feasibility}. 
We next show how to satisfy the above conditions.
$\bullet$ Set the cost of $I_{\mathrm{all}}$ to $D = b > 3$ (as in Fig.~\ref{fig:np-hard-appendix}(b), where $b = 10$ and $I_{\mathrm{all}}$ contains vectors covering all labels in $\Sigma$). 
$\bullet$ Let $A = b + 1$, $B = 2b$, and set the elastic–factor threshold to $c = \tfrac{2b}{N}$ with $N > 4b$. 
If $N \le 4b$, we pad $I_{\top}$ with vectors carrying no labels to increase $N$ (as in Fig.~\ref{fig:np-hard-appendix}(b), where $N > 40$).
$\bullet$ Each element index $I_{u_i}$ is built from the $b$ vectors in $I_{\mathrm{all}}$ plus one additional vector whose label set is $s_1 \cdots s_x u_i$, 
where $s_1, \dots, s_x$ are the subsets containing $u_i$. 
Thus $|I_{u_i}| = b + 1 = A$, created on the query–label set $L_q = \{s_1, \dots, s_x, u_i\}$.
For this additional vector, we also create another vector with label set $s_1 \cdots s_x u'_i$, 
where $u'_i$ is a duplicate label equivalent in meaning to $u_i$ 
(i.e., if $u_i \in s_j$ then $u'_i \in s_j$). 
Accordingly, we introduce another element index $I_{u'_i}$ of size $b + 1$, 
consisting of the $b$ vectors from $I_{\mathrm{all}}$ plus one vector with labels $s_1 \cdots s_x u'_i$, created on the query–label set $L_q = \{s_1, \dots, s_x, u'_i\}$.  
$\bullet$ Each subset index $I_{s_j}$ is created on the query–label set $L_q = \{s_j\}$. 
Since each subset $s_j$ contains at most three elements $u_i$ (and their duplicates $u'_i$), 
it covers at most $3 \times 2$ vectors of the form $s_1 \cdots s_x u_i$ or $s_1 \cdots s_x u'_i$, 
together with the $b$ vectors in $I_{\mathrm{all}}$. 
We then append up to $2b - 6$ extra vectors labeled $s_j$ so that the total number of vectors matching $s_j$ equals $B = 2b$.  

In summary, the query workload is $\mathcal{L} = \{\{s_1\}, \dots, \{s_l\}, \{s_1 \cdots s_x u_i\}, \dots, \{s_1 \cdots s_x u'_i\}, \emptyset, \Sigma\}$, and the corresponding universal index set is $\mathcal{I} = \{I_{s_1}, \dots, I_{s_l}, I_{u_1}, \dots, I_{u_p}, I_{\top}, I_{\mathrm{all}}\}$.

\sstitle{Conditional Checking.}  
Note that $c = \frac{2b}{N} < \frac{2b}{4b} = \tfrac{1}{2}$.  
For Condition (1), when $u_i \in s_j$, the overlap between $I_{u_i}$ and $I_{s_j}$ equals $|I_{u_i}|$,  
so the overlap ratio is $\frac{|I_{u_i}|}{|I_{s_j}|} = \frac{A}{B} = \frac{b+1}{2b} > \tfrac{1}{2} > c$.  
For Condition (2), for each $s_j$, the overlap ratio with $I_{\top}$ is $\frac{|I_{s_j}|}{|I_{\top}|} = \frac{B}{N} = \frac{2b}{N} = c$.  
For each $u_i$, it is $\frac{|I_{u_i}|}{|I_{\top}|} = \frac{A}{N} = \frac{b+1}{N}$.  
Since $2b > b + 1$ and $b > 3$, we have $\frac{b+1}{N} < c$, satisfying the condition.  
For Condition (3), to cover vectors containing label $u_i$, selecting a subset index $I_{s_j}$ with $u_i \in s_j$ costs $B = 2b$.  
Selecting the element index $I_{u_i}$ also requires selecting $I_{u’_i}$ (otherwise, the vector with label set $s_1 \cdots s_x u’_i$ would not be included),
for a total cost of $2A = 2(b + 1)$.  
As $2b < 2(b + 1)$, choosing $I_{s_j}$ is cheaper, so this condition holds.

\stitle{Correctness.}
Note that $I_{\top}$ must always be selected but incurs zero cost,  
while $I_{\mathrm{all}}$ need not be selected since it is already covered by other indexes.
We set the space budget $\tau = k \cdot B$, which permits selecting at most $k$ subset indexes (by Condition~(3)).

\sstitle{3-$\mathsf{SC} \Rightarrow \mathsf{EIS}$.}  
If the 3-$\mathsf{SC}$ instance admits a cover $\mathbb{S} \subseteq \mathcal{S}$ with $|\mathbb{S}| \le k$,  
select the corresponding subset indexes $\{ I_{s_j} : s_j \in \mathbb{S} \}$.  
Each $I_{s_j}$ is created on the query–label set $s_j$ and covers all vectors whose label set is $u_i$ for every $u_i \in s_j$.  
The overlap ratio is at least $c$, so the elastic factor between $I_{s_j}$ and both $s_j$ and $u_i$ is at least $c$. 
This solves $\mathsf{EIS}$ because (1) the top index $I_{\top}$ covers the query-label sets $\mathcal{S}$ (as well as $\emptyset$) in $\mathcal{L}$ with elasticity factor $c$, and (2) the index built on the selected $\mathbb{S}$ covers the query-label sets $\{\{s_1 \cdots s_x u_i\}, \dots, \{s_1 \cdots s_x u'_i\}$ (as well as $\Sigma$, since $|I_{\mathrm{all}|/|I_{s_j}|}| = b/2b > c$) in $\mathcal{L}$ with elasticity factor $c$.

\sstitle{3-$\mathsf{SC} \Leftarrow \mathsf{EIS}$.}  
Conversely, suppose the $\mathsf{EIS}$ instance admits a feasible selection $\mathbb{I}$.  
By Condition~(3), any $I_{u_i} \in \mathbb{I}$ can be replaced by some $I_{s_j}$ containing $u_i$  
without violating constraints or increasing total cost.  
Thus we may assume the solution consists solely of subset indexes after such replacements.  
Let $\mathbb{S} = \{ s_j : I_{s_j} \in \mathbb{I} \}$.  
Because every $u_i$ (in the query workload $\mathcal{U} \subseteq \mathcal{L}$) must be covered by some subset index (with overlap ratio at least $c$), we have  
$\bigcup_{s_j \in \mathbb{S}} s_j = \mathcal{U}$.  
The budget enforces $|\mathbb{S}| \le \tau / B = k$, yielding a valid 3-$\mathsf{SC}$ solution.

\section*{A.2 Extra Discussions}

\stitle{Other Types of Label Constraints.}  
Our work primarily addresses the \emph{label-containment} constraint,  
where a vector's label set must contain the query label set.  
Other label constraints can also be supported:

\noindent
\sstitle{(1) Label equality.}  
Here, we require base vectors whose label set $L_i$ exactly matches the query label set $L_q$.  
To handle AKNN search under this constraint, we group base vectors by their label sets to form a label-set inverted list.  
An index is then built for each group, and the overall dataset cardinality remains unchanged.  
For example, an index for $L_q = \{A\}$ is built over base vectors $x_2$ and $x_5$, both labeled $\{A\}$.  
When using a graph-based index—the most efficient structure to date—the total space complexity is $O(NM)$,  
where $M$ is the maximum node degree constrained by edge occlusion.

\noindent
\sstitle{(2) Label overlap.}  
Here we require that a vector's label set intersects the query label set, i.e., $L_i \cap L_q \neq \emptyset$.  
To support AKNN search under this type of label constraint, we decompose the query into $|L_q|$ subqueries,  
each with a single-label containment constraint.  
For example, a label-overlap query with label set $L_q = \{AB\}$ can be answered by merging the results of  
label-containment queries with label sets $L_q = \{A\}$ and $L_q = \{B\}$ and taking their union.
These extensions show that our approach naturally generalizes beyond simple label containment,
while we focus on the containment case throughout this paper.

\stitle{Vector Similarity Search.}  
Vector similarity search is closely related to AKNN search and has been extensively studied.  
Most prior work targets approximate search~\cite{LSH-1999},  
since exact search is prohibitively expensive in high-dimensional spaces~\cite{Curse-of-dim-1998}.  
Among these methods, $c$-approximate nearest neighbor search returns results within a factor $c$ of the true distance  
and can be solved in sublinear time using locality-sensitive hashing (LSH)  
~\cite{LSH-Pstable-2004,RPLSH-1995,LSH-1999,SRS-yifang-2014,PMLSH-bolong-2020,C2LSH-2012-SIGMOD,QALSH-2015-VLDB}.  
For the AKNN search problem in this paper, graph-based vector indexes 
~\cite{NSG-VLDB-2019-deng-cai,SSG-PAMI-2022-deng-cai,Starling-SIGMOD-2024-Mengzhao,tMRNG:journals/pacmmod/PengCCYX23,Diskann-NIPS-2019,HNSW-PAMI-2020,HVS-VLDB-2021-kejing-lu,SymphonyQG-SIGMOD-2025}  
represent the current state of the art and are significantly more efficient than LSH according to benchmarks  
~\cite{ANNSurvey-TKDE-2020-Wei-Wang}.  
In addition, inverted-list indexes~\cite{PQ-PAMI-2014,IMI-PAMI-2015} 
and quantization-based methods  
~\cite{PQ-PAMI-2014,OPQ-PAMI-2014,LOPQ-2014-CVPR,ITQ:/pami/GongLGP13,Rabitq-SIGMOD-2024,ExRaBitQ-arxiv-2024,MRQ-RESQ-arxiv-2024-Mingyu}  are widely used for AKNN search.  
Inverted indexes provide low space overhead, while quantization techniques  
speed up distance computations~\cite{BSA-DDC-ICDE-2024-Mingyu,Graph-Index-Accelearte-SIGMOD-2025-Mengzhao}.  
In this paper, we employ the widely adopted $\HNSW$ algorithm~\cite{hnswlib}  
as the underlying index for each vector group, although other optimized AKNN libraries  
~\cite{Diskann-NIPS-2019,VSAG-arxiv-2025}  
can be used as drop-in replacements.

\subsection*{A.3 Extra Experiments}

\stitle{Exp-7: Test of Label-Overlap Constraints.}  
We previously discussed how our method for AKNN search with label-containment constraints  
can be extended to handle label-overlap constraints.  
Specifically, a query with label overlap is decomposed into multiple single-label containment sub-queries;  
We solve each sub-query independently and then merge the results.
We evaluate this approach on both synthetic SIFT data with a Zipf distribution  
and the real-world \textsc{Paper} dataset, as shown in Fig.~\ref{fig:Overlap}.  
The figure shows that, despite requiring multiple search calls for single-label containment sub-queries,  
our methods $\ELI$-0.2 and $\ELI$-2.0 achieve search performance comparable to $\ACORN$-$\gamma$,  
while delivering $2\times$–$10\times$ the efficiency of $\ACORN$ and $\UNG$.
Since our current framework does not treat overlap queries as part of the optimization workload, we plan to explore extensions of our greedy method to support such queries  
and, more broadly, other complex label constraints.

\begin{figure}[t]
\centering
\begin{footnotesize}
\begin{tikzpicture}
    \begin{customlegend}[legend columns=5,
        legend entries={$\UNG$,$\ELI$-0.2,$\ELI$-2.0,$\ACORN$-1,$\ACORN$-$\gamma$},
        legend style={at={(0.45,1.15)},anchor=north,draw=none,font=\scriptsize,column sep=0.1cm}]
    \addlegendimage{line width=0.15mm,color=norm1,mark=o,mark size=0.5mm}
    \addlegendimage{line width=0.15mm,color=norm2,mark=triangle,mark size=0.5mm}
    \addlegendimage{line width=0.15mm,color=norm3,mark=square,mark size=0.5mm}
    \addlegendimage{line width=0.15mm,color=norm4,mark=otimes,mark size=0.5mm}
    \addlegendimage{line width=0.15mm,color=norm5,mark=star,mark size=0.5mm}
    \end{customlegend}
\end{tikzpicture}
\\[-\lineskip]

\subfloat[SIFT]{\vgap

\begin{tikzpicture}[scale=1]
\begin{axis}[
height=\columnwidth/2.60,
width=\columnwidth/1.80,
xlabel=recall@10(\%),
ylabel=Qps,
ymode=log,
title={Label Overlap(\textbf{Zipf},$|\Sigma|$=12)},
label style={font=\scriptsize},
tick label style={font=\scriptsize},
title style={font=\scriptsize},
title style={yshift=-2.5mm},
ymajorgrids=true,
xmajorgrids=true,
grid style=dashed,
]
\addplot[line width=0.15mm,color=norm1,mark=o,mark size=0.5mm]
plot coordinates {
    ( 83.64, 606.907 )
    ( 88.165, 388.304 )
    ( 92.409, 207.68 )
    ( 95.464, 117.499 )
    ( 96.752, 79.956 )
    ( 98.017, 49.388 )
};
\addplot[line width=0.15mm,color=norm2,mark=triangle,mark size=0.5mm]
plot coordinates {
    ( 87.898, 4748.34 )
    ( 94.962, 2698.33 )
    ( 98.466, 1484.12 )
    ( 99.645, 815.993 )
    ( 99.945, 454.36 )
};
\addplot[line width=0.15mm,color=norm3,mark=square,mark size=0.5mm]
plot coordinates {
    ( 88.703, 2710.03 )
    ( 95.412, 1564.46 )
    ( 98.567, 926.698 )
    ( 99.663, 513.242 )
    ( 99.945, 284.228 )
};
\addplot[line width=0.15mm,color=norm4,mark=otimes,mark size=0.5mm]
plot coordinates {
    ( 80.344, 879.894 )
    ( 88.326, 453.885 )
    ( 89.007, 409.87 )
    ( 93.769, 226.552 )
    ( 96.506, 125.302 )
    ( 96.99, 110.305 )
    ( 98.699, 49.425 )
    ( 99.524, 20.15 )
};
\addplot[line width=0.15mm,color=norm5,mark=star,mark size=0.5mm]
plot coordinates {
    ( 73.989, 5636.98 )
    ( 86.349, 3270.11 )
    ( 92.924, 889.759 )
    ( 96.362, 1055.41 )
    ( 98.218, 726.639 )
    ( 98.358, 682.64 )
    ( 99.194, 386.16 )
    ( 99.594, 220.332 )
    ( 99.66, 192.145 )
    ( 99.885, 85.581 )
    ( 99.965, 32.468 )
};

\end{axis}
\end{tikzpicture}\hspace{0.5mm}
}
\subfloat[PAPER]{\vgap

\begin{tikzpicture}[scale=1]
\begin{axis}[
height=\columnwidth/2.60,
width=\columnwidth/1.80,
xlabel=recall@10(\%),
ylabel=Qps,
ymode=log,
title={Label Overlap(\textbf{Real},$|\Sigma|$=7)},
label style={font=\scriptsize},
tick label style={font=\scriptsize},
title style={font=\scriptsize},
title style={yshift=-2.5mm},
ymajorgrids=true,
xmajorgrids=true,
grid style=dashed,
]
\addplot[line width=0.15mm,color=norm1,mark=o,mark size=0.5mm]
plot coordinates {
    ( 99.852, 647.04 )
    ( 99.944, 161.689 )
    ( 99.958, 106.631 )
    ( 99.956, 59.865 )
    ( 99.971, 33.113 )
    ( 99.978, 23.311 )
    ( 99.968, 15.334 )
};
\addplot[line width=0.15mm,color=norm2,mark=triangle,mark size=0.5mm]
plot coordinates {
    ( 97.024, 2500.0 )
    ( 99.097, 1376.84 )
    ( 99.709, 756.773 )
    ( 99.893, 418.866 )
    ( 99.956, 237.057 )
};
\addplot[line width=0.15mm,color=norm3,mark=square,mark size=0.5mm]
plot coordinates {
    ( 96.964, 2432.5 )
    ( 99.133, 1263.26 )
    ( 99.756, 657.635 )
    ( 99.915, 372.038 )
};
\addplot[line width=0.15mm,color=norm4,mark=otimes,mark size=0.5mm]
plot coordinates {
    ( 72.122, 2955.96 )
    ( 86.819, 1688.33 )
    ( 93.743, 957.304 )
    ( 97.059, 534.274 )
    ( 97.278, 501.555 )
    ( 98.644, 284.164 )
    ( 99.301, 161.687 )
    ( 99.396, 139.661 )
    ( 99.712, 63.719 )
    ( 99.851, 24.769 )
};
\addplot[line width=0.15mm,color=norm5,mark=star,mark size=0.5mm]
plot coordinates {
    ( 88.136, 6816.63 )
    ( 94.791, 4142.5 )
    ( 97.69, 2253.27 )
    ( 98.967, 1439.06 )
    ( 99.585, 810.636 )
    ( 99.626, 707.714 )
    ( 99.837, 232.11 )
    ( 99.925, 206.509 )
};

\end{axis}
\end{tikzpicture}\hspace{0.5mm}
}
\caption{ The Test of Label-Overlap Constraints}\label{fig:Overlap}

\end{footnotesize}
\end{figure}

\stitle{Exp-8: Scalability to Large Datasets.}  
To validate the scalability of our algorithm on large-scale data,  
we adopt the \textsc{DEEP} dataset containing 100M vectors of 96 dimensions.  
Labels follow a Zipf distribution with varying query selectivities, and the results are shown in Fig.~\ref{fig:scale}.  
Because brute-force search achieves only 0.2–0.6 Qps, we omit it from the figure. 
Moreover, $\UNG$ fails to scale, encountering a core dump during index construction on the \textsc{DEEP}100M dataset.
From Fig.~\ref{fig:scale}, $\ACORN$ exhibits $4\times$ worse performance than our $\ELI$-0.2 method  
at 90\% recall and $3\times$ worse than $\ELI$-2.0.  
Our methods maintain consistent performance as the alphabet size $|\Sigma|$ increases,  
whereas $\ACORN$ fails to reach 90\% recall for large $|\Sigma|$.  

We further analyze index-selection statistics of our methods on this dataset by varying the alphabet size,  
with results shown in Fig.~\ref{fig:large-Selection}.  
$\ELI$-2.0 indexes only an additional 100 million vectors with 351 selected index sets---about 1/12 of all base vectors.
In contrast, $\ELI$-0.2 selects more indexes, covering roughly 400 million vectors (about one-third of the 1.2 billion candidates).  
This larger index set pays off: $\ELI$-0.2 achieves $4\times$–$6\times$ higher query efficiency than $\ELI$-2.0.  
These results underscore the trade-off between index size and query efficiency in real-world large-scale scenarios.

\begin{figure}[t]
\centering
\begin{footnotesize}
\begin{tikzpicture}
    \begin{customlegend}[legend columns=4,
        legend entries={$\ELI$-0.2,$\ELI$-2.0,$\ACORN$-1,$\ACORN$-$\gamma$},
        legend style={at={(0.45,1.15)},anchor=north,draw=none,font=\scriptsize,column sep=0.1cm}]
    \addlegendimage{line width=0.15mm,color=norm2,mark=triangle,mark size=0.5mm}
    \addlegendimage{line width=0.15mm,color=norm3,mark=square,mark size=0.5mm}
    \addlegendimage{line width=0.15mm,color=norm4,mark=otimes,mark size=0.5mm}
    \addlegendimage{line width=0.15mm,color=norm5,mark=star,mark size=0.5mm}
    \end{customlegend}
\end{tikzpicture}
\\[-\lineskip]
\subfloat[DEEP (High-Selectivity)]{\vgap
\begin{tikzpicture}[scale=1]
\begin{axis}[
height=\columnwidth/2.60,
width=\columnwidth/1.80,
xlabel=recall@10(\%),
ylabel=Qps,
ymode=log,
title={ Ave. Select 24\% (\textbf{Zipf},$|\Sigma|$=8)},
label style={font=\scriptsize},
tick label style={font=\scriptsize},
title style={font=\scriptsize},
title style={yshift=-2.5mm},
ymajorgrids=true,
xmajorgrids=true,
grid style=dashed,
]
\addplot[line width=0.15mm,color=norm2,mark=triangle,mark size=0.5mm]
plot coordinates {
    ( 75.81, 1834.86 )
    ( 85.69, 1051.52 )
    ( 92.5, 603.865 )
    ( 96.38, 334.672 )
    ( 98.58, 186.047 )
    ( 98.66, 176.243 )
    ( 99.41, 102.733 )
    ( 99.63, 69.099 )
    ( 99.67, 62.617 )
    ( 99.76, 55.916 )
};
\addplot[line width=0.15mm,color=norm3,mark=square,mark size=0.5mm]
plot coordinates {
    ( 75.55, 584.454 )
    ( 85.76, 302.572 )
    ( 92.68, 202.963 )
    ( 96.66, 137.646 )
    ( 98.58, 104.888 )
    ( 98.7, 154.226 )
    ( 99.46, 80.276 )
    ( 99.66, 62.123 )
    ( 99.67, 58.028 )
    ( 99.75, 51.763 )
};
\addplot[line width=0.15mm,color=norm4,mark=otimes,mark size=0.5mm]
plot coordinates {
    ( 71.95, 59.927 )
    ( 79.59, 41.254 )
    ( 85.33, 30.0 )
    ( 88.25, 26.584 )
    ( 88.94, 29.983 )
    ( 89.62, 26.392 )
};
\addplot[line width=0.15mm,color=norm5,mark=star,mark size=0.5mm]
plot coordinates {
    ( 76.16, 160.875 )
    ( 83.81, 94.126 )
    ( 88.52, 66.525 )
    ( 91.49, 46.179 )
    ( 92.66, 43.326 )
    ( 92.92, 54.544 )
    ( 93.22, 50.284 )
};

\end{axis}
\end{tikzpicture}\hspace{0.5mm}
\begin{tikzpicture}[scale=1]
\begin{axis}[
height=\columnwidth/2.60,
width=\columnwidth/1.80,
xlabel=recall@10(\%),
ylabel=Qps,
ymode=log,
title={ Ave. Select 18\% (\textbf{Zipf},$|\Sigma|$=12)},
label style={font=\scriptsize},
tick label style={font=\scriptsize},
title style={font=\scriptsize},
title style={yshift=-2.5mm},
ymajorgrids=true,
xmajorgrids=true,
grid style=dashed,
]
\addplot[line width=0.15mm,color=norm2,mark=triangle,mark size=0.5mm]
plot coordinates {
    ( 77.48, 1076.43 )
    ( 86.86, 584.112 )
    ( 93.38, 325.627 )
    ( 96.73, 180.278 )
    ( 98.41, 98.551 )
    ( 98.56, 92.687 )
    ( 99.36, 53.706 )
    ( 99.64, 36.438 )
    ( 99.69, 32.983 )
    ( 99.77, 29.465 )
};
\addplot[line width=0.15mm,color=norm3,mark=square,mark size=0.5mm]
plot coordinates {
    ( 80.83, 449.236 )
    ( 89.11, 265.393 )
    ( 94.44, 141.143 )
    ( 97.4, 88.3 )
    ( 98.68, 33.392 )
    ( 98.8, 23.85 )
    ( 99.46, 30.24 )
    ( 99.7, 21.357 )
    ( 99.73, 19.418 )
    ( 99.76, 17.332 )
};
\addplot[line width=0.15mm,color=norm4,mark=otimes,mark size=0.5mm]
plot coordinates {
    ( 77.14, 79.974 )
    ( 81.64, 42.503 )
    ( 84.15, 29.358 )
    ( 84.96, 23.754 )
    ( 85.45, 12.018 )
};
\addplot[line width=0.15mm,color=norm5,mark=star,mark size=0.5mm]
plot coordinates {
    ( 77.88, 229.674 )
    ( 82.61, 125.125 )
    ( 83.08, 146.37 )
    ( 85.79, 69.793 )
    ( 87.5, 58.928 )
    ( 87.7, 57.307 )
};

\end{axis}
\end{tikzpicture}\hspace{0.5mm}
}
\\
\subfloat[DEEP (Low-Selectivity)]{\vgap
\begin{tikzpicture}[scale=1]
\begin{axis}[
height=\columnwidth/2.60,
width=\columnwidth/1.80,
xlabel=recall@10(\%),
ylabel=Qps,
ymode=log,
title={ Ave. Select 11\% (\textbf{Zipf},$|\Sigma|$=24)},
label style={font=\scriptsize},
tick label style={font=\scriptsize},
title style={font=\scriptsize},
title style={yshift=-2.5mm},
ymajorgrids=true,
xmajorgrids=true,
grid style=dashed,
]
\addplot[line width=0.15mm,color=norm2,mark=triangle,mark size=0.5mm]
plot coordinates {
    ( 80.88, 303.214 )
    ( 89.82, 186.637 )
    ( 94.84, 105.341 )
    ( 97.69, 54.969 )
    ( 99.04, 26.882 )
    ( 99.14, 24.991 )
    ( 99.56, 13.24 )
    ( 99.81, 8.811 )
    ( 99.83, 7.953 )
    ( 99.86, 7.08 )
};
\addplot[line width=0.15mm,color=norm3,mark=square,mark size=0.5mm]
plot coordinates {
    ( 86.95, 54.954 )
    ( 92.86, 24.119 )
    ( 96.41, 21.573 )
    ( 98.28, 13.293 )
    ( 99.21, 7.058 )
    ( 99.27, 6.626 )
    ( 99.63, 3.693 )
    ( 99.77, 2.467 )
    ( 99.81, 2.218 )
    ( 99.86, 1.979 )
};
\addplot[line width=0.15mm,color=norm4,mark=otimes,mark size=0.5mm]
plot coordinates {
    ( 70.42, 29.727 )
    ( 72.71, 26.555 )
    ( 73.22, 29.888 )
    ( 73.75, 26.611 )
};
\addplot[line width=0.15mm,color=norm5,mark=star,mark size=0.5mm]
plot coordinates {
    ( 70.45, 48.96 )
    ( 70.81, 63.613 )
    ( 71.21, 57.343 )
};

\end{axis}
\end{tikzpicture}\hspace{0.5mm}
\begin{tikzpicture}[scale=1]
\begin{axis}[
height=\columnwidth/2.60,
width=\columnwidth/1.80,
xlabel=recall@10(\%),
ylabel=Qps,
ymode=log,
title={ Ave. Select 9\% (\textbf{Zipf},$|\Sigma|$=32)},
label style={font=\scriptsize},
tick label style={font=\scriptsize},
title style={font=\scriptsize},
title style={yshift=-2.5mm},
ymajorgrids=true,
xmajorgrids=true,
grid style=dashed,
]
\addplot[line width=0.15mm,color=norm2,mark=triangle,mark size=0.5mm]
plot coordinates {
    ( 80.53, 618.047 )
    ( 88.43, 265.816 )
    ( 93.25, 154.847 )
    ( 95.54, 83.119 )
    ( 96.69, 45.463 )
    ( 96.8, 42.772 )
    ( 97.19, 24.698 )
    ( 97.37, 16.216 )
    ( 97.41, 14.653 )
    ( 97.43, 12.918 )
};
\addplot[line width=0.15mm,color=norm3,mark=square,mark size=0.5mm]
plot coordinates {
    ( 86.27, 67.272 )
    ( 91.63, 24.628 )
    ( 94.65, 17.472 )
    ( 96.11, 12.639 )
    ( 96.92, 7.001 )
    ( 96.99, 6.606 )
    ( 97.3, 3.804 )
    ( 97.44, 2.553 )
    ( 97.47, 2.303 )
    ( 97.48, 2.05 )
};
\addplot[line width=0.15mm,color=norm4,mark=otimes,mark size=0.5mm]
plot coordinates {
};

\end{axis}
\end{tikzpicture}\hspace{0.5mm}
}

\caption{The Test of Scalability}\label{fig:scale}
\end{footnotesize}
\end{figure}

\begin{figure}[!t]
    \centering
    \includegraphics[width=0.99\linewidth]{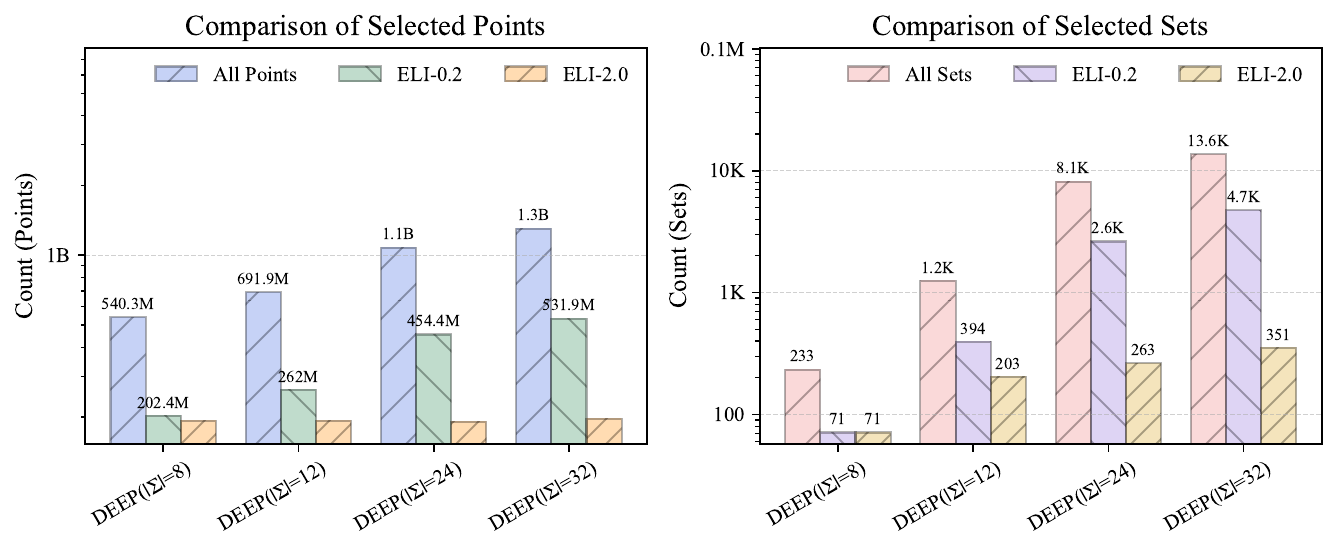}
\caption{The Index Select Statistic of Large-Scale Data}
    \label{fig:large-Selection}
\end{figure}

\begin{figure}[t]
\centering
\begin{footnotesize}
\begin{tikzpicture}
    \begin{customlegend}[legend columns=5,
        legend entries={$\UNG$,$\ELI$-0.2},
        legend style={at={(0.45,1.15)},anchor=north,draw=none,font=\scriptsize,column sep=0.1cm}]
    \addlegendimage{line width=0.15mm,color=norm1,mark=o,mark size=0.5mm}
    \addlegendimage{line width=0.15mm,color=norm2,mark=triangle,mark size=0.5mm}
    \end{customlegend}
\end{tikzpicture}
\subfloat[SIFT $|\Sigma|$=64,128]{\vgap
\begin{tikzpicture}[scale=1]
\begin{axis}[
height=\columnwidth/2.60,
width=\columnwidth/1.80,
xlabel=recall@10(\%),
ylabel=Qps,
ymode=log,
title={ Ave. Select 6.3\% (\textbf{Zipf},$|\Sigma|$=64)},
label style={font=\scriptsize},
tick label style={font=\scriptsize},
title style={font=\scriptsize},
title style={yshift=-2.5mm},
ymajorgrids=true,
xmajorgrids=true,
grid style=dashed,
]
\addplot[line width=0.15mm,color=norm1,mark=o,mark size=0.5mm]
plot coordinates {
    ( 78.877, 1486.55 )
    ( 83.607, 1206.71 )
    ( 88.906, 853.024 )
    ( 90.03, 771.724 )
    ( 94.517, 423.191 )
    ( 95.614, 339.455 )
    ( 96.307, 280.183 )
    ( 97.536, 174.685 )
    ( 97.993, 136.038 )
    ( 98.957, 60.63 )
    ( 99.651, 19.325 )
    ( 99.828, 10.74 )
};
\addplot[line width=0.15mm,color=norm2,mark=triangle,mark size=0.5mm]
plot coordinates {
    ( 96.256, 7633.59 )
    ( 98.729, 5339.03 )
    ( 99.689, 3480.68 )
    ( 99.942, 1813.89 )
};

\end{axis}
\end{tikzpicture}\hspace{0.5mm}
\begin{tikzpicture}[scale=1]
\begin{axis}[
height=\columnwidth/2.60,
width=\columnwidth/1.80,
xlabel=recall@10(\%),
ylabel=Qps,
ymode=log,
title={ Ave. Select 4.7\% (\textbf{Zipf},$|\Sigma|$=128)},
label style={font=\scriptsize},
tick label style={font=\scriptsize},
title style={font=\scriptsize},
title style={yshift=-2.5mm},
ymajorgrids=true,
xmajorgrids=true,
grid style=dashed,
]
\addplot[line width=0.15mm,color=norm1,mark=o,mark size=0.5mm]
plot coordinates {
    ( 75.904, 1365.19 )
    ( 80.727, 1161.44 )
    ( 86.414, 878.889 )
    ( 87.791, 803.019 )
    ( 92.921, 472.077 )
    ( 94.174, 383.098 )
    ( 95.025, 318.644 )
    ( 96.593, 201.098 )
    ( 97.119, 157.176 )
    ( 98.418, 68.861 )
    ( 99.418, 21.273 )
    ( 99.686, 11.628 )
};
\addplot[line width=0.15mm,color=norm2,mark=triangle,mark size=0.5mm]
plot coordinates {
    ( 97.106, 9199.63 )
    ( 99.03, 6631.3 )
    ( 99.743, 4448.4 )
    ( 99.946, 2716.65 )
};

\end{axis}
\end{tikzpicture}\hspace{0.5mm}
}
\\
\subfloat[SIFT $|\Sigma|$=256,512]{\vgap
\begin{tikzpicture}[scale=1]
\begin{axis}[
height=\columnwidth/2.60,
width=\columnwidth/1.80,
xlabel=recall@10(\%),
ylabel=Qps,
ymode=log,
title={ Ave. Select 3.7\% (\textbf{Zipf},$|\Sigma|$=256)},
label style={font=\scriptsize},
tick label style={font=\scriptsize},
title style={font=\scriptsize},
title style={yshift=-2.5mm},
ymajorgrids=true,
xmajorgrids=true,
grid style=dashed,
]
\addplot[line width=0.15mm,color=norm1,mark=o,mark size=0.5mm]
plot coordinates {
    ( 74.83, 1288.83 )
    ( 79.621, 1135.2 )
    ( 85.222, 895.656 )
    ( 86.437, 825.832 )
    ( 91.883, 506.637 )
    ( 93.226, 418.41 )
    ( 94.188, 349.675 )
    ( 95.77, 219.375 )
    ( 96.399, 171.35 )
    ( 98.026, 73.658 )
    ( 99.219, 22.331 )
    ( 99.584, 11.422 )
};
\addplot[line width=0.15mm,color=norm2,mark=triangle,mark size=0.5mm]
plot coordinates {
    ( 97.57, 10319.9 )
    ( 99.18, 7788.16 )
    ( 99.818, 5096.84 )
    ( 99.961, 3591.95 )
};

\end{axis}
\end{tikzpicture}\hspace{0.5mm}
\begin{tikzpicture}[scale=1]
\begin{axis}[
height=\columnwidth/2.60,
width=\columnwidth/1.80,
xlabel=recall@10(\%),
ylabel=Qps,
ymode=log,
title={ Ave. Select 3.5\% (\textbf{Zipf},$|\Sigma|$=512)},
label style={font=\scriptsize},
tick label style={font=\scriptsize},
title style={font=\scriptsize},
title style={yshift=-2.5mm},
ymajorgrids=true,
xmajorgrids=true,
grid style=dashed,
]
\addplot[line width=0.15mm,color=norm1,mark=o,mark size=0.5mm]
plot coordinates {
    ( 74.176, 1179.8 )
    ( 78.767, 1061.23 )
    ( 84.407, 851.209 )
    ( 85.75, 800.256 )
    ( 91.22, 507.099 )
    ( 92.561, 416.354 )
    ( 93.474, 346.272 )
    ( 95.172, 218.079 )
    ( 95.825, 169.019 )
    ( 97.506, 70.334 )
    ( 98.906, 22.042 )
    ( 99.356, 11.423 )
};
\addplot[line width=0.15mm,color=norm2,mark=triangle,mark size=0.5mm]
plot coordinates {
    ( 97.879, 10729.6 )
    ( 99.306, 8130.08 )
    ( 99.841, 5740.53 )
    ( 99.965, 3809.52 )
};

\end{axis}
\end{tikzpicture}\hspace{0.5mm}
}
\caption{Varying the Alphabet Size $|\Sigma|$}\label{fig:recall-var-large}

\end{footnotesize}
\end{figure}
\begin{figure}[t]
\centering
\begin{footnotesize}
\begin{tikzpicture}
    \begin{customlegend}[legend columns=5,
        legend entries={$\UNG$,$\ELI$-0.2,$\ELI$-2.0,$\ACORN$-1,$\ACORN$-$\gamma$},
        legend style={at={(0.45,1.15)},anchor=north,draw=none,font=\scriptsize,column sep=0.1cm}]
    \addlegendimage{line width=0.15mm,color=norm1,mark=o,mark size=0.5mm}
    \addlegendimage{line width=0.15mm,color=norm2,mark=triangle,mark size=0.5mm}
    \addlegendimage{line width=0.15mm,color=norm3,mark=square,mark size=0.5mm}
    \addlegendimage{line width=0.15mm,color=norm4,mark=otimes,mark size=0.5mm}
    \addlegendimage{line width=0.15mm,color=norm5,mark=star,mark size=0.5mm}
    \end{customlegend}
\end{tikzpicture}
\subfloat[MSMARCO-Fewer Processes]{\vgap
\begin{tikzpicture}[scale=1]
\begin{axis}[
height=\columnwidth/2.50,
width=\columnwidth/1.80,
xlabel=recall@10(\%),
ylabel=QPS,
title={thread@4},
label style={font=\scriptsize},
tick label style={font=\scriptsize},
title style={font=\scriptsize},
title style={yshift=-2.5mm},
ymajorgrids=true,
xmajorgrids=true,
grid style=dashed,
ymode=log,
]
\addplot[line width=0.15mm,color=norm1,mark=o,mark size=0.5mm]
plot coordinates {
    ( 72.18, 2824.9 )
    ( 91.16, 1026.7 )
    ( 94.13, 813.0 )
    ( 96.86, 458.9 )
    ( 97.07, 405.8 )
    ( 98.58, 200.8 )
};
\addplot[line width=0.15mm,color=norm2,mark=triangle,mark size=0.5mm]
plot coordinates {
    ( 89.88, 4098.4 )
    ( 95.4, 2525.3 )
    ( 97.9, 1319.3 )
    ( 99.13, 783.1 )
    ( 99.69, 492.4 )
    ( 99.7, 509.2 )
    ( 99.92, 333.4 )
};
\addplot[line width=0.15mm,color=norm3,mark=square,mark size=0.5mm]
plot coordinates {
    ( 90.92, 4032.3 )
    ( 95.93, 2320.2 )
    ( 98.26, 1353.2 )
    ( 99.33, 801.9 )
    ( 99.72, 417.0 )
    ( 99.73, 385.7 )
    ( 99.92, 211.5 )
};
\addplot[line width=0.15mm,color=norm4,mark=otimes,mark size=0.5mm]
plot coordinates {
    ( 72.26, 1642.0 )
    ( 80.08, 915.8 )
    ( 84.9, 452.3 )
    ( 85.29, 472.1 )
    ( 87.71, 301.2 )
    ( 88.82, 190.4 )
    ( 88.99, 176.0 )
    ( 89.19, 151.2 )
};
\addplot[line width=0.15mm,color=norm5,mark=star,mark size=0.5mm]
plot coordinates {
    ( 76.77, 3225.8 )
    ( 82.5, 1474.9 )
    ( 85.73, 786.2 )
    ( 85.89, 783.1 )
    ( 87.24, 507.6 )
    ( 87.86, 343.5 )
    ( 87.95, 286.0 )
    ( 88.02, 213.5 )
};

\end{axis}
\end{tikzpicture}\hspace{0.5mm}
\begin{tikzpicture}[scale=1]
\begin{axis}[
height=\columnwidth/2.50,
width=\columnwidth/1.80,
xlabel=recall@10(\%),
ylabel=QPS,
title={thread@8},
label style={font=\scriptsize},
tick label style={font=\scriptsize},
title style={font=\scriptsize},
title style={yshift=-2.5mm},
ymajorgrids=true,
xmajorgrids=true,
grid style=dashed,
ymode=log,
]
\addplot[line width=0.15mm,color=norm1,mark=o,mark size=0.5mm]
plot coordinates {
    ( 72.18, 9803.9 )
    ( 91.16, 2288.3 )
    ( 94.13, 1540.8 )
    ( 96.86, 909.1 )
    ( 97.07, 772.8 )
    ( 98.58, 348.2 )
};
\addplot[line width=0.15mm,color=norm2,mark=triangle,mark size=0.5mm]
plot coordinates {
    ( 89.88, 8849.6 )
    ( 95.4, 5263.2 )
    ( 97.9, 2785.5 )
    ( 99.13, 1536.1 )
    ( 99.69, 1105.0 )
    ( 99.7, 910.7 )
    ( 99.92, 615.0 )
};
\addplot[line width=0.15mm,color=norm3,mark=square,mark size=0.5mm]
plot coordinates {
    ( 90.92, 8064.5 )
    ( 95.93, 4504.5 )
    ( 98.26, 2341.9 )
    ( 99.33, 1272.3 )
    ( 99.72, 729.9 )
    ( 99.73, 622.3 )
    ( 99.92, 323.0 )
};
\addplot[line width=0.15mm,color=norm4,mark=otimes,mark size=0.5mm]
plot coordinates {
    ( 72.26, 3144.7 )
    ( 80.08, 1733.1 )
    ( 84.9, 1004.0 )
    ( 85.29, 885.0 )
    ( 87.71, 517.1 )
    ( 88.82, 326.6 )
    ( 88.99, 288.2 )
    ( 89.19, 274.3 )
};
\addplot[line width=0.15mm,color=norm5,mark=star,mark size=0.5mm]
plot coordinates {
    ( 76.77, 6410.3 )
    ( 82.5, 3484.3 )
    ( 85.73, 1901.1 )
    ( 85.89, 2057.6 )
    ( 87.24, 1150.8 )
    ( 87.86, 700.8 )
    ( 87.95, 555.2 )
    ( 88.02, 446.4 )
};

\end{axis}
\end{tikzpicture}\hspace{0.5mm}
}
\\
\subfloat[MSMARCO-More Processes]{\vgap
\begin{tikzpicture}[scale=1]
\begin{axis}[
height=\columnwidth/2.50,
width=\columnwidth/1.80,
xlabel=recall@10(\%),
ylabel=QPS,
title={thread@16},
label style={font=\scriptsize},
tick label style={font=\scriptsize},
title style={font=\scriptsize},
title style={yshift=-2.5mm},
ymajorgrids=true,
xmajorgrids=true,
grid style=dashed,
ymode=log,
]
\addplot[line width=0.15mm,color=norm1,mark=o,mark size=0.5mm]
plot coordinates {
    ( 72.19, 10309.3 )
    ( 91.16, 2445.0 )
    ( 94.13, 1610.3 )
    ( 96.86, 979.4 )
    ( 97.07, 979.4 )
    ( 98.58, 613.5 )
};
\addplot[line width=0.15mm,color=norm2,mark=triangle,mark size=0.5mm]
plot coordinates {
    ( 89.88, 11494.3 )
    ( 95.4, 6896.6 )
    ( 97.9, 4237.3 )
    ( 99.13, 2564.1 )
    ( 99.69, 1225.5 )
    ( 99.7, 1246.9 )
    ( 99.92, 877.2 )
};
\addplot[line width=0.15mm,color=norm3,mark=square,mark size=0.5mm]
plot coordinates {
    ( 90.92, 11111.1 )
    ( 95.93, 6329.1 )
    ( 98.26, 3759.4 )
    ( 99.33, 2057.6 )
    ( 99.72, 1168.2 )
    ( 99.73, 1046.0 )
    ( 99.92, 630.1 )
};
\addplot[line width=0.15mm,color=norm4,mark=otimes,mark size=0.5mm]
plot coordinates {
    ( 72.26, 5263.2 )
    ( 80.08, 3003.0 )
    ( 84.9, 1683.5 )
    ( 85.29, 1730.1 )
    ( 87.71, 974.7 )
    ( 88.82, 648.5 )
    ( 88.99, 552.5 )
    ( 89.19, 485.4 )
};
\addplot[line width=0.15mm,color=norm5,mark=star,mark size=0.5mm]
plot coordinates {
    ( 76.77, 8474.6 )
    ( 82.5, 4830.9 )
    ( 85.73, 2673.8 )
    ( 85.89, 2433.1 )
    ( 87.24, 1440.9 )
    ( 87.86, 1030.9 )
    ( 87.95, 941.6 )
    ( 88.02, 837.5 )
};

\end{axis}
\end{tikzpicture}\hspace{0.5mm}
\begin{tikzpicture}[scale=1]
\begin{axis}[
height=\columnwidth/2.50,
width=\columnwidth/1.80,
xlabel=recall@10(\%),
ylabel=QPS,
title={thread@32},
label style={font=\scriptsize},
tick label style={font=\scriptsize},
title style={font=\scriptsize},
title style={yshift=-2.5mm},
ymajorgrids=true,
xmajorgrids=true,
grid style=dashed,
ymode=log,
]
\addplot[line width=0.15mm,color=norm1,mark=o,mark size=0.5mm]
plot coordinates {
    ( 72.17, 16949.2 )
    ( 91.16, 4184.1 )
    ( 94.13, 2907.0 )
    ( 96.86, 1742.2 )
    ( 97.07, 1582.3 )
    ( 98.58, 826.4 )
};
\addplot[line width=0.15mm,color=norm2,mark=triangle,mark size=0.5mm]
plot coordinates {
    ( 89.88, 13157.9 )
    ( 95.4, 6944.4 )
    ( 97.9, 3984.1 )
    ( 99.13, 2688.2 )
    ( 99.69, 1227.0 )
    ( 99.7, 1280.4 )
    ( 99.92, 742.4 )
};
\addplot[line width=0.15mm,color=norm3,mark=square,mark size=0.5mm]
plot coordinates {
    ( 90.92, 7299.3 )
    ( 95.93, 5618.0 )
    ( 98.26, 2915.4 )
    ( 99.33, 1451.4 )
    ( 99.72, 1176.5 )
    ( 99.73, 991.1 )
    ( 99.92, 736.9 )
};
\addplot[line width=0.15mm,color=norm4,mark=otimes,mark size=0.5mm]
plot coordinates {
    ( 72.26, 7194.2 )
    ( 80.08, 4587.2 )
    ( 84.9, 2590.7 )
    ( 85.29, 2320.2 )
    ( 87.71, 1597.4 )
    ( 88.82, 1003.0 )
    ( 88.99, 903.3 )
    ( 89.19, 810.4 )
};
\addplot[line width=0.15mm,color=norm5,mark=star,mark size=0.5mm]
plot coordinates {
    ( 76.77, 17241.4 )
    ( 82.5, 9803.9 )
    ( 85.73, 5555.6 )
    ( 85.89, 5128.2 )
    ( 87.24, 3012.1 )
    ( 87.86, 2053.4 )
    ( 87.95, 1883.2 )
    ( 88.02, 1618.1 )
};

\end{axis}
\end{tikzpicture}\hspace{0.5mm}
}
\vgap\vgap
\caption{Varying the Thread Number (on MSMARCO)}\label{fig:recall-var-threads}

\end{footnotesize}
\end{figure}
\begin{figure}[t]
\centering
\begin{footnotesize}
\begin{tikzpicture}
    \begin{customlegend}[legend columns=5,
        legend entries={$\ELI$-2.0,$\ELI$-0.2,$\ELI$-0.5,$\ELI$-Opt},
        legend style={at={(0.45,1.15)},anchor=north,draw=none,font=\scriptsize,column sep=0.1cm}]
    \addlegendimage{line width=0.15mm,color=blue1,mark=square,mark size=0.5mm}
    \addlegendimage{line width=0.15mm,color=blue2,mark=triangle,mark size=0.5mm}
    \addlegendimage{line width=0.15mm,color=blue3,mark=o,mark size=0.5mm}
    \addlegendimage{line width=0.15mm,color=blue4,mark=otimes,mark size=0.5mm}
    \end{customlegend}
\end{tikzpicture}
\subfloat[MSMARC]{\vgap
\begin{tikzpicture}[scale=1]
\begin{axis}[
height=\columnwidth/2.50,
width=\columnwidth/1.80,
xlabel=recall@10(\%),
ylabel=QPS,
title={contain@32},
label style={font=\scriptsize},
tick label style={font=\scriptsize},
title style={font=\scriptsize},
title style={yshift=-2.5mm},
ymajorgrids=true,
xmajorgrids=true,
grid style=dashed,
ymode=log,
]
\addplot[line width=0.15mm,color=blue3,mark=o,mark size=0.5mm]
plot coordinates {
    ( 88.9, 3012.1 )
    ( 93.85, 2207.5 )
    ( 95.46, 1792.1 )
    ( 96.16, 1555.2 )
    ( 97.06, 1022.5 )
    ( 97.11, 922.5 )
    ( 97.44, 635.3 )
    ( 97.5, 568.2 )
};
\addplot[line width=0.15mm,color=blue2,mark=triangle,mark size=0.5mm]
plot coordinates {
    ( 90.08, 2227.2 )
    ( 94.26, 1529.0 )
    ( 95.82, 1169.6 )
    ( 96.42, 1014.2 )
    ( 97.04, 612.7 )
    ( 97.14, 548.8 )
    ( 97.45, 372.9 )
    ( 97.49, 237.5 )
};
\addplot[line width=0.15mm,color=blue1,mark=square,mark size=0.5mm]
plot coordinates {
    ( 92.68, 954.2 )
    ( 95.67, 556.8 )
    ( 96.83, 312.8 )
    ( 97.32, 172.4 )
    ( 97.55, 96.2 )
    ( 97.59, 90.3 )
    ( 97.67, 54.3 )
    ( 97.7, 37.8 )
};
\addplot[line width=0.15mm,color=blue4,mark=otimes,mark size=0.5mm]
plot coordinates {
    ( 87.79, 2747.3 )
    ( 93.1, 2045.0 )
    ( 94.92, 1712.3 )
    ( 95.77, 1642.0 )
    ( 96.89, 1010.1 )
    ( 96.97, 1020.4 )
    ( 97.44, 701.3 )
    ( 97.5, 666.7 )
};

\end{axis}
\end{tikzpicture}\hspace{0.5mm}
}
\subfloat[OpenAI-1536]{\vgap

\begin{tikzpicture}[scale=1]
\begin{axis}[
height=\columnwidth/2.50,
width=\columnwidth/1.80,
xlabel=recall@10(\%),
ylabel=QPS,
title={contain@32},
label style={font=\scriptsize},
tick label style={font=\scriptsize},
title style={font=\scriptsize},
title style={yshift=-2.5mm},
ymajorgrids=true,
xmajorgrids=true,
grid style=dashed,
ymode=log,
]
\addplot[line width=0.15mm,color=blue3,mark=o,mark size=0.5mm]
plot coordinates {
    ( 89.66, 2100.8 )
    ( 94.3, 1605.1 )
    ( 96.02, 1287.0 )
    ( 96.62, 1108.7 )
    ( 97.35, 709.2 )
    ( 97.4, 631.7 )
    ( 97.55, 434.2 )
    ( 97.61, 386.0 )
};
\addplot[line width=0.15mm,color=blue2,mark=triangle,mark size=0.5mm]
plot coordinates {
    ( 90.88, 1562.5 )
    ( 94.57, 1103.7 )
    ( 96.13, 836.8 )
    ( 96.58, 717.4 )
    ( 97.35, 430.3 )
    ( 97.42, 389.1 )
    ( 97.56, 259.5 )
    ( 97.59, 230.7 )
};
\addplot[line width=0.15mm,color=blue1,mark=square,mark size=0.5mm]
plot coordinates {
    ( 93.27, 554.9 )
    ( 95.73, 325.3 )
    ( 97.02, 186.4 )
    ( 97.43, 104.5 )
    ( 97.61, 59.1 )
    ( 97.62, 55.8 )
    ( 97.7, 33.6 )
    ( 97.71, 23.7 )
    ( 97.72, 21.7 )
};
\addplot[line width=0.15mm,color=blue4,mark=otimes,mark size=0.5mm]
plot coordinates {
    ( 88.51, 2325.6 )
    ( 93.6, 1814.9 )
    ( 95.63, 1481.5 )
    ( 96.38, 1290.3 )
    ( 97.25, 858.4 )
    ( 97.38, 771.6 )
    ( 97.46, 543.5 )
    ( 97.56, 482.6 )
};

\end{axis}
\end{tikzpicture}\hspace{0.5mm}
}

\vgap\caption{The Comparison with the Exhaustive Method}\label{fig:compare-to-optimal}

\end{footnotesize}
\end{figure}
\begin{figure}[!t]
\centering
\begin{footnotesize}
\begin{tikzpicture}
    \begin{customlegend}[legend columns=2,
        legend entries={$\ELI$-0.2,$\ELI$-2.0},
        legend style={at={(0.45,1.15)},anchor=north,draw=none,font=\scriptsize,column sep=0.1cm}]
    \addlegendimage{line width=0.15mm,color=norm2,mark=triangle,mark size=0.5mm}
    \addlegendimage{line width=0.15mm,color=norm3,mark=square,mark size=0.5mm}
    \end{customlegend}
\end{tikzpicture}
\\[-\lineskip]

\subfloat[SIFT]{\vgap

\begin{tikzpicture}[scale=1]
\begin{axis}[
height=\columnwidth/2.60,
width=\columnwidth/1.80,
xlabel=Ratio,
ylabel=Qps,
ymode=log,
xtick={1.0, 1.001, 1.002, 1.003},
xticklabels={1.0, 1.001, 1.002, 1.003},
title={ Ave Select 18\% (\textbf{Zipf},$|\Sigma|$=12)},
label style={font=\scriptsize},
tick label style={font=\scriptsize},
title style={font=\scriptsize},
title style={yshift=-2.5mm},
ymajorgrids=true,
xmajorgrids=true,
grid style=dashed,
]
\addplot[line width=0.15mm,color=norm2,mark=o,mark size=0.5mm]
plot coordinates {
    ( 1.00427, 4616.81 )
    ( 1.0014, 2637.13 )
    ( 1.00039, 999.3 )
    ( 1.00004, 760.341 )
    ( 1.0, 422.797 )
};
\addplot[line width=0.15mm,color=norm3,mark=triangle,mark size=0.5mm]
plot coordinates {
    ( 1.00237, 3245.7 )
    ( 1.00065, 1883.95 )
    ( 1.00016, 933.881 )
    ( 1.00003, 633.032 )
    ( 1.0, 363.003 )
};

\end{axis}
\end{tikzpicture}\hspace{0.5mm}

\begin{tikzpicture}[scale=1]
\begin{axis}[
height=\columnwidth/2.60,
width=\columnwidth/1.80,
xlabel=Ratio,
ylabel=Qps,
ymode=log,
xtick={1.0, 1.001, 1.002, 1.003},
xticklabels={1.0, 1.001, 1.002, 1.003},
title={ Ave Select 11\% (\textbf{Zipf},$|\Sigma|$=24)},
label style={font=\scriptsize},
tick label style={font=\scriptsize},
title style={font=\scriptsize},
title style={yshift=-2.5mm},
ymajorgrids=true,
xmajorgrids=true,
grid style=dashed,
]
\addplot[line width=0.15mm,color=norm2,mark=o,mark size=0.5mm]
plot coordinates {
    ( 1.00201, 7818.61 )
    ( 1.00062, 5175.98 )
    ( 1.00014, 3217.5 )
    ( 1.00004, 1955.03 )
    ( 1.00002, 1164.69 )
};
\addplot[line width=0.15mm,color=norm3,mark=triangle,mark size=0.5mm]
plot coordinates {
    ( 1.00137, 3277.61 )
    ( 1.00041, 1927.9 )
    ( 1.0001, 1128.41 )
    ( 1.00003, 663.13 )
};

\end{axis}
\end{tikzpicture}\hspace{0.5mm}
}
\\
\subfloat[Arxiv]{\vgap

\begin{tikzpicture}[scale=1]
\begin{axis}[
height=\columnwidth/2.60,
width=\columnwidth/1.80,
xlabel=Ratio,
ylabel=Qps,
ymode=log,
xtick={1.0, 1.0003, 1.0006, 1.001},
xticklabels={1.0, 1.0003, 1.0006, 1.001},
title={ Ave Select 18\% (\textbf{Zipf},$|\Sigma|$=12)},
label style={font=\scriptsize},
tick label style={font=\scriptsize},
title style={font=\scriptsize},
title style={yshift=-2.5mm},
ymajorgrids=true,
xmajorgrids=true,
grid style=dashed,
]
\addplot[line width=0.15mm,color=norm2,mark=o,mark size=0.5mm]
plot coordinates {
    ( 1.00142, 706.215 )
    ( 1.00011, 506.073 )
    ( 1.00003, 357.91 )
    ( 1.0, 235.793 )
};
\addplot[line width=0.15mm,color=norm3,mark=triangle,mark size=0.5mm]
plot coordinates {
    ( 1.00143, 906.618 )
    ( 1.00048, 703.73 )
    ( 1.00009, 488.281 )
    ( 1.00002, 319.387 )
    ( 1.0, 187.617 )
};

\end{axis}
\end{tikzpicture}\hspace{0.5mm}

\begin{tikzpicture}[scale=1]
\begin{axis}[
height=\columnwidth/2.60,
width=\columnwidth/1.80,
xlabel=Ratio,
ylabel=Qps,
ymode=log,
xtick={1.0, 1.0003, 1.0006, 1.001},
xticklabels={1.0, 1.0003, 1.0006, 1.001},
title={ Ave Select 11\% (\textbf{Zipf},$|\Sigma|$=24)},
label style={font=\scriptsize},
tick label style={font=\scriptsize},
title style={font=\scriptsize},
title style={yshift=-2.5mm},
ymajorgrids=true,
xmajorgrids=true,
grid style=dashed,
]
\addplot[line width=0.15mm,color=norm2,mark=o,mark size=0.5mm]
plot coordinates {
    ( 1.00099, 924.214 )
    ( 1.00024, 756.43 )
    ( 1.00005, 617.665 )
    ( 1.00002, 463.607 )
    ( 1.00001, 326.264 )
};
\addplot[line width=0.15mm,color=norm3,mark=triangle,mark size=0.5mm]
plot coordinates {
    ( 1.00072, 866.551 )
    ( 1.00017, 703.73 )
    ( 1.00005, 530.504 )
    ( 1.00001, 371.195 )
};

\end{axis}
\end{tikzpicture}\hspace{0.5mm}
}
\caption{ The Effect of the Distance Ratio}\label{fig:ratio}

\end{footnotesize}
\end{figure}
\begin{figure}[!t]
\centering
\begin{footnotesize}
\begin{tikzpicture}
    \begin{customlegend}[legend columns=2,
        legend entries={$\ELI$-0.2,$\ELI$-2.0},
        legend style={at={(0.45,1.15)},anchor=north,draw=none,font=\scriptsize,column sep=0.1cm}]
    \addlegendimage{line width=0.15mm,color=norm2,mark=triangle,mark size=0.5mm}
    \addlegendimage{line width=0.15mm,color=norm3,mark=square,mark size=0.5mm}
    \end{customlegend}
\end{tikzpicture}
\\[-\lineskip]

\subfloat[SIFT]{\vgap

\begin{tikzpicture}[scale=1]
\begin{axis}[
height=\columnwidth/2.60,
width=\columnwidth/1.80,
xlabel=recall@10(\%),
ylabel=Qps,
ymode=log,
title={ Ave Select 18\% (\textbf{Zipf},$|\Sigma|$=12)},
label style={font=\scriptsize},
tick label style={font=\scriptsize},
title style={font=\scriptsize},
title style={yshift=-2.5mm},
ymajorgrids=true,
xmajorgrids=true,
grid style=dashed,
]
\addplot[line width=0.15mm,color=norm2,mark=o,mark size=0.5mm]
plot coordinates {
    ( 91.212, 6920.42 )
    ( 96.571, 5042.86 )
    ( 98.86, 2897.71 )
    ( 99.515, 1654.81 )
    ( 99.635, 946.88 )
    ( 99.637, 893.495 )
    ( 99.655, 533.049 )
    ( 99.657, 369.467 )
};
\addplot[line width=0.15mm,color=norm3,mark=triangle,mark size=0.5mm]
plot coordinates {
    ( 92.411, 2936.0 )
    ( 97.069, 2305.74 )
    ( 98.956, 1302.59 )
    ( 99.52, 739.481 )
    ( 99.641, 417.502 )
    ( 99.642, 393.066 )
    ( 99.655, 235.627 )
    ( 99.657, 163.199 )
};

\end{axis}
\end{tikzpicture}\hspace{0.5mm}

\begin{tikzpicture}[scale=1]
\begin{axis}[
height=\columnwidth/2.60,
width=\columnwidth/1.80,
xlabel=recall@10(\%),
ylabel=Qps,
ymode=log,
title={ Ave Select 11\% (\textbf{Zipf},$|\Sigma|$=24)},
label style={font=\scriptsize},
tick label style={font=\scriptsize},
title style={font=\scriptsize},
title style={yshift=-2.5mm},
ymajorgrids=true,
xmajorgrids=true,
grid style=dashed,
]
\addplot[line width=0.15mm,color=norm2,mark=o,mark size=0.5mm]
plot coordinates {
    ( 93.155, 8665.51 )
    ( 97.101, 6230.53 )
    ( 98.768, 3815.34 )
    ( 99.24, 2284.15 )
    ( 99.343, 1333.51 )
    ( 99.348, 1273.07 )
    ( 99.363, 774.353 )
    ( 99.365, 549.269 )
};
\addplot[line width=0.15mm,color=norm3,mark=triangle,mark size=0.5mm]
plot coordinates {
    ( 95.133, 3680.53 )
    ( 97.886, 2163.57 )
    ( 98.993, 1248.13 )
    ( 99.287, 710.126 )
    ( 99.351, 401.171 )
    ( 99.352, 374.925 )
    ( 99.362, 228.206 )
    ( 99.363, 159.862 )
};

\end{axis}
\end{tikzpicture}\hspace{0.5mm}
}
\\
\subfloat[Arxiv]{\vgap
\begin{tikzpicture}[scale=1]
\begin{axis}[
height=\columnwidth/2.60,
width=\columnwidth/1.80,
xlabel=recall@10(\%),
ylabel=Qps,
ymode=log,
title={ Ave Select 18\% (\textbf{Zipf},$|\Sigma|$=12)},
label style={font=\scriptsize},
tick label style={font=\scriptsize},
title style={font=\scriptsize},
title style={yshift=-2.5mm},
ymajorgrids=true,
xmajorgrids=true,
grid style=dashed,
]
\addplot[line width=0.15mm,color=norm2,mark=o,mark size=0.5mm]
plot coordinates {
    ( 90.08, 1067.24 )
    ( 94.26, 943.396 )
    ( 97.78, 768.049 )
    ( 99.39, 568.828 )
    ( 99.84, 395.57 )
    ( 99.94, 256.016 )
};
\addplot[line width=0.15mm,color=norm3,mark=triangle,mark size=0.5mm]
plot coordinates {
    ( 90.95, 1010.1 )
    ( 94.49, 873.362 )
    ( 97.97, 551.268 )
    ( 99.52, 290.023 )
    ( 99.95, 194.856 )
};

\end{axis}
\end{tikzpicture}\hspace{0.5mm}
\begin{tikzpicture}[scale=1]
\begin{axis}[
height=\columnwidth/2.60,
width=\columnwidth/1.80,
xlabel=recall@10(\%),
ylabel=Qps,
ymode=log,
title={ Ave Select 11\% (\textbf{Zipf},$|\Sigma|$=24)},
label style={font=\scriptsize},
tick label style={font=\scriptsize},
title style={font=\scriptsize},
title style={yshift=-2.5mm},
ymajorgrids=true,
xmajorgrids=true,
grid style=dashed,
]
\addplot[line width=0.15mm,color=norm2,mark=o,mark size=0.5mm]
plot coordinates {
    ( 93.67, 786.164 )
    ( 98.74, 582.751 )
    ( 99.94, 470.146 )
};
\addplot[line width=0.15mm,color=norm3,mark=triangle,mark size=0.5mm]
plot coordinates {
    ( 94.31, 719.942 )
    ( 96.67, 598.086 )
    ( 99.74, 515.73 )
    ( 99.94, 368.732 )
};

\end{axis}
\end{tikzpicture}\hspace{0.5mm}
}

\caption{The Test of Maximum Inner Product Search}\label{fig:IP}

\end{footnotesize}
\end{figure}

\stitle{Exp-9: Varying Alphabet Size.}  
In Exp-1 we varied the alphabet size over a limited range;  
here we further test the search efficiency of different methods---particularly $\UNG$, $\ACORN$, and our $\ELI$-0.2---under much larger alphabets.  
We increase $|\Sigma|$ from 64 to 128, 256, and 512 to evaluate $\UNG$ and $\ELI$-0.2.  
The $\ACORN$ method fails to achieve 80\% recall at these scales and is therefore omitted from Fig.~\ref{fig:recall-var-large}.  
Results show that our method achieves nearly an $800\times$ speedup in search efficiency at 99\% recall when $|\Sigma| = 512$.  
This improvement arises because the fixed-efficiency design of $\ELI$-0.2 exploits the lower selectivity of large alphabets.  
In contrast, $\UNG$ suffers substantial degradation in search efficiency as $|\Sigma|$ grows.
We also compare index-build times.  
For $|\Sigma| = 512$, $\ELI$-0.2 builds the index in 155 seconds,  
whereas $\UNG$ requires 2,091 seconds—over $13\times$ longer---demonstrating the superior scalability of $\ELI$-0.2.

\stitle{Exp-10: Varying Thread Number.}  
During search, we apply parallelism to improve performance.  
To evaluate scalability, we vary the number of threads from 4 to 32 with $|\Sigma| = 12$  
and report results on the MSMARCO dataset (other datasets show similar trends).  
Figure~\ref{fig:recall-var-threads} presents the results.
Our methods achieve the highest search performance across all thread settings,  
demonstrating robustness in multi-threaded environments.  
Moreover, our architecture invokes only a single sub-index for each query,  
making it well suited for distributed systems—a direction we plan to explore in future work.

\stitle{Exp-11: Comparison with the Exhaustive Method.}  
Because our method selects a subset of indexes from the universal set corresponding to the query workload,  
we compare it with an \emph{exhaustive} method that builds indexes for \emph{all} possible query–label sets  
(i.e., selects the entire universal index set).  
We also introduce $\ELI$-0.5, an efficiency-constrained variant of $\mathsf{EIS}$  
with the elastic-factor threshold increased to $0.5$.
Figure~\ref{fig:compare-to-optimal} shows the search efficiency of  
$\ELI$-2.0, $\ELI$-0.2, $\ELI$-0.5, and the exhaustive method ($\ELI$-Opt)  
under a Zipf distribution with $|\Sigma|=32$ on the MSMARCO and OpenAI-1536 datasets.  
The results indicate that $\ELI$-0.2 and $\ELI$-0.5 achieve near-optimal search efficiency at high recall,  
closely matching $\ELI$-Opt.  
$\ELI$-2.0 delivers roughly $3\times$ lower Qps than the exhaustive method but requires far less space.
In terms of index size, $\ELI$-2.0, $\ELI$-0.2, $\ELI$-0.5, and $\ELI$-Opt consume  
383\,MB, 634\,MB, 812\,MB, and 1,280\,MB, respectively.  
Excluding the mandatory 192\,MB top index, $\ELI$-2.0 uses only about one-sixth of the space of the exhaustive approach  
while maintaining acceptable search performance.  
Moreover, $\ELI$-0.5 requires only about half the space required by $\ELI$-Opt
to achieve nearly identical search efficiency,  
highlighting the advantage of our methods in efficiency-oriented scenarios.

\stitle{Exp-12: Effect of Distance Ratio.}  
To further assess search performance, we measure the \emph{distance ratio}  
~\cite{C2LSH-2012-SIGMOD,SRP-STOC-2002,PMLSH-bolong-2020}:  the ratio of the average distance between a query and the returned objects  
to the average distance between the query and its ground-truth $k$ nearest neighbors.  
A ratio of $1.0$ indicates that the exact $k$NN results are found;  
larger ratios allow more flexibility in the returned objects.  
We report Qps at varying distance ratios on the SIFT and \textsc{Arxiv} datasets,  
with results shown in Fig.~\ref{fig:ratio}.  
Our methods require more time (i.e., achieve lower Qps) to reach very small distance ratios,  
but performance improves as the ratio increases.  
Notably, our methods often achieve an average distance ratio near $1.001$,  
substantially outperforming prior LSH-based approaches  
~\cite{C2LSH-2012-SIGMOD,LSH-Pstable-2004,QALSH-2015-VLDB,SRS-yifang-2014,PMLSH-bolong-2020},  
demonstrating our superiority from another perspective.

\stitle{Exp-13: Test of Maximum Inner Product Search.}  
Our default experiments use Euclidean distance to measure similarity between queries and base vectors.  
Here we replace Euclidean distance with the inner product and evaluate whether our method still performs well.  
This setting corresponds to Maximum Inner Product Search (MIPS).  
We test this extension on the SIFT and \textsc{Arxiv} datasets, with results shown in Fig.~\ref{fig:IP}.  
As illustrated, MIPS delivers computational performance comparable to the Euclidean metric under our framework,  
maintaining high Qps even at high recall.  
These findings indicate that our method is broadly applicable across common vector similarity measures  
and align with prior evidence~\cite{SymphonyQG-SIGMOD-2025,ann-benchmakrs}  
that graph-based indexes (e.g., $\HNSW$) sustain high search efficiency across different similarity metrics.


\end{document}